\documentclass[%
reprint,
superscriptaddress, 
notitlepage,
nofootinbib,
 amsmath,amssymb,
 aps, prx, longbibliography
]{revtex4-1}
\usepackage{etoolbox}
\usepackage{comment}
\usepackage{lmodern}
\usepackage{graphicx}
\usepackage{mathtools}
\usepackage{amsmath}
\usepackage{yhmath}
\usepackage{amssymb}
\usepackage[export]{adjustbox}
\usepackage{dcolumn}% Align table columns on decimal point
\usepackage{bm}% bold math
\usepackage{hyperref}% add hypertext capabilities
\hypersetup{linktocpage,colorlinks,citecolor={blue},pdfdisplaydoctitle=true,pdfpagemode=UseOutlines,bookmarksnumbered=true}
\usepackage{mathrsfs,dsfont}
\usepackage{xcolor}
\usepackage[normalem]{ulem}

\usepackage{orcidlink}
\usepackage{bbold} % identity operator
\usepackage{gensymb}
\usepackage{csquotes}
\usepackage{float}
\usepackage{braket}

\newcommand{\mb}[1]{\mathbf{#1}}
\renewcommand{\min}{\mathrm{Min}}
\renewcommand{\max}{\mathrm{Max}}
\renewcommand{\Re}{\mathrm{Re}}
\renewcommand{\Im}{\mathrm{Im}}

\definecolor{cbl}{rgb}{0,0,1}
 
\definecolor{crd}{rgb}{1,0,0}
 
\definecolor{Blue}{rgb}{0.0, 0.0, 0.5}

\newcommand{\be}{\begin{equation}}
\newcommand{\ee}{\end{equation}}
\newcommand{\bea}{\begin{eqnarray}}
\newcommand{\eea}{\end{eqnarray}}
\newcommand{\bsplit}{\begin{split}}
\newcommand{\esplit}{\end{split}}

\newcommand{\up}{\uparrow}
\newcommand{\down}{\downarrow}

\DeclarePairedDelimiter\floor{\lfloor}{\rfloor}

\graphicspath{{Figure/PNG/}{Figure/PDF/}{Figure/EPS/}{Figure/TEX/}{Figure/}}

\usepackage{cancel}

\begin{document}

\newcommand{\titleinfo}{
{Theory of robust quantum many-body scars in long-range interacting systems}
}

\title{\titleinfo}

\author{Alessio Lerose~\orcidlink{0000-0003-1555-5327}}
\email{alessio.lerose@gmail.com}
\thanks{Corresponding author.}
\affiliation{Department of Theoretical Physics, University of Geneva, Quai Ernest-Ansermet 30, 1205 Geneva, Switzerland}
\affiliation{Rudolf Peierls Centre for Theoretical Physics, Clarendon Laboratory, Oxford OX1 3PU, United Kingdom}

\author{Tommaso Parolini}
%\thanks{Present address: CGnal S.r.l. Corso Venezia 43, 20121 Milano, Italy}
\affiliation{The Abdus Salam International Centre for Theoretical Physics (ICTP), Strada Costiera 11, 34151 Trieste, Italy}
\affiliation{Scuola Internazionale di Studi Superiori Avanzati (SISSA), Via Bonomea 265, 34136 Trieste, Italy}

\author{Rosario Fazio~\orcidlink{0000-0002-7793-179X}}
\affiliation{The Abdus Salam International Centre for Theoretical Physics (ICTP), Strada Costiera 11, 34151 Trieste, Italy}
\affiliation{Dipartimento di Fisica, Universit\`a di Napoli “Federico II”, Monte S. Angelo, I-80126 Napoli, Italy}

\author{Dmitry A. Abanin}
\affiliation{Department of Theoretical Physics,
University of Geneva, Quai Ernest-Ansermet 30,
1205 Geneva, Switzerland}
\affiliation{Department of Physics, Princeton University, Princeton, New Jersey 08544, USA}

\author{Silvia Pappalardi~\orcidlink{0000-0001-6931-8736}}
\affiliation{Institut f\"ur Theoretische Physik, Universit\"at zu K\"oln, Z\"ulpicher Straße 77, 50937 K\"oln, Germany}

\begin{abstract}
{Quantum many-body scars (QMBS) are exceptional energy eigenstates of quantum many-body systems associated with violations of thermalization for special non-equilibrium initial states.
Their various systematic constructions require fine-tuning of local Hamiltonian parameters.}
In this work we demonstrate that %the setting of 
\emph{long-range} interacting quantum spin systems generically host \emph{robust} QMBS. We analyze spectral properties upon raising the power-law decay exponent $\alpha$ of spin-spin interactions from the solvable permutationally-symmetric limit $\alpha=0$. First, we numerically establish that despite spectral signatures of chaos appear for infinitesimal $\alpha$, the towers of $\alpha=0$ energy eigenstates with large collective spin are smoothly deformed as $\alpha$ is increased, and exhibit characteristic QMBS features.
To elucidate the nature and fate of these states in larger systems, we introduce an analytical approach based on mapping the spin Hamiltonian onto a relativistic quantum rotor non-linearly coupled to an extensive set of bosonic modes. 
We analitycally solve for the eigenstates of this interacting impurity model {by means of a novel polaron-type canonical transformation}, and show their self-consistent localization in large-spin sectors of the original Hamiltonian for $0<\alpha<d$ {(with $d = \text{spatial dimension of the lattice}$)}.
Our theory unveils the stability mechanism of such QMBS for arbitrary system size and predicts instances of its breakdown, e.g. near dynamical critical points or in presence of semiclassical chaos, which we verify numerically in long-range quantum Ising chains.
As a byproduct, we find a predictive criterion for presence or absence of heating under periodic driving for $0<\alpha<d$, beyond existing Floquet-prethermalization theorems. 
\end{abstract}

\date{\today}
\maketitle

\section{Introduction}

While open long-standing problems in condensed-matter physics contributed to fuel the development of quantum simulators~\cite{feynman2018simulating},
it quickly became clear that these {new experimental setups}  allow to investigate a separate set of fundamental questions on quantum matter out of equilibrium. 
A primary question concerns the mechanism of thermalization in isolated quantum many-particle systems -- a central tenet of macroscopic thermodynamics~\cite{polkovnikov2011colloquium}.  The possibility of violating this behavior has recently attracted  great interest, as it underpins the ongoing quest for design and coherent control of non-thermal states of matter. In this context  
several mechanisms have been discovered and characterized, including various instances of many-body localization phenomena~\cite{abanin2019colloquium,DeRoeck1,YaoQuasiMBLWithoutDisorder,Schiulaz1,Michailidis,SmithDisorderFreeLocalization,BrenesConfinementLGT,RefaelStarkMBL,SchultzStarkMBL,RobinsonNonthermalStatesShort,LeroseSuraceQuasilocalization}, prethermalization arising from weak integrability breaking~\cite{bertini2015prethermalization,mallayya2019prethermalization,AbaninRigorousPrethermalization,DeRoeckVerreet}, and Hilbert space fragmentation~\cite{SalaPRX20,KhemaniHermeleNandkishorePRB20}. 
\smallskip

Certain quantum many-body systems may fail to thermalize only when initialized in a specific class of simple far-from-equilibrium states, whereas all other initial states give ordinary thermalization dynamics. 
This phenomenon has been associated with the existence of anomalous highly excited energy eigenstates known as \emph{quantum many-body scars} (QMBS) --- a term coined in Ref.~\cite{turner2018weak} by analogy to wavefunction scarring in  single-particle semiclassical machanics~\cite{heller1984}. 
While distinct mechanisms fall under the umbrella of QMBS~\cite{
serbyn2021quantum,moudgalya2022quantum,chandran2023qmbsreview}, their common 
hallmark is a vanishing fraction of eigenstates violating the strong eigenstate thermalization hypothesis (ETH)~\cite{kim2014testing,dalessio2016from}, i.e. exhibiting non-thermal values of local observables and anomalously low entanglement, embedded in an otherwise ETH-compatible spectrum. 
The discovery of long-lived coherent oscillations in Rydberg-atom quantum simulations of dynamics far from equilibrium~\cite{BernienRydberg} triggered a huge theoretical interest in QMBS~\cite{turner2018quantum, shiraishi2017systematic, ho2019periodic,  khemani2019signatures, choi2019emergent, iadecola2019uantum, schecter2019weak, mark2020unified, turner2021correspondence,  moudgalya2022exhaustive, moudgalya2028exact, omiya2023quantum, gotta2023asymptotic, hummel2023genuine, surace2023quantum, logaric2023quantum}, and
it is even more remarkable considering that 
all the work thus far points to fragility of QMBS to pertubations~\cite{lin2020slow,surace2021exact,serbyn2021quantum,moudgalya2022quantum,chandran2023qmbsreview,gotta2023asymptotic}.
Identifying  Hamiltonians with \emph{robust} QMBS  within an experimentally sensible subclass of interactions would be an important breakthrough for both theory and experiments.\smallskip

With very few exceptions~\cite{bull2022tuning,desaules2022hypergrid}, the vast majority of work on QMBS concerned local Hamiltonians. \emph{Long-range} interactions represent however a promising avenue of investigation: on the experimental side, they are naturally present in several analog quantum-simulation platforms such as trapped ions~\cite{blatt2012quantum,britton2012engineered,richerme2014nonlocal}, dipolar gases~\cite{Chomaz_2023}, polar molecules~\cite{yan2013observation}, cold atoms in cavities~\cite{baumann2010dicke}, and even solid-state spinful defects~\cite{kuckso2018critical}; on the theory side, they are known to give rise to anomalous dynamical phenomena~\cite{KastnerPRL11_Diverging,mori2018prethermalization,neyenhuis2017observation,LeroseShort,LeroseLong,DefenuPNAS,LeroseDWLR,GorshkovConfinement,Verdel19_ResonantSB,SacredLog,defenu2023out}.
This behavior can be traced back to a peculiar form of integrability breaking. 
In the limit of all-to-all (infinite-range) interactions the full permutational symmetry  allows to reduce the quantum many-body dynamics to the semiclassical dynamics of {a} few collective degrees of freedom.
As the range of interactions is decreased,  
all other degrees of freedom associated with  gradually shorter wavelengths 
{acquire non-trivial dynamics, coupled with those of the collective degrees of freedom}. 
This tunable decoupling gives rise a scenario that interpolates between few-body and many-body physics. 
As a consequence, 
non-equilibrium behavior  
shows resilience 
over long prethermal stages of dynamics~\cite{ZunkovicPRL18_Merging,HalimehPRB17_PersistentOrder,BrescianiCooperativeShielding, GorshkovConfinement,collura2022discrete}, 
whose duration increases with the interaction range and grows unbounded in the thermodynamic limit when interactions decay slower than  $1/r^d$ ($d=$ system dimensionality)~\cite{LeroseKapitza,mori2018prethermalization,SacredLog}. The occurrence of thermalization at longer times remains, however, an open question.
\smallskip

In this paper 
we demonstrate the occurrence of 
\emph{robust QMBS} arising from isotropic long-range interactions in quantum spin lattices. This finding implies \emph{thermalization breakdown at arbitrarily long times} in this large class of quantum many-body systems.
As {a} key point of our analysis, we unveil an emergent exact solvability of {the small fraction of} eigenstates originating from high-permutational-symmetry sectors in these systems.
We show that such eigenstates get smoothly deformed upon changing Hamiltonian parameters as long as the decay of spin-spin interactions remains sufficiently slow, {in contrast with the exponential sensitivity of all the other eigenstates}.
The energy eigenvalues associated with such robust QMBS form {unequally} spaced regular towers, labelled by the collective spin quantum numbers as well as by the occupation numbers of  quasiparticle excitations emerging from our analytical construction. 
Together with the numerical observation of quantum chaos signatures in level statistics for arbitrary decaying interactions, our findings establish \emph{robust QMBS as a generic feature of long-range interacting spin lattices}.
Our theory gives stringent quantitative criteria for the stability of QMBS, which we numerically illustrate in the variable-range quantum Ising chain.

\section{Overview}

The results of this paper apply to fairly general quantum spin lattices with interactions depending algebraically on the distance $r$ as $1/r^\alpha$.
For the sake of concreteness we work with a variable-range Ising-like quantum spin chain, introduced in Sec.~\ref{sec_model}.
The following Sections present the main contributions of this work:

\begin{figure*}
\centering
\includegraphics[width=0.96\textwidth]{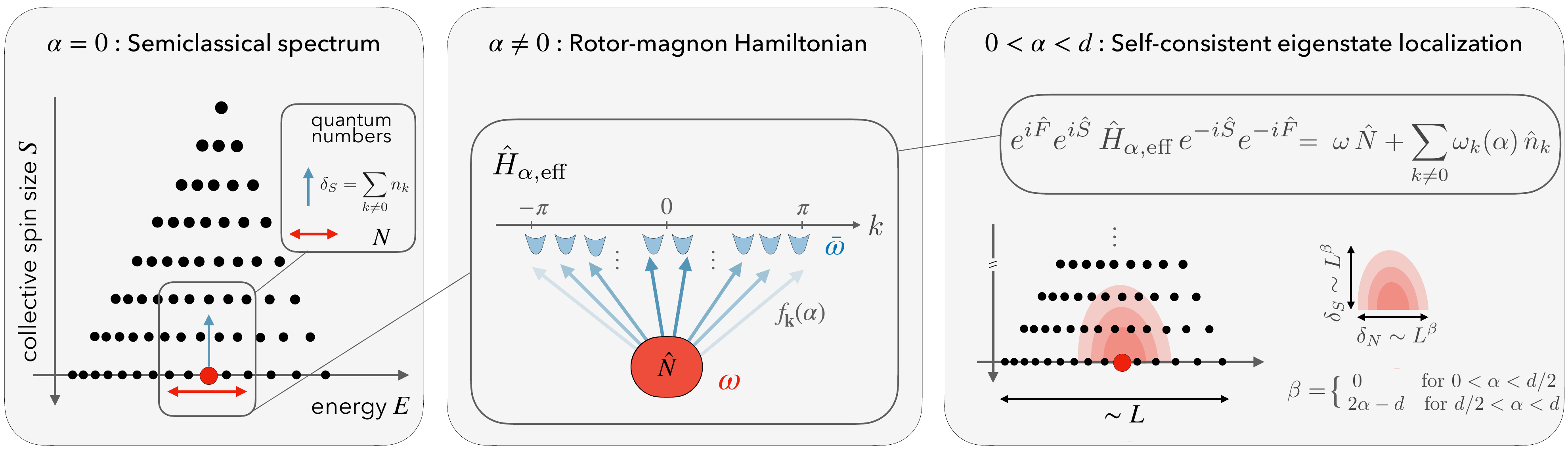}
\caption{
Graphical summary of the theory of robust quantum many-body scars developed in this work.
(Left panel) The starting point is the solvable semiclassical spectrum of the mean-field limit of interacting spin lattices ($\alpha=0$), labelled by the exact quantum numbers $S=s L^d-\delta_S$ (collective spin size) and $N$ (quantized classical action). Spin depletion $\delta_S$ is associated with increasing degeneracy, which can be resolved in terms of magnons with non-zero momentum.
(Central panel) We express the finite-range spin Hamiltonian with $\alpha\ne0$ in terms of these degrees of freedom. 
This yields a reformulation as an effective quantum impurity model: a relativistic quantum rotor parametrically coupled to an ensemble of bosons.
In the dilute sector $\delta_S \ll L/2$ of a given energy shell the effective rotor-magnon Hamiltonian simplifies to the extent that we can find a (non-trivial) exact solution for its spectrum and eigenstates.
(Right panel) Self-consistency of this solution is satisfied only for $0<\alpha<d$, in agreement with our numerical data analysis for small system size.
Our theory unveils the mechanism of eigenstate localization underpinning the robust quantum many-body scars. It also predicts and classifies instances of its breakdown, which we verify numerically.
}
\label{fig_summary1}
\end{figure*}

 \begin{figure*}
\includegraphics[width=0.94\textwidth]{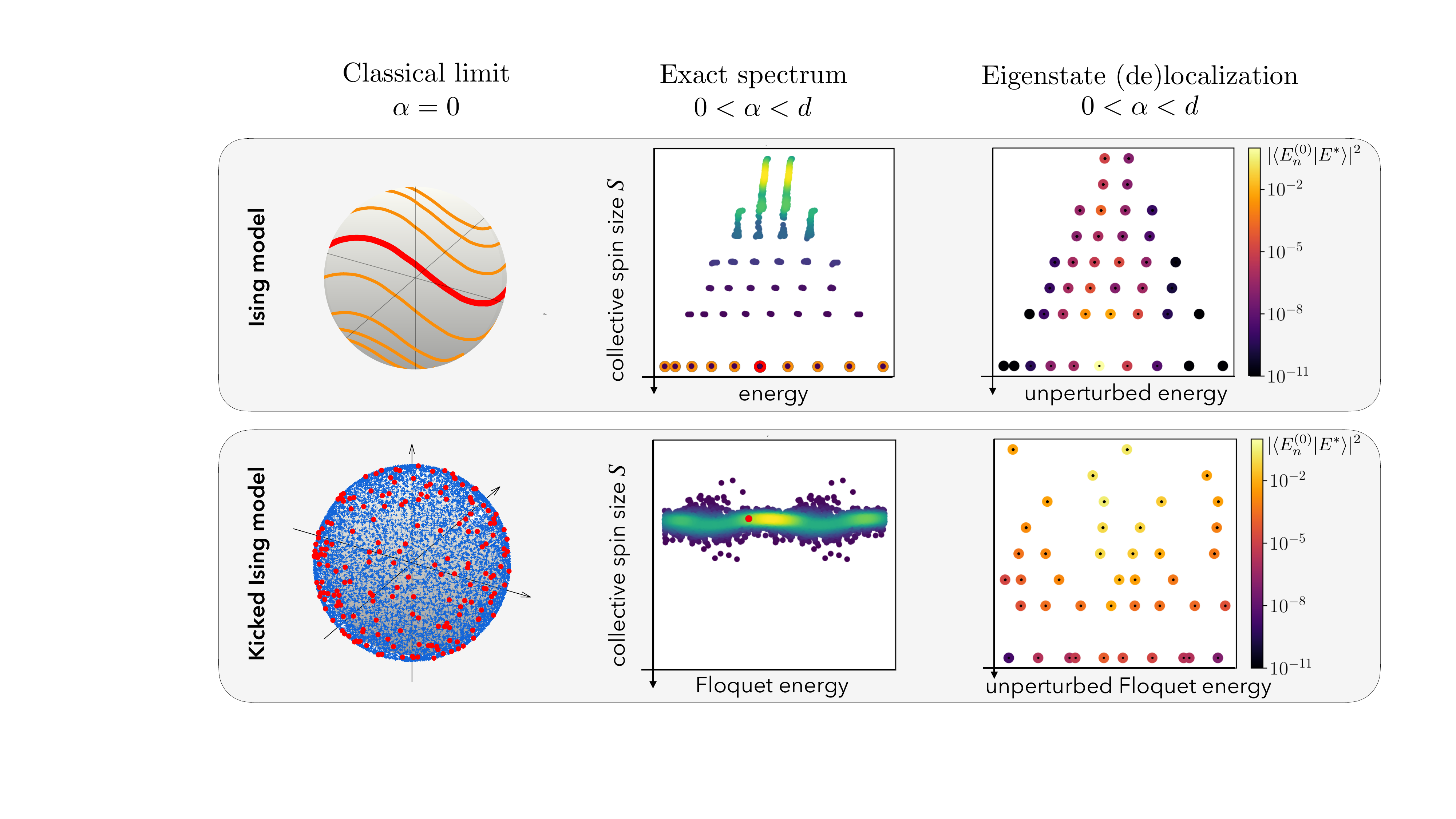}
\caption{
Relation between classical mean-field trajectories for $\alpha=0$ (left column) and many-body energy eigenstates for $\alpha\ne0$ (center and right column). 
Classical integrability gives room to robust quantum many-body scars localized in the large-spin sectors of Hilbert space, coexisting with a bulk of repelling energy levels for $0<\alpha<d$ (top row). Classical chaos, instead, prevents quantum many-body scarring for arbitrarily small $\alpha>0$ (bottom row).
This dichotomy is illustrated here with data for the quantum Ising model, Eq.~\eqref{eq_lrxy} with $\gamma=1$ {(cf. Figs.~\ref{fig_stable_L18} and \ref{fig_kick_L18} below)}. We report: classical collective-spin trajectories $\langle \hat{\bold{S}}(t) \rangle/L$ for $\alpha=0$ and $L\to\infty$ (left); exact energy spectra resolved in terms of the average collective spin size $S=\langle |\hat{\bold{S}}| \rangle$  for $\alpha=0.5$ and $L=18$ (center); weights on the unperturbed $\alpha=0$ basis $\{|E^{(0)}_n\rangle\}$ of the representative eigenstate $|E^*\rangle$ highlighted in red in the center column (right); with time-independent Hamiltonian (top) or subject to strong periodic driving (bottom) --- in the latter case we report stroboscopic trajectories and Floquet spectra.
Quantum many-body scarring is witnessed by wavefunction concentration in large collective-spin size sectors
{(details are provided below)}. 
}
\label{fig_summary2}
\end{figure*}

\begin{itemize}
\item  Performing an extensive study of numerical energy spectra of variable-range quantum Ising chains, we show that random-matrix-like level repulsion appears for infinitesimal $\alpha$  along with resilient anomalous eigenstates, characterized by large collective spin size, low entanglement, and high overlap with product states (Sec.~\ref{sec_numerics}). We identify these eigenstates as candidate QMBS-like states, smoothly deformed from the large-spin eigenstates of the permutationally-symmetric (solvable) mean-field limit $\alpha=0$.
Using a refined numerical approach which involves a projection of the Hamiltonian onto large-spin subspaces (App.~\ref{app_approximethod}), we find evidence for robustness of the QMBS for arbitrary $L$ when $0<\alpha\lesssim d$, and for their eventual hybridization when $\alpha\gg d$.

\item We formulate a self-consistent analytical theory of eigenstate localization in Hilbert space that fully backs our numerical observations (Sec.~\ref{sec_theory}). First, we introduce a procedure to map an interacting spin lattice onto a collective spin coupled to an ensemble of bosonic modes, physically representing magnon excitations with non-vanishing momenta.  The coupling strength is controlled by $\alpha$ and $d$ and it is vanishingly small for $\alpha\to0$.
In a Hilbert space shell with fixed energy density and large collective spin, the exact spin-boson description reduces to a simpler but still interacting quantum impurity model: a relativistic quantum rotor parametrically coupled to a quadratic bosonic bath. We find an exact solution to this effective Hamiltonian, which allows us to compute the degree of eigenstate delocalization in Hilbert space and hence to check self-consistency.
We find that the eigenstates of this rotor-magnon Hamiltonian are self-consistent QMBS of the original spin Hamiltonian only for $0<\alpha<d$.
Figure~\ref{fig_summary1} visually summarizes the steps of our eigenstate localization theory, which relies on two  fundamental ingredients: 
\begin{enumerate}
    \item sufficiently slow interaction decay $0<\alpha<d$;
    \item classical  integrability of the %underlying few-body 
    mean-field limit.%Hamiltonian.
\end{enumerate}
These QMBS thus possess, by construction, a clear fingerprint from semiclassical periodic orbits  -- sometimes considered a defining feature of QMBS. 
\item We predict and classify instabilities of the QMBS, arising from violations of the above conditions. Eigenstate delocalization most notably occurs in presence of discrete-symmetry breaking or of mean-field chaotic dynamics. We accurately test these predictions for the long-range quantum Ising chain, respectively in the ordered phase or in presence of periodic kicks (Sec.~\ref{sec_predictions}). Figure~\ref{fig_summary2} reports representative examples.
\end{itemize}

The results of this work have a broader impact on quantum many-body physics and its applications, which we briefly elaborate on in the conclusive Sec.~\ref{sec_conclusions}.

\section{The model}
\label{sec_model}

In this paper we consider a $d$-dimensional lattice of $V=L^d$ quantum spins governed by the Hamiltonian
\begin{widetext}
\be
\label{eq_xxzlrd}
\hat H_{\alpha} = -   \sum_{\mb{i},\mb{j}} J_{\mb{i},\mb{j}} \bigg( \frac{1+\gamma} 2 \hat\sigma^x_{\mb{i}} \hat\sigma^x_{\mb{j}}
+ \frac{1-\gamma} 2 \hat\sigma^y_{\mb{i}} \hat\sigma^y_{\mb{j}}
+ \frac \Delta 2 \hat\sigma^z_{\mb{i}} \hat\sigma^z_{\mb{j}} \bigg) - h  \sum_{\mb{j}}  \hat\sigma^z_{\mb{j}} 
\ee
\end{widetext}
where $\gamma$  and $\Delta$ parametrize the XY and XXZ anisotropies, respectively. In this equation $\hat\sigma^{x,y,z}_{\mb{r}}=\hat s^{x,y,z}_{\mb{j}}/s$ are rescaled spin-$s$ operators acting on the spin at site $\mb{j}=(j_1,\dots,j_d)$ of the lattice, where $j_a=1,\dots,L$ for $a=1,\dots,d$. 
For spins-$1/2$ these are the standard Pauli matrices.

We will consider interactions depending only on the distance between spins, with algebraic decay:
\be
J_{\mb{i},\mb{j}} \equiv J_{\mb{i}-\mb{j}}  = \frac{J}{\lvert \lvert \mb{i}-\mb{j} \rvert \rvert^\alpha}.
\ee
where $\lvert \lvert \mb{i} - \mb{j} \rvert \rvert \equiv \sqrt{\sum_{a=1}^d [\min(|i_a-j_a|, L-|i_a-j_a|)]^2}$ realizes periodic boundary conditions.
The constant $J$ is chosen 
to make the mean-field  energy independent of the decay exponent $\alpha$ (\textit{Kac normalization}~\cite{KacNormalization}),
\be
J = \frac{J_0}{2\mathcal{N}_{\alpha,L}}, \qquad \mathcal{N}_{\alpha,L} = \frac 1 2 \sum_{\mb{r}\neq\mathbf{0}}  \frac{1}{\lvert \lvert \mb{r} \rvert \rvert^\alpha} \, .
\ee
Such prescription is necessary to have a well defined thermodynamic limit for $\alpha \le d$: The system-size divergent rescaling factor $\mathcal{N}_{\alpha,L} \sim L^{d-\alpha} $   ensures that the energy per spin remains finite. 
For definiteness we will assume ferromagnetic interactions $J_0>0$ throughout (spectral properties do not change under $H\mapsto -H$). We also set $s=1/2$.\footnote{For $s=1/2$ the terms $\mb{i}=\mb{j}$ produce an inconsequential additive constant $E=\sum_{\mb{j}} J_{\mb{i},\mb{j}}  (1+\Delta/2)$, as Pauli matrices square to $1$; we will thus set $J_{\mb{i},\mb{i}}=0$.}

{Increasing the range of interactions (i.e. decreasing $\alpha$) weakens spatial fluctuations, leading the system toward its mean-field limit, analogously to increasing the system dimensionality~\cite{dutta2001phase,defenu2017criticality,defenu2023review}.} It is thus convenient to perform our theoretical analysis with fixed $d$ and study the properties of the model upon varying $\alpha$.
As the effects of spatial fluctuations are strongest in one dimension, 
we will work with 
a variable-range XY quantum spin chain, 
{i.e. we choose $d=1$~\footnote{{For dimensions $d>1$, different lattice structures generally alter the shape of the function $f_k(\alpha)$ [defined in Eq.~\eqref{eq_fkalpha} below]. However, the basic properties at small $k\simeq 0$ (responsible for the peculiar behavior of these systems) remain the same for $\alpha\leq d$.}} and set $\Delta=0$~\footnote{{Note that beyond-nearest-neighbor interactions break integrability regardless of the value of $\Delta$, see below.}}}
\begin{widetext}
\be
\label{eq_lrxy}
\hat H_{\alpha}= - \frac{J_0}{\mathcal{N}_{\alpha,L}}  \sum_{j=1}^L \sum_{ r=1}^{ \floor{L/2}} {\vphantom{\sum}}'  \frac{1}{ r^\alpha} \bigg( \frac{1+\gamma} 2 \hat\sigma^x_{j} \hat\sigma^x_{j+r}
+ \frac{1-\gamma} 2 \hat\sigma^y_{j} \hat\sigma^y_{j+r} \bigg)
 - h  \sum_{j=1}^L  \hat\sigma^z_{j} 
\ee
\end{widetext}
(periodic boundary conditions are understood; {the primed sum denotes that terms $r=L/2$ are counted only once for $L$ even}).
We anticipate that our results do not rely at all on any of the model restrictions above, as we will explicitly show in Sec.~\ref{subsec_generality}.

{The Hamiltonian~\eqref{eq_lrxy} has translation, spatial reflection, 
and spin-flip ($\prod_{j=1}^L \hat{\sigma}^z_j$) symmetries.}
For $\alpha=\infty$ the Hamiltonian~\eqref{eq_lrxy} reduces to the  XY quantum spin chain with nearest-neighbor interactions. This model is exactly solvable by mapping to a quadratic fermionic chain via Jordan-Wigner transformation~\cite{SchultzMattisLieb}. 
For finite $\alpha$ integrability is broken by couplings beyond the nearest neighbors, as the associated Jordan-Wigner string turns into multi-fermion interactions.  
As $\alpha$ is decreased, 
longer-range interactions cooperate to suppress spatial fluctuations, resulting in a qualitative enhancement of the system's ability to order %: % for sufficiently small $\alpha$:
in \emph{excited} states 
for $\alpha<2$~\cite{dyson1969existence}.

The tendency 
to sustain collective spin alignment 
becomes increasingly prominent as $\alpha$ is decreased.  
This is conveniently understood by viewing a long-range interacting system as a ``perturbation'' to the infinite-range limit with all-to-all interactions ($\alpha=0$).
For $\alpha \to 0$ 
Eq.~\eqref{eq_lrxy} reduces to a Hamiltonian describing dynamics of a single collective spin, completely decoupled from all the other degrees of freedom --- 
 realizing the mean-field description of the system.
The effect of spatially modulated interactions $\alpha\neq0$ is then to dynamically couple this collective degree of freedom to all the other finite-wavelength modes describing spatial fluctuations, resulting in complex quantum many-body dynamics.

For later purpose we show how to make this picture explicit. We express the finite-range Hamiltonian $\hat H_\alpha$ as a perturbation to $\hat H_{\alpha=0}$ by rewriting it in momentum space. Let us define the Fourier transform the spin operators (recall $\hat s^{\mu}_j = \hat \sigma^{\mu}_j/2$ for $\mu=x,y,z$)~\footnote{{To lighten the notation we drop the operator hats on top of the Fourier-transform tildes.}}
\be
\label{eq_fourierspin}
 \tilde S^{\mu}_k= \sum_{j=1}^L e^{ikj} \hat s^{\mu}_j \, ,
\ee
with 
\be
k\equiv k_\ell=2\pi \ell / L, \quad \ell=0,\pm1,\pm2,\dots,\pm \floor{L/2}
\ee
(for $L$ even $k_{\pm L/2}$ coincide);
$ \tilde {\bold{S}}_{k=0} \equiv \hat {\bold{S}} = \sum_j \hat {\bold{s}}_j $ is the system's collective spin.
In terms of momentum-space operators, the variable-range XY spin chain reads
\begin{widetext}
\be
\label{eq_lrxyfourier}
\hat H_\alpha = - \frac{J_0}{L}  \sum_{k} f_{k}(\alpha) \Big[ \big(\tilde S^+_k  \tilde S^-_{-k} + \tilde S^-_k  \tilde S^+_{-k}\big)
+\gamma \big(\tilde S^+_k  \tilde S^+_{-k} + \tilde S^-_k  \tilde S^-_{-k}\big) \Big]
 - 2 h  \tilde S^z_{k=0} \, , 
\ee
\end{widetext}
where $\tilde S^{\pm}_k=\tilde S^{x}_k \pm i\tilde S^{y}_k$.
(Note that in this expression the various $k$-modes are \emph{not} dynamically decoupled, as $[\tilde S^\alpha_k,\tilde S^\beta_q] = i \epsilon^{\alpha\beta\gamma} \tilde S^\gamma_{k+q}$.)
In Eq.~\eqref{eq_lrxyfourier} we defined the function
\be
\label{eq_fkalpha}
f_{k}(\alpha) = \frac 1 {\mathcal{N}_{\alpha,L}} \sum_{r=1}^{L/2} \frac {\cos(kr)}{ r^\alpha} = 
\frac { \sum_{r=1}^{L/2} \frac {\cos(kr)}{ r^\alpha} }{ \sum_{r=1}^{L/2} \frac {1}{ r^\alpha} }.
\ee
By construction, $f_{k=0}(\alpha)=1$.
\begin{figure*}
\centering
\includegraphics[width=0.42\textwidth]{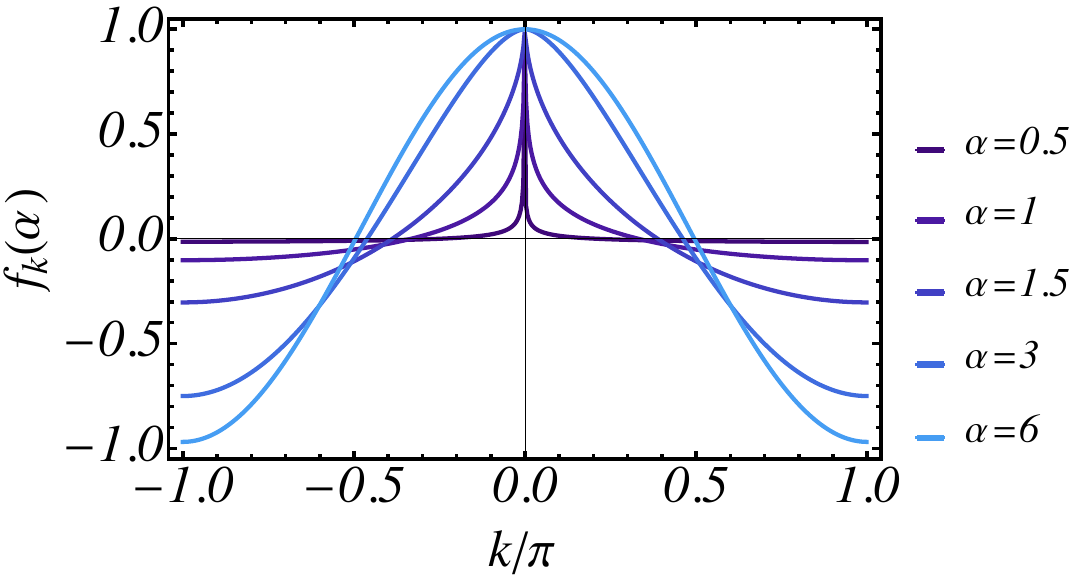} \hspace{0.25cm}
\includegraphics[width=0.42\textwidth]{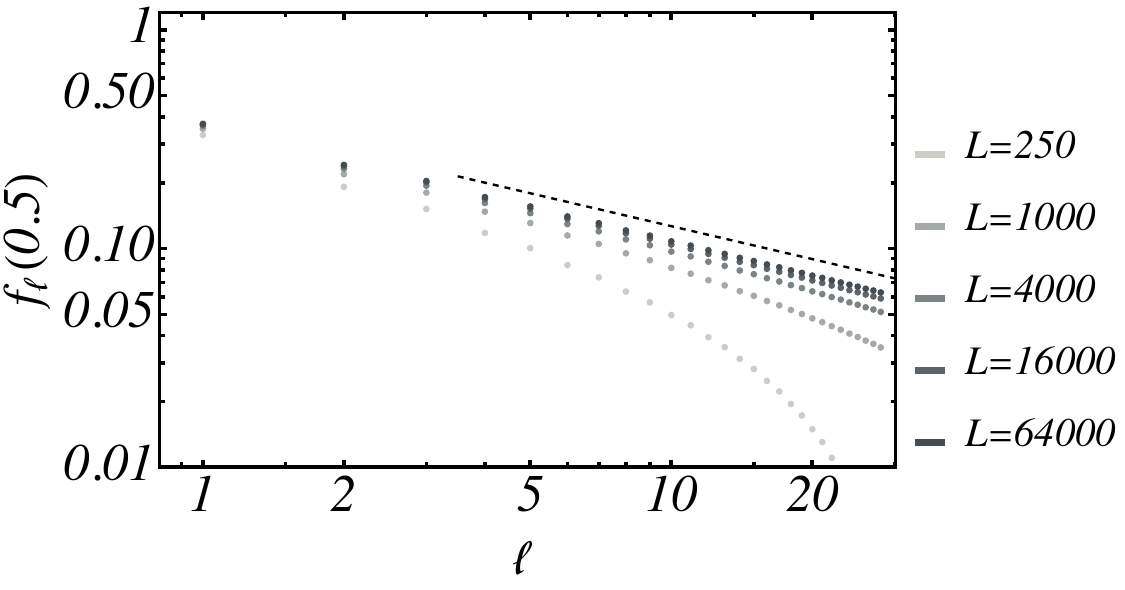}
\caption{Left panel: Couplings $f_k$ in Eq.~\eqref{eq_fkalpha} for a range of values of $\alpha$; here we set $L=1000$. Right panel: convergence of the discrete sequence $f_\ell(\alpha)$ for $\alpha=0.5$ as $L\to\infty$; the dashed line highlights the asymptotics $\sim \ell^{-(1-\alpha)}$ from Eq.~\eqref{eq_fkalphaasympt}.}
\label{fig_fkalpha}
\end{figure*}
We now identify the variable-range perturbation by singling out the $k=0$ part of the Hamiltonian: 
 \begin{widetext}
\be
\label{eq_Valpha}
\hat H_\alpha = \hat H_{\alpha=0}(\hat{\bold{S}}) + \hat V_\alpha \, , \qquad \text{with } \; \hat V_{\alpha} = - \frac{J_0}{L}  \sum_{k\neq0} f_{k}(\alpha) \Big[ \big(\tilde S^+_k  \tilde S^-_{-k} + \tilde S^-_k  \tilde S^+_{-k}\big)
+\gamma \big(\tilde S^+_k  \tilde S^+_{-k} + \tilde S^-_k  \tilde S^-_{-k}\big) \Big].
\ee
\end{widetext}
Note that the term $\hat V_\alpha$ is in general an extensive perturbation\footnote{The factor $1/L$ is due to the convention chosen for the  spin Fourier transform in Eq.~\eqref{eq_fourierspin}.}.

The quantity $f_{k}(\alpha)$ controls how strongly spin fluctuation modes with various wavelengths come into play, thus determining the physical properties of $\hat H_\alpha$.
Its behavior upon varying $\alpha$ is summarized in Fig.~\ref{fig_fkalpha}. Its shape shrinks from $f_{k}(\alpha\to\infty)=\cos k$ to 
\be
\label{eq_falpha0}
f_{k}(\alpha\to0)=\delta_{k,0}\, ,
\ee
becoming increasingly singular at $k=0$ as $\alpha$ is decreased: 
\be
\label{eq_fkalphaasympt}
\left\{
\begin{split}
f_{k}(\alpha) 
&
\sim 1-c(\alpha) k^{2}  
& 
\text{for } 
&
\alpha>3 ;
\\
f_{k}(\alpha) 
&
\sim 1-c(\alpha) |k|^{\alpha-1}  
&
\text{for } 
& 
1<\alpha<3 ;
\\
f_{k_\ell}(\alpha)  \equiv f_{\ell}(\alpha)
& 
\sim c(\alpha) |\ell|^{-(1-\alpha)} 
&
\text{for } 
& 
\alpha<1 .\\
\end{split}
\right.
\ee
These and other properties of $f_{k}(\alpha)$ are derived  in App.~\ref{app_fkalpha}.
The last case is the most interesting for this paper: For $\alpha < 1$, $f_{k}(\alpha)$  squeezes onto the vertical axis as  $L\to\infty$; upon zooming near $k=0$ one finds that the values $f_{2\pi\ell/L}$ converge as $L\to\infty$ to \emph{a discrete sequence of finite values,} denoted $f_\ell$, which fall off as $|\ell|^{-(1-\alpha)}$, cf. the right panel of Fig.~\ref{fig_fkalpha}~\cite{SacredLog,DefenuPNAS}. 
Thus, for small $0<\alpha<1$, only modes with extensive wavelengths $k_\ell \propto 1/L$ do non-trivially participate in dynamics and interact;
physically, 
in this regime interactions vary so slowly with the spatial distance that the system behaves as a permutationally invariant system over finite length scales, and it is thus unable to resolve finite wavelengths. 
As $\alpha$ is increased further, all $k\neq0$ eventually {acquire non-trivial dynamics}.  

The splitting in Eq.~\eqref{eq_Valpha} thus gives us an intuitive physical picture of dynamics for $\alpha>0$ in terms of a collective spin weakly coupled to finite-wavelength spin excitations. This representation can be expected to provide a starting point for a meaningful perturbation theory when $f_{k\neq0}(\alpha)$ is small. The precise notion of smallness will be clarified in this paper.

\section{Numerical analysis}
\label{sec_numerics}

We analyze \emph{spectral} properties of highly excited energy eigenstates
\begin{equation}
    \hat H_{\alpha} |E_n\rangle = E_n |E_n\rangle  
\end{equation}
 for finite $0<\alpha<\infty$. 
 While dynamical relaxation to thermal equilibrium is known to become slow as $\alpha$ is decreased (see the Introduction), it is generally believed that thermalization is ultimately attained at long times for any $0<\alpha<\infty$, as the system is non-integrable. Accordingly, the energy spectrum is believed to satisfy the eigenstate thermalization hypothesis (ETH)~\cite{DeutschETH,SrednickiETH}. In this Section, we report extensive exact diagonalization (ED) numerical results which show that the actual scenario may be subtler. While we find level statistics compatible with ETH down to very small values of $\alpha\gtrsim 0.001$ (in agreement with Ref.~\cite{russomanno2021quantum}, see also Ref.~\cite{SredinickiETHinLR}), we will show that a relevant subset of eigenstates exhibits remarkable deviations, characterized by anomalous QMBS-like properties such as large collective spin size, low entanglement, and large overlap with product states. Such deviations are increasingly pronounced as $\alpha$ is decreased. Our data suggest  the permanence of such anomalous properties for arbitrary $L$ at least for~$\alpha\lesssim 1$. \\

For definiteness, in this Section, we set $\gamma=1$ in Eq.~\eqref{eq_lrxy}  (quantum Ising chain).
  We further set $h=2J_0=1$ so as to avoid signatures of the thermal phase transition (or of the excited-state quantum phase transition~\cite{cejnar2021excited}) in highly excited states. In Sec.~\ref{sec_ESQPT} below we will comment on how these phenomena affect our results.
 
The quantum numbers of the Hamiltonian~\eqref{eq_lrxy} associated with translation, spatial reflection, 
and spin-flip symmetries are $K=0,2\pi/L,\dots,2\pi (L-1)/L$, $I=\pm 1$, and $P_z=\pm 1$ respectively. 
In our full ED computations we fix for definiteness $K=0$, $I=1$, and $P_z=1$.

\subsection{Level repulsion}
We study the energy eigenvalue statistics of the model 
by probing the level spacings $s_n = E_n - E_{n-1}$ via the ratio between nearby energy gaps \cite{oganesyan2007localization}
\begin{equation}
\label{eq:rdef}
r_{n} = \frac{\min \{s_n, s_{n+1}\}}{\max\{s_n, s_{n+1}\} }\ .
\end{equation}
Generic non-integrable models are expected to obey 
{Wigner-Dyson} level statistics \cite{bohigas1984characterization}, characterized by a distribution of the ratios $p_{\text{WD}}(r) = \frac {27}4 \, r\, \frac{1+r}{(1+r+r^2)^{5/2}}$ and average {level spacing ratio} $\braket{r}_{\text{WD}} = 0.5295$. 
On the other hand, integrable Hamiltonians are characterized by {Poisson} statistics \cite{berry1977level}, with $p_{\text{P}}(r) = \frac{2}{(1+r)^2}$ and average $\braket{r}_{\text{P}} = 0.386$.

\begin{figure}[t]
\includegraphics[width=.48\textwidth]{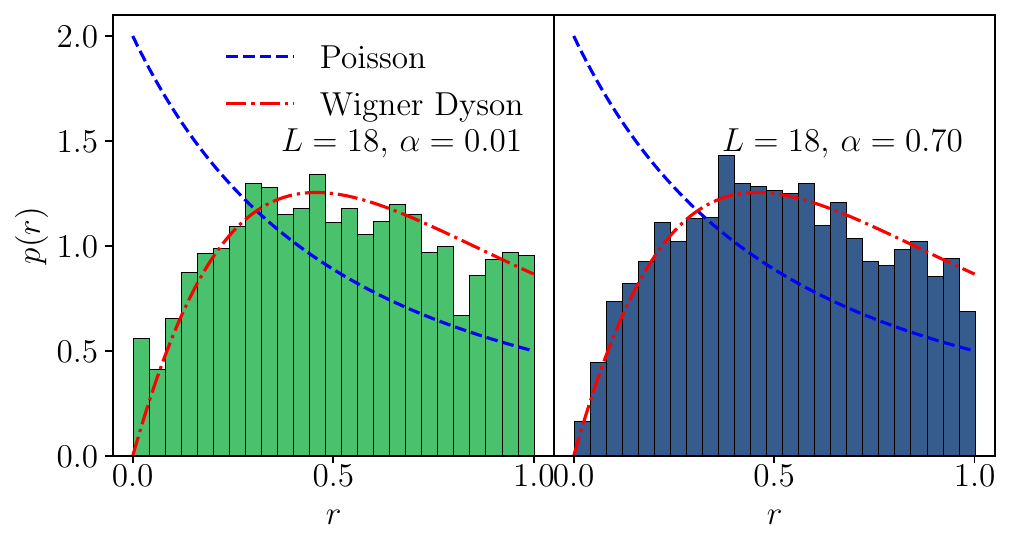}
\includegraphics[width=.48\textwidth]{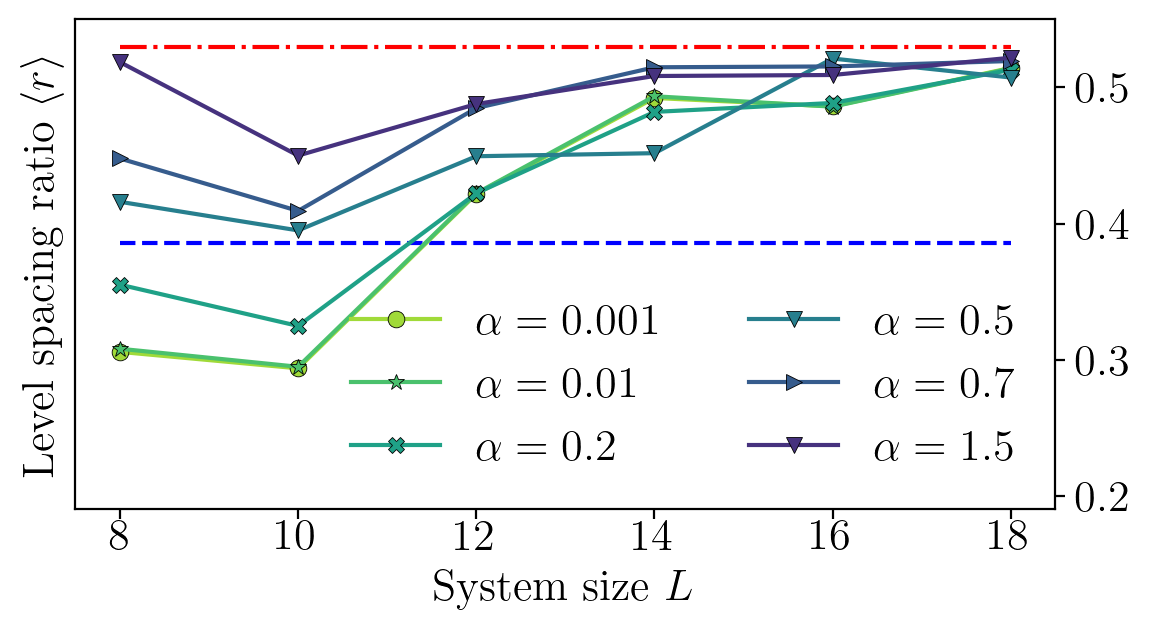}
\caption{Level repulsion for the long-range Hamiltonian{~\eqref{eq_lrxy} $\gamma=1$ and ${h=2J_0=1}$}. The {dot-dashed red lines indicate the Wigner--Dyson spectral statistics, while the dashed blue ones indicate Poisson statistics}.  (Top) Distribution of the level spacing ratio $p(r)$ for $L=18$ for $\alpha=0.01$ (left) and $\alpha=0.7$ (right). (Bottom) Average level spacing ratio as a function of the system size $L$ for different $\alpha=0.001, \,0.01, \,0.2,\, 0.5, \,0.8, \,1.5, \,3$. }
\label{fig_r}
\end{figure}

Our data indicate that the mean level spacing ratio is compatible with the Wigner--Dyson prediction for all $\alpha >0$. The numerical distribution of $p(r)$ shows a very good agreement with the Wigner--Dyson prediction ({dot-dashed} red line) and clear disagreement from the Poisson one ({dashed} blue line), as displayed in Fig.~\ref{fig_r} for $\alpha=0.01$ ({top} left) and $\alpha=0.7$ ({top} right) for a system of $L=18$ spins.
This is well confirmed by the trend of $\braket{r}$ 
with system size. {The bottom panel of} Fig.~\ref{fig_r} shows  convergence to the Wigner--Dyson value $\braket{r}_{\text{WD}}$ upon increasing system size $L$. This holds for $\alpha$ as small as $0.001$, signalling high sensitivity of spectral statistics to the interaction range.
These results complement those of Ref.~\cite{russomanno2021quantum}.

\begin{figure*}
\centering
\includegraphics[width=1\textwidth]{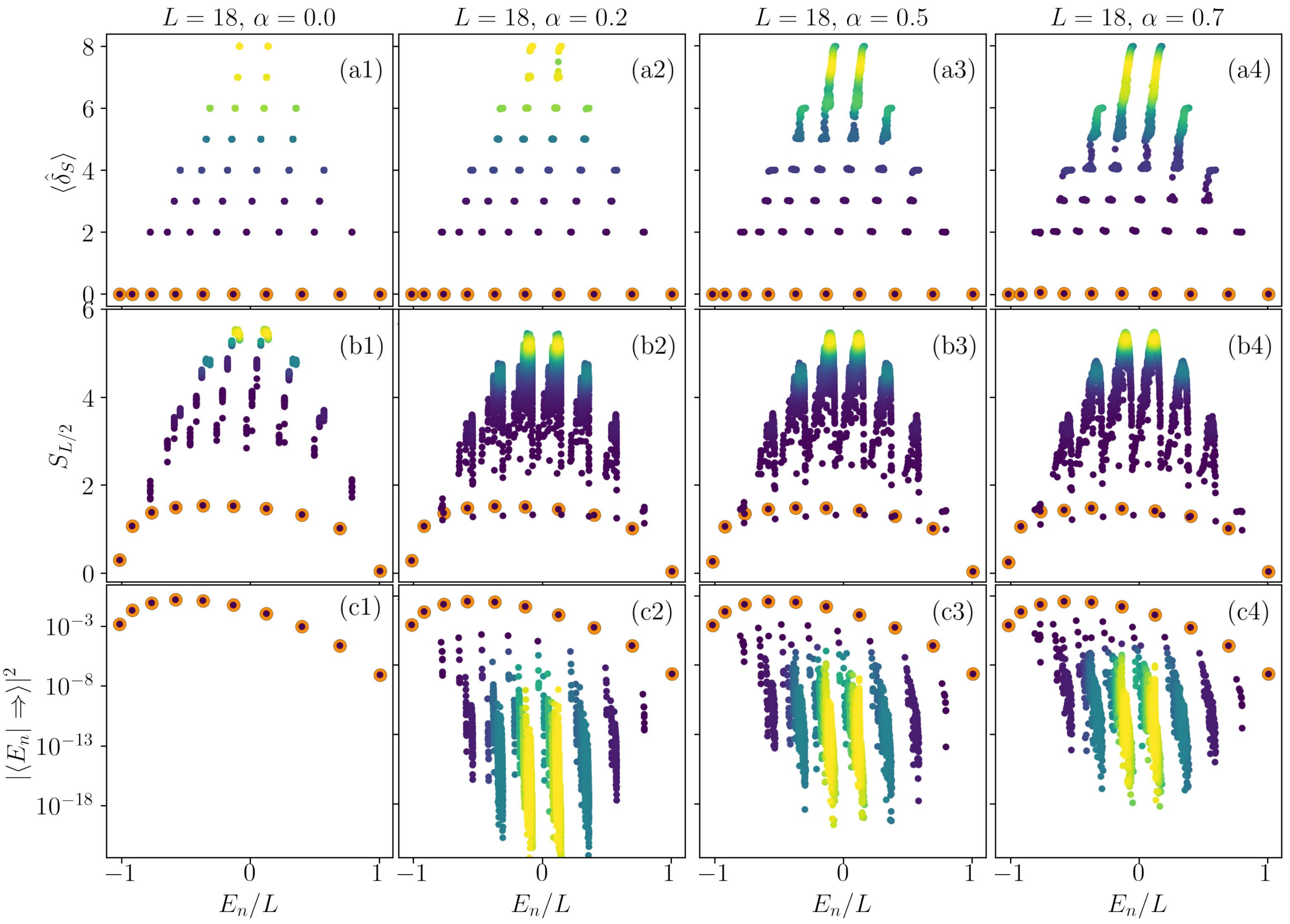}
\caption{QMBS in the long-range quantum Ising chain. We report full ED numerical energy spectra of the model in Eq.~\eqref{eq_lrxy} with $L=18$, $\gamma=1$ and ${h=2J_0=1}$, restricted to the translation-invariant, parity and spin-flip symmetric sector.  Scatter plot of the observables $\langle \hat\delta_S\rangle_{E_n}$ (row a), half-chain entanglement entropy $S_{L/2}(E_n)$ (row b), and overlap with the fully polarized state {along x $|\langle E_n| \Rightarrow\rangle|^2$} (row c) of each eigenstate $|E_n\rangle$ as a function of the energy density $E_n/L$. Columns (1-4) correspond to increasing values of $\alpha=0$, $0.2$, $0.5$, $0.7$.
{The heatmap color illustrates the density of eigenstates, from darkest (isolated eigenstates) to lightest (maximum density).
}
}
\label{fig_stable_L18}
\end{figure*}

\subsection{Anomalous eigenstates}
\label{sec_numerics_scars}

In spite of the level repulsion, we find that the spectrum of the long-range quantum Ising Hamiltonian has a distinctive structure, featuring a thermodynamically negligible fraction of atypical non-thermal eigenstates running througout the many-body energy spectrum. 

In Fig.~\ref{fig_stable_L18} we plot, for all energy eigenstates $\{\ket{E_n}\}$:\\ \textit{a)}  the average depletion of the total spin 
$  \hat {\mathbf S}^2  =  (\frac L 2 - \hat\delta_S)(\frac L 2 - \hat\delta_S+1) $
from its maximal value, i.e.
\begin{align}
\label{eq:totalS}
\langle \hat{\delta}_{S}\rangle_{E_n} = \braket{E_n|\hat {\delta}_S|E_n} 
\end{align}
[{more precisely, we parameterize the eigenvalues of $\hat {\bold S}^2$ as $S(S+1)$ with $S=\frac L 2 - \delta_S$,  $\delta_S=0,1,\dots,\floor{\frac L2}$, and introduce the operator $\hat \delta_S$ with eigenvalues $\delta_S$ associated with the corresponding eigenvectors of $\hat {\bold S}^2$}]; \\
\textit{b)} the half-chain entanglement entropy: 
\be
S_{L/2} = \text{Tr}(\hat\rho_{L/2} \ln \hat\rho_{L/2}), \quad \hat\rho_{L/2} = \text{Tr}_{L/2+1,\dots,L} |E_n\rangle\langle E_n| \ ;
\ee
\textit{c)} the overlap with the fully polarized state\footnote{{Because we restrict the Hamiltonian to the symmetry sector $P_z=+1$, the overlaps in Eq.~\eqref{eq:overlap} are computed as one half of the overlaps with the even component $\frac 1 {\sqrt{2}}( |\Rightarrow\rangle+ |\Leftarrow\rangle$) of the state.}}
\begin{equation}
\label{eq:overlap}
|\langle E_n|\Rightarrow\rangle|^2 \ ,
\end{equation}
where {$| \boldsymbol{\Rightarrow} \, \rangle  \equiv | \rightarrow_1\rightarrow_2\dots \rightarrow_L\, \rangle$ is the product state of spins pointing along $x$ ($\hat\sigma^x_j | \boldsymbol{\Rightarrow} \, \rangle = | \boldsymbol{\Rightarrow} \, \rangle$ for all $j$)}. \\
The anomalous structure of the many-body energy spectrum is highlighted by either of these three quantities in Fig.~\ref{fig_stable_L18}. 

In the limit $\alpha=0$ the Hamiltonian conserves the collective spin size $[\hat H_{\alpha=0}, \hat {\mathbf{S}}^2 ]=0$ and hence $\delta_S = L/2-S$ is an exact quantum number. The energy spectrum forms unequally spaced towers (\emph{Peres lattice}~\cite{peres1984new}) labelled by $\delta_S$ -- the horizontal rows in Fig.~\ref{fig_stable_L18}a1.
The degeneracy of the eigenspaces increases exponentially with $\delta_S$, i.e., as the collective spin size gets smaller. 
The so-called \emph{Dicke manifold}~\cite{dicke1954coherence} spanned by the $L+1$ non-degenerate eigenstates with maximum collective spin size $\delta_S=0$ is the permutationally symmetric subspace.
This tower of eigenstates corresponds to the large orange dots in the figure. Their bipartite entanglement entropy is known to scale logarithmically with system size, i.e., $S_{L/2}\sim \ln L$, away from the edges of the spectrum~\cite{latorre2005entanglement}. The fully polarized state $|\Rightarrow\rangle$ has non-vanishing overlap with them only. Note that some states are missing in the plot due to our symmetry restrictions\footnote{{Let us recall that we diagonalize the Hamiltonian in the symmetry sector $I=1$ (reflection-symmetric about the center of the chain) and $K=0$ (translationally invariant). For this reason, there are only $L/2+1$ collective states $\ket N$ rather than $L+1$, and we do not access states with $\langle \hat\delta_S\rangle=1$, which have non-zero momentum (cf. the discussion in Sec.~\ref{subsec_bosonlabelling} below).}}.

As $\alpha$ is increased from $0$ the degeneracies are lifted. While the splitting of the exponentially degenerate sectors having $\delta_S = \mathcal{O}(L)$ dominate the level spacing statistics and induces chaotic features (see discussion above), scatter plots of the quantities~\eqref{eq:totalS}-\eqref{eq:overlap} in Fig.~\ref{fig_stable_L18} indicate a strong deviation from the standard ETH scenario for $\alpha>0$. Deviation is strongest for the eigenstates originating from sectors with small $\delta_S \ll L/2$ (i.e. large collective spin), as
these towers of states are remarkably resilient to the finite range of interactions.  This is shown in columns 2--4 of Fig.~\ref{fig_stable_L18} corresponding to $\alpha=0.2$, $0.5$, $0.7$, respectively, obtained by full ED of a chain of $L=18$ spins.
The states with smallest $\braket{\hat\delta_S}$ (large orange dots) are still characterized by entanglement entropy much lower than the rest of the eigenstates and by very large overlap with fully polarized states such as $\ket{\Rightarrow}$ (Fig.~\ref{fig_stable_L18}c).\footnote{We note the presence of another set of eigenstates with anomalous small entanglement entropy, see Fig.~\ref{fig_stable_L18}b2,b3,b4. These possess a defined total spin, i.e.\ $\braket{\hat\delta_S}=2n$ with $n=1, 2, \dotsc,  L/4$ and can be identified as the top energy state of each ``stripe''. Analysis of these states is beyond the scope of this work.} 
\\

In order to better characterize the stability properties of these  large-spin eigenstates with small $\braket{\hat{\delta}_S}$ for larger system sizes, we resorted to a refined numerical method drawing on Refs.~\cite{protopopov2017effect, protopopov2020nonabelian}. 
This consists in \emph{restricting the Hilbert space to a subspace $\mathcal H_{N_\text{{shell}}}$ with large collective spin size}, where $N_{\rm shell}$ is the number of kept sectors $\delta_S=0,\dots,N_{\rm shell}-1$. 
For instance, ${N_\text{{shell}}}=1$ corresponds to the Dicke manifold $\delta_S=0$, while the full Hilbert space is retrieved for ${N_\text{{shell}}=L/2+1}$.
Constructing an orthonormal basis in $\mathcal H_{N_\text{{shell}}}$ carrying the quantum number $\delta_S$ is not an elementary task, as eigenstates of $\hat{\bold S}^2$ are typically highly entangled.
We used the construction of Ref.~\cite{protopopov2017effect,protopopov2020nonabelian} based on regular-Cayley-tree fusing rules.
The crucial advantage of this method is that the matrix elements of the basic spin operators $\hat \sigma^\mu_j$ within $\mathcal H_{N_\text{{shell}}}$ can be computed \emph{analytically}. Thus, as the dimension of $\mathcal{H}_{N_\text{{shell}}}$ scales only as $L^{N_\text{{shell}}}/(N_\text{{shell}}-1)!$, for not too large $N_\text{{shell}}$  this technique allowed us to construct a sparse matrix representation of the truncated Hamiltonian $\hat H_\alpha \rvert_{\mathcal{H}_{N_\text{{shell}}}}$ for system sizes $L\gg18$ much larger than reachable by full ED --- see Fig.~\ref{fig_projectionmethod} in App.~\ref{app_approximethod} for precise estimates.
The validity of this truncation is \textit{a posteriori} justified  by studying the convergence of the various observables [listed as \textit{a)}, \textit{b)}, \textit{c)} above] upon increasing the cutoff ${N_\text{shell}}$. Here convergence was achieved for $L>18$ when $\alpha\lesssim 1$, as reported in App.~\ref{app_approximethod}. There we also describe 
the details of our method, which may be of independent interest.

\begin{figure}[t]
\centering
\includegraphics[width=.46\textwidth]{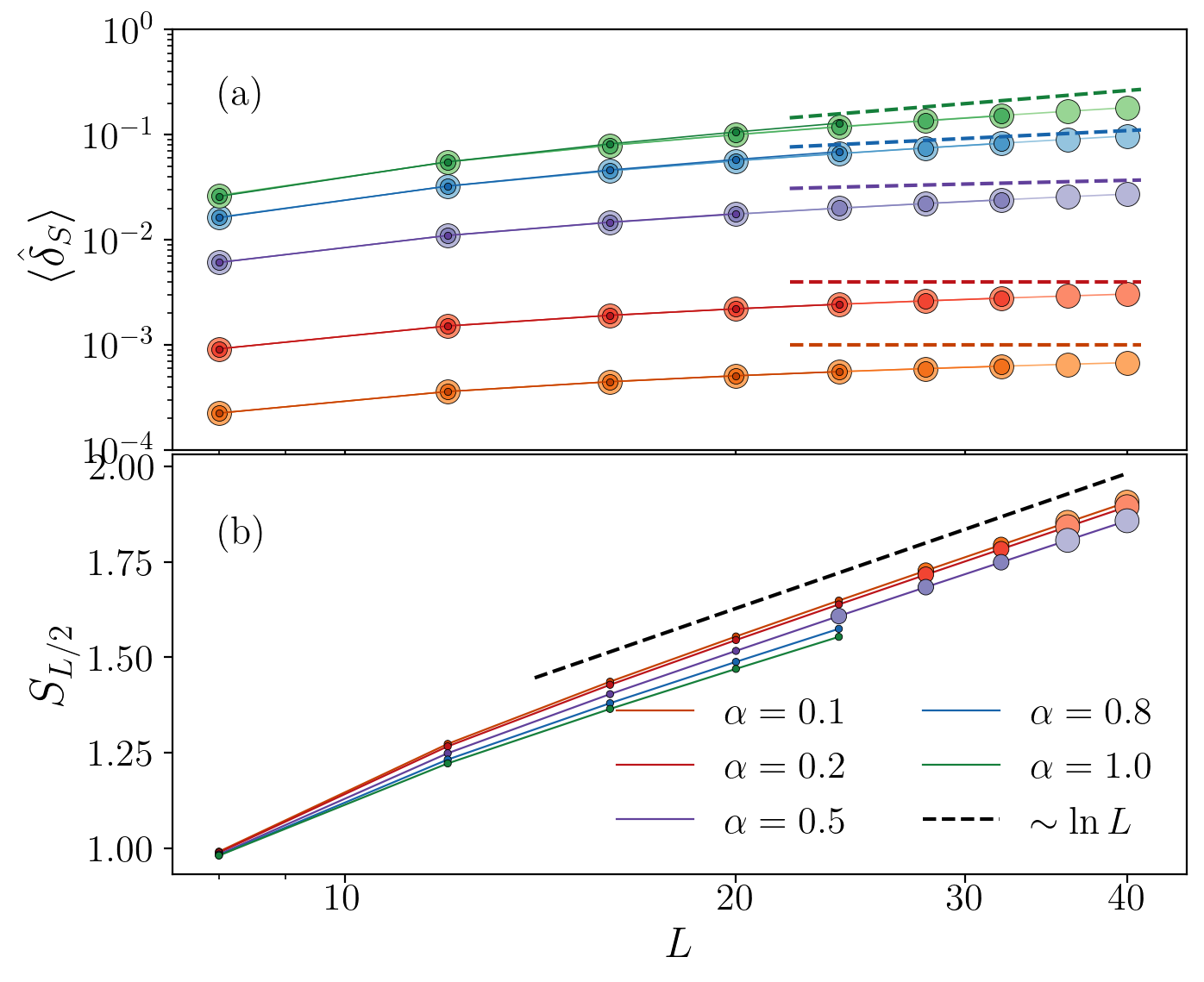}
\caption{
Observables for the large-spin eigenstate closest to zero energy for $\alpha=0.1, 0.2, 0.5, 0.8, 1$ in orange, red, purple, blue, and green respectively. (Top) Spin depletion $\langle\hat\delta_S\rangle$ \eqref{eq:totalS} as a function of system size $L$ for increasing $N_\text{{shell}}=1, 3, 4$ represented by dots with decreasing sizes. {Slight discrepancies between $N_\text{{shell}}=3$ and $N_\text{{shell}}=4$ are visible for~$\alpha=0.8$ and~$1$. 
} The dashed lines are given by the theory prediction in Eq.~\eqref{eq_depletionestimate} below. 
(Bottom) Half-chain entanglement entropy scaling $S_{L/2}$ with $L$. Only data converged with $N_\text{shell}$ are reported.  
}
\label{fig_proj_obs}
\end{figure}

Our numerical data suggest that the large-spin eigenstates exhibit a \emph{sub-extensive growth of the spin depletion $\langle \hat\delta_ S\rangle $} and a \emph{logarithmic scaling of the half-chain entanglement entropy} $S_{L/2}$  with system size $L$ for $\alpha\lesssim1$. 
{In Fig.~\ref{fig_proj_obs}a we plot $\langle \hat\delta _S\rangle_{E_{n(L)}}$ as a function of $L$ for the large-spin eigenstate 
$|E_{n(L)}\rangle$ with energy density $E_{n(L)}/L$ closest to the infinite-temperature value $0$, for various $0<\alpha\le1$.}
We compute these quantities up to $L=40$. Data for increasing $N_\text{{shell}}$ are represented by dots with decreasing size.
Convergence is excellent  even with very small $N_\text{{shell}}$ for $\alpha\lesssim 0.5$, while it is slower for larger $\alpha$. Results strongly suggest a sublinear scaling of $\langle \hat\delta_S\rangle$ with $L$, consistent with saturation to a finite value for small $\alpha$.  
Data for the entanglement entropy, shown in Fig.~\ref{fig_proj_obs}b, are consistent with a logarithmic scaling with system size $L$, {indicative of non-thermal long-range quantum correlations in these eigenstates}. 
Note that for each value of $L$ we used the largest $N_\text{{shell}}$ for which observables are converged. This constrained us to smaller $L$ for larger~$\alpha$.\\

\begin{figure}[t]
\centering
\includegraphics[width=.46\textwidth]{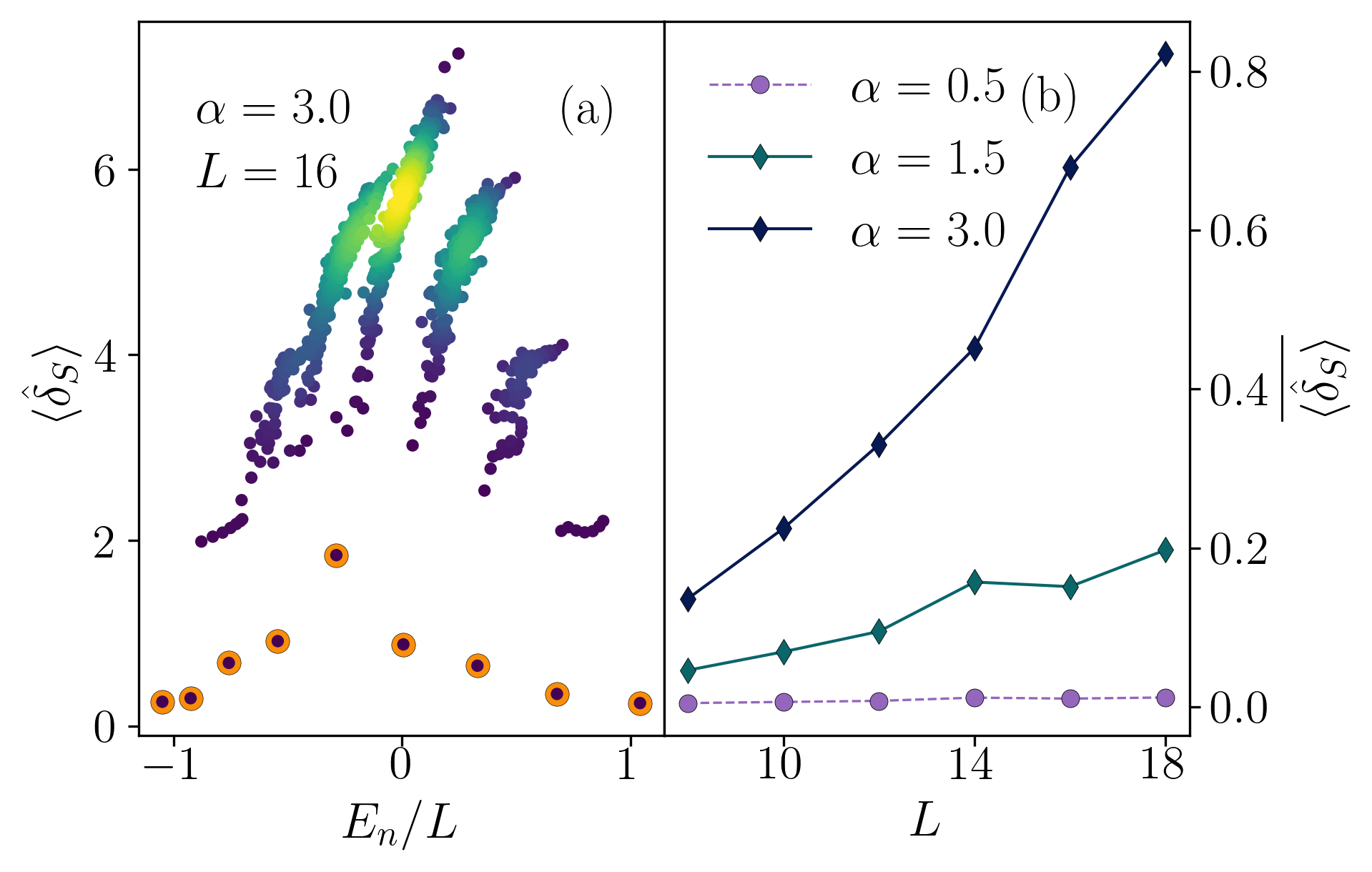}
\caption{
Eigenstates' spin depletion $\langle \hat\delta_S\rangle $ in the quantum Ising chain for shorter interaction range $\alpha>1$. (a) Scatter plot of the exact energy spectrum for $\alpha=3$ with $L=16$ and ${h=2J_0=1}$. (b) Average spin depletion $\langle \delta_S\rangle_{E_n} $ of the $L/2+1$ eigenstates with smallest value [cf. large orange dots in panel (a)]  as a function of the system size $L$ for increasing $\alpha =0.5,1.5, %2.6, 
3$. {The heatmap color illustrates the density of eigenstates, from darkest (isolated eigenstates) to lightest (maximum density).} 
}
\label{fig_alpha_large}
\end{figure}

These results must be contrasted to the behavior found for $\alpha\gg 1$, where large-spin eigenstates exhibit a tendency to delocalize through the many-body spectrum. This is illustrated in Fig.~\ref{fig_alpha_large}. In panel (a) we report the eigenstates' spin depletion $\langle \hat\delta_S\rangle_{E_n} $ for $\alpha=3$ and $L=16$. While largest-spin eigenstates can still be singled out (large orange dots), the spin depletion is much larger than for $0<\alpha<1$, cf. Fig.~\ref{fig_stable_L18} row a. Analysis for various sizes $L$ suggests, furthermore, that the separation of these eigenstates from the thermal bulk is only a finite-size effect (Fig.~\ref{fig_alpha_large}b). The spin depletion $\langle \hat\delta_S\rangle_{E_n}$ averaged over the $L/2+1$ eigenstates with smallest value increases approximately linearly with $L$ for $\alpha>1$, in contrast with the strong suppression found for $\alpha<1$. \\

In summary,  we have shown that the numerical energy spectrum of the long-range quantum Ising chain exhibits non-thermal features compatible with a QMBS scenario. The system has a vanishing fraction of eigenstates characterized by \textit{a)} the depletion from maximal collective spin size $\langle \hat\delta_S\rangle $ scaling sub-extensively with $L$, \textit{b)} logarithmic scaling of the half-chain entanglement entropy $S_{L/2}\sim \ln L$ [cf. Fig.~\ref{fig_proj_obs}] and \textit{c)} large overlap with highly polarized states [cf. Fig.~\ref{fig_stable_L18}]. 
Our numerical data suggest that these non-thermal features may persist for arbitrarily large system size when $\alpha\lesssim1$.
Below we will back this numerical evidence with an analytical construction of these  QMBS eigenstates, which confirms the suggested stability and precisely delimits their robustness. 

{As was done in Fig.~\ref{fig_stable_L18}, in all plots of numerical spectra reported in this work we will tag by large orange dots the primary set of QMBS, defined as the (at most $L/2+1$) energy eigenstates with smallest $\langle \hat \delta_S\rangle$, that are \emph{smoothly} connected with the Dicke manifold $\delta_S=0$ as $\alpha\to0$. }

We note that in the limit $\gamma=0$ of Eq.~\eqref{eq_lrxy} the QMBS smoothly reduce to the well-known \textit{Anderson tower of states} associated with the spontaneous breaking of the continuous rotational $U(1)$ symmetry~\cite{anderson1952approximate,tasaki2019long}. In this limit, however, their stability for arbitrary $L$ is \textit{a priori} guaranteed by the symmetry, as they are the {ground states} in the various symmetry sectors.

\subsection{Initial state dependence in dynamics}
\label{sec_inistate}

\begin{figure}[t]
\centering
\hspace{-0.5cm}
\includegraphics[width=.5\textwidth]{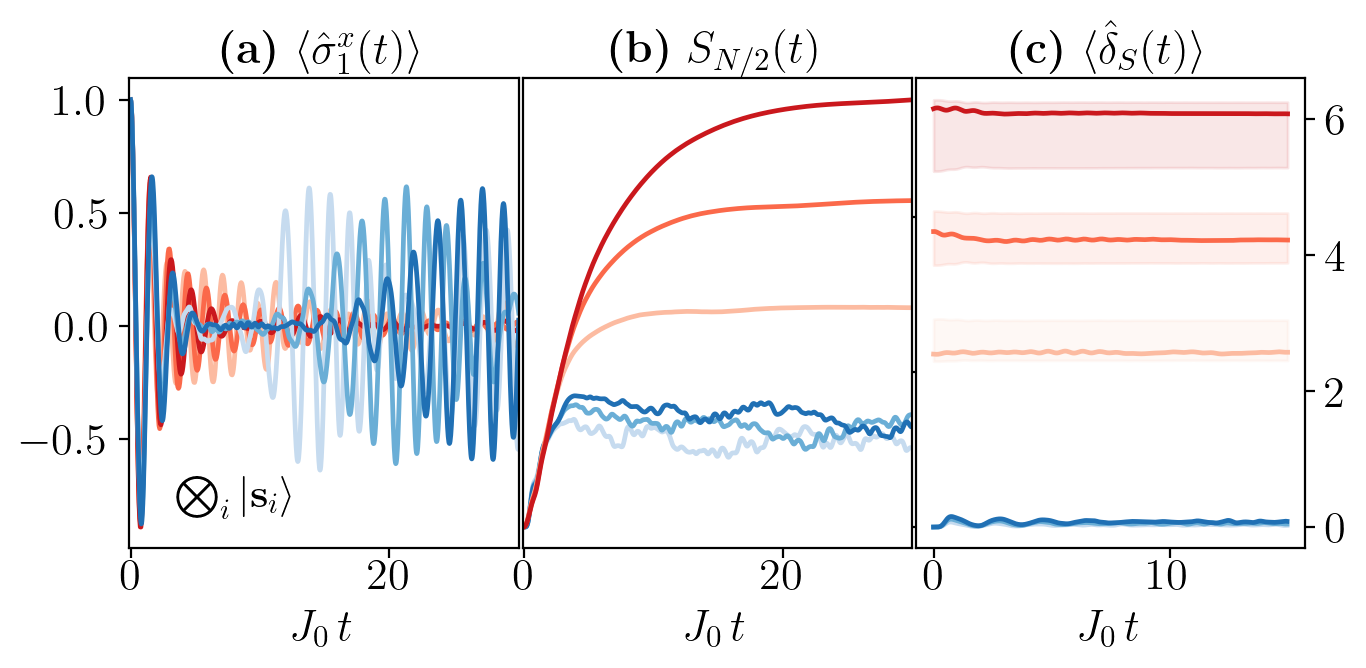}
\caption{Dynamics of a long-range quantum Ising chain [Eq.~(4) with $\gamma=h=1$,  {$J_0=1/2$}, $\alpha=0.7$] initialized in the fully polarized state $|\Rightarrow\rangle$ (blue lines) and in random product states defined in the main text (red lines) for system sizes $L=10,14,18$ in increasing color intensity. (a) Single-site spin polarization dynamics {$\langle \psi(t) |\hat \sigma^x_1|\psi(t)\rangle $}; {here the red lines correspond to individual typical instances of initial random product states}.
(b) Entanglement entropy growth $S_{L/2}(t)$; {here the red  lines are obtained by averaging over $100$ initial random product states}. {(c) Collective spin depletion $\langle \hat \delta _S(t)\rangle$; here the red lines correspond to individual instances of random initial product states, and the shaded areas correspond to the standard deviations over $100$ random instances.
} 
}
\label{fig_dyn}
\end{figure}

The dynamical counterpart of the spectral structure of QMBS found above is given by out-of-equilibrium dynamics depending strongly on the initial condition. For completeness we report evidence of this behavior, exemplified by Fig.~\ref{fig_dyn}, where we contrast initialization in a polarized state (blue) to a random product state (red), subject to evolution governed by a long-range quantum Ising Hamiltonian, Eq.~\eqref{eq_lrxy} with $\gamma=1$ and $\alpha=0.7$.

Polarized states, such as {$\ket{\Rightarrow}=\ket{\rightarrow \dots \rightarrow}$}, have a large overlap with the QMBS [cf. Fig.~\ref{fig_stable_L18}]. Consequently, they
display anomalous non-equilibrium dynamics qualitatively similar to the collective-spin dynamics of the limit $\alpha=0$. Local observables such as the single-site spin polarization {$\braket{\hat \sigma_1^x(t)}$ exhibit oscillations, associated with classical non-linear collective precession, damped over a time scale growing as a fractional power of $L$, followed by sizable revivals over a time scale proportional to $L$;} see blue curves in Fig.~\ref{fig_dyn}a.
The {transient damping, governed by spin-spin interactions, is accompanied by a logarithmic growth of the entanglement entropy in time}, see blue curves in Fig.~\ref{fig_dyn}b, while {the dynamical collective-spin depletion $\langle \hat \delta_S(t)\rangle \gtrsim 0$ remains weak}, see blue curves in Fig.~\ref{fig_dyn}c. This transient phenomenology was numerically observed in Refs.~\cite{schama2013entro,buyskikh2016entanglement} and eventually quantitatively explained via a semiclassical analytical theory~\cite{SacredLog}. (Qualitatively different behavior occurs in proximity to classical phase-space instabilities \cite{SacredLog,mori2018prethermalization}, cf. Sec.~\ref{sec_ESQPT} below.) The stability of QMBS for arbitrary system size --- surmised by our numerical results above and corroborated by our analytical results below --- implies persistence of the anomalous dynamical features for all times.

We contrast this behavior with dynamics initialized in a random product state such as $|\psi_0\rangle  ={ |\rightarrow_1\rangle }|\mathbf{s}_2\rangle \dots |\mathbf{s}_L\rangle $
where the first spin points along $z$ while other spins
 point in random uncorrelated directions on the Bloch sphere, $|\mathbf{s}_i\rangle = \cos(\theta_i/2) |\rightarrow_i\rangle + e^{i \phi_i}\sin(\theta_i/2) |\leftarrow_i\rangle $ with $\theta_i\in[0, \pi)$ and $\phi_i\in [0, 2 \pi)$. 
In this case, the initial collective spin size is typically low ($\langle\hat\delta_S\rangle$ large)
and, accordingly, dynamics are expected to be compatible with a standard thermalization scenario. Here the local spin polarization 
attains the thermal value after a few oscillations, without revivals; the entanglement entropy displays rapid (linear) growth in time with volume-law saturation; {the collective spin depletion $\langle\hat\delta_S(t)\rangle$ remains large at all times}  
[see {the red curves in the corresponding panels of Fig.~\ref{fig_dyn}a-c}].

\section{Quantum rotor-magnon theory}
\label{sec_theory}

 In this Section we analyze spectral properties at finite~$\alpha$ using an analytical approach we develop. This allows us to formulate a predictive theory for the stability of non-thermal eigenstates observed in numerics.
 
 Taking the exact solution of the $\alpha=0$ model as starting point (Secs.~\ref{subsec_alphazero}), we interpret its spectral degeneracy in terms of magnon excitations (Sec.~\ref{subsec_bosonlabelling}). Hence, we provide basic intuition for the persistence of a few non-thermal eigenstates  for small finite $\alpha$ (Sec.~\ref{subsec_intuition}).
 Building on this intuition we systematically map the original spin Hamiltonian to that of a collective spin interacting with $L-1$ bosonic modes (Sec.~\ref{subsec_matrixelements}), and
 we demonstrate that this mapping takes a particularly simple and physically suggestive form in the large-spin sectors (Sec.~\ref{subsec_rotorsw}).
 Remarkably, we show that the resulting effective model is an interacting impurity model, for which we find an exact solution  (Sec.~\ref{subsec_exactsol}), and we compute the spreading of energy eigenstates in Hilbert space as $\alpha$ is increased (Sec.~\ref{subsec_selfconsistency}). 
 These results show that \emph{non-thermal eigenstates are stable for arbitrary system size $L$ when $0<\alpha<1$ in absence of semi-classical chaos}.
 We conclude this Section by expounding the generality of our theory (Sec.~\ref{subsec_generality}). We make reference to Fig.~\ref{fig_summary1} for a visual summary of the main steps of this Section.
 
 We note that a related rotor-magnon description of the \emph{low-energy} spectrum  has been recently developed by Roscilde et al.~\cite{roscilde2023entangling,roscilde2023rotor}.

\subsection{Spectrum and eigenstates in the infinite-range limit ($\alpha=0$)}
\label{subsec_alphazero}

 {The Hamiltonian~\eqref{eq_lrxy} with infinite interaction range ($\alpha=0$) is  the much-studied Lipkin-Meshkov-Glick model of nuclear physics~\cite{lipkin1965validity,Romera_2014}.} In this Subsection we review its exact spectrum~\cite{bapst2012quantum,mazza2012dynamical,SciollaBiroliMF,ribeiro2008exact}, which plays a central role in the following theory.

The Hamiltonian $\hat H_{\alpha=0}$  is invariant under all spin permutations. It reduces to a function of the collective spin components $\hat S^{x,y,z}= \frac 1 2 \sum_{j=1}^L \hat \sigma^{x,y,z}_j$:\footnote{{Note that in the literature the Lipkin-Meshkov-Glick model is often parametrized as $\hat H = \epsilon\left[-\frac{2\xi}{S} (\hat S^x)^2 + (1-\xi)(\hat S^z + S )\right]$ (see e.g. Ref.~\cite{santos2016excited}); up to an overall energy shift this corresponds to our Eq.~\eqref{eq_lrxy0} upon identifying $h/J_0=\frac{\xi-1}{2\xi}$.}}
\begin{equation}
\label{eq_lrxy0}
\hat H_{\alpha=0} %&
= - \frac{4 J_0}{L}   \bigg[ \frac{1+\gamma} 2 \big(\hat S^x\big)^2
+ \frac{1-\gamma} 2 \big(\hat S^y\big)^2 \bigg]
 - 2 h \hat S^z \, .
\end{equation}
This Hamiltonian manifestly conserves the collective spin size $\hat {\bold S}^2 = (\hat S^x)^2 + (\hat S^y)^2 + (\hat S^z)^2$.
The eigenvalues of $\hat {\bold S}^2$ can be written as $S(S+1)$, with 
\be
S=L/2-\delta_S, \qquad \delta_S=0,1,2,\dots,\floor[\bigg]{ \frac L 2 }.
\ee
 The good quantum number $\delta_S$ quantifies the discrepancy of the collective spin size from its maximal value (cf. Fig.~\ref{fig_stable_L18}a1).

Hilbert space sector associated with $\delta_S$ contains $d_{L,\delta_S}$ identical copies of a spin-$S$ representation of $SU(2)$, where
\be
d_{L,\delta_S} =  \binom{L}{\delta_S} - \binom{L}{\delta_S-1} . 
\ee
In each such $(2S+1)$-dimensional space the Hamiltonian acts as Eq.~\eqref{eq_lrxy0} thought as the Hamiltonian of a single spin of size $S$.
We denote by 
\be
\label{eq_spectrumlabel}
E_{\delta_S,N}, \qquad N=0,1,\dots,L-2\delta_S,
\ee
 the ``tower'' of $L-2\delta_S+1=2S+1$ single-spin energy eigenvalues in the block $\delta_S$. 
 We label the eigenstates as
 \be 
\label{eq_eigenstatelabel}
 | \delta_S,\kappa\, ;N \rangle 
\ee
where the additional %``silent'' 
quantum number $\kappa$ parametrizes the $d_{L,\delta_S}$-fold degeneracy. Thus, in this expression, the pair $(\delta_S,\kappa)$ identifies a $SU(2)$ tower and $N$ identifies the eigenstate within that tower.

In all the towers with large collective spin $S = \rho L/2$ ($0<  \rho\le 1$)
 the single-spin Hamiltonian approaches a semiclassical limit as $L\to\infty$. To see this, observe that a general infinite-range spin-$1/2$ Hamiltonian can be written as
\be
\hat H_{\alpha=0} = S \, \mathcal{H}( \hat{\bold{s}};\rho ), \qquad  \hat{\bold{s}}\equiv \bigg(\frac{\hat S^x} S,\frac{\hat S^y} S, \frac{\hat S^z} S\bigg)
\ee
where $\mathcal{H}$ is a classical smooth function and $\rho$ enters as a parameter.
[For our model,  $\mathcal{H}(x,y,z;\rho)= - 2 J_0 \rho    \Big( \frac{1+\gamma} 2 x^2
+ \frac{1-\gamma} 2 y^2 \Big)
 - 2 h z $.]
 The Schr\"odinger equation reads
 \be
 i \frac 1 S \partial_t |\psi \rangle = \mathcal{H}( \hat{\bold{s}};\rho )| \psi \rangle,
 \ee
 where $|\psi\rangle$ is a state in the $(2S+1)$-dimensional tower,  and the rescaled spin variables satisfy 
 \be
 [\hat s^\alpha, \hat s^\beta] = i \frac 1 S \epsilon^{\alpha\beta\gamma} \hat s^\gamma.
 \ee
 Thus, the system has an effective $\hbar_{\rm {eff}}= 1 / S$, and the thermodynamic limit coincides with the classical limit.
The leading behavior of spectrum, eigenstates, and dynamics as $L\to\infty$ can be computed by standard semiclassical tools applied to $\mathcal{H}$, i.e. Bohr-Sommerfeld quantization rule, Wenzel-Kramers-Brillouin (WKB) and truncated Wigner approximation, respectively.

For the purposes of this paper, the following considerations suffice.
The single-spin classical Hamiltonian $\mathcal{H}$ is (trivially) Liouville-integrable, and it can be recast to action-angle variables.
For a given value of the spin length $\rho$ we parametrize the classical phase space with canonical variables $q,p$ [for example
$
\bold{s} = \;\big ( \sqrt{1-p^2} \cos q, \sqrt{1-p^2} \sin q, p\big)
$].
Hence, we can rewrite
\be
\label{eq_aarepr}
\mathcal{H}(\bold{s};\rho) \mapsto \mathcal{H}(n;\rho)
\ee
where 
$
n = \frac 1 {2\pi} \oint_E p \,dq 
$
is the classical action variable, corresponding to the phase-space area enclosed by a trajectory.  Parametrizing trajectories by their energy $E$, the representation~\eqref{eq_aarepr} can be found by inverting the equation 
 $n=n(E;\rho)$.
By construction, $\mathcal{H}$ does not depend on the angle $\varphi$ canonically conjugated to $n$.
\begin{figure}
\centering
\includegraphics[width=0.44\textwidth]{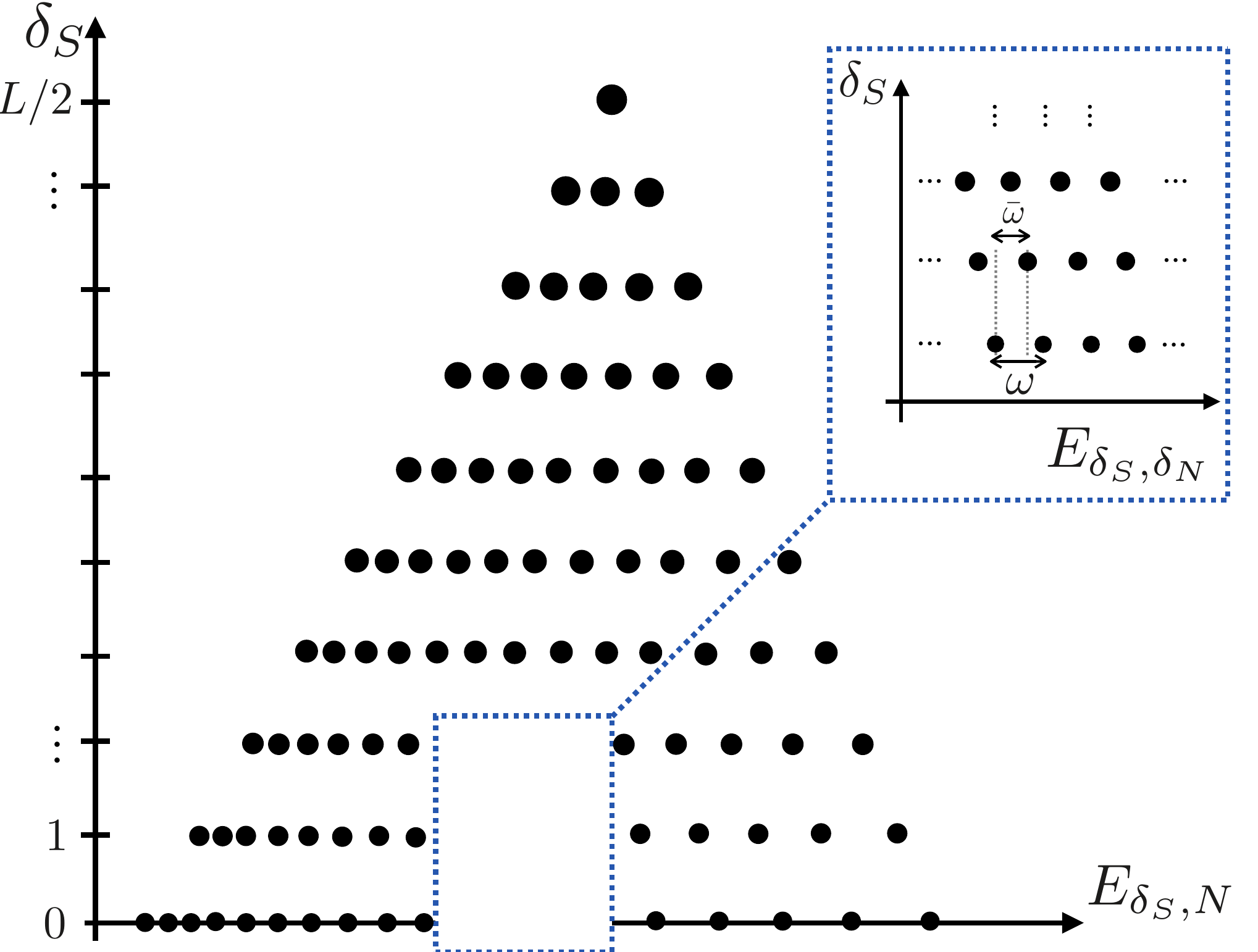}
\caption{Sketch of the unperturbed spectrum of $\hat H_{\alpha=0}$, cf. Eq.~\eqref{eq_fullalpha0spectrum}. 
Towers of energy levels associated with distinct $SU(2)$ irreps appear as rows;  
 level thickness graphically denotes the degeneracy $d_{L,\delta_S}$. The inset highlights the locally regular 2d lattice at a given energy density and $\delta_S \ll L/2$, characterized by the two frequencies $\omega$ and $\bar\omega$, cf.  Eq.~\eqref{eq_alpha0spectrum}.
}
\label{fig_alpha0spectrum}
\end{figure}
According to the Bohr-Sommerfeld quantization rule the quantum energy levels can be found by quantizing the classical action in integer multiples of $\hbar_{\rm eff}$. We thus set $n=N/S=N/(\rho L/2)$, where $N=0,1,2,\dots,\rho L$ is the integer quantum number labelling energy levels within the considered tower [cf. Eq.~\eqref{eq_spectrumlabel}].  
The dimensionless {quantum operator $\hat N$ associated with the classical action} is simply defined by $\hat N |N\rangle = N |N\rangle$, and the quantum operators  
 $e^{\pm i\hat\varphi}$ 
 associated with the classical conjugated angle act as raising and lowering operators on the ladder of eigenstates.
On the other hand, the collective spin size $\rho$ is also similarly quantized, $\rho = S/(L/2) = 1- 2\delta_S/L$. Taking $\delta_S \ll L/2$ we can consistently neglect $\mathcal{O}(1/L^2)$ corrections in $n$ and obtain a formula for the semiclassical spectrum to leading order in $\hbar_{\rm {eff}}= 2/L$:
\be
\label{eq_fullalpha0spectrum}
E_{\delta_S,N} \sim \frac L 2  \, \mathcal{H}\bigg( \frac {2N} L ; 1- \frac{2\delta_S} L \bigg).
\ee
This is illustrated in Fig.~\ref{fig_alpha0spectrum}.

The level spacing $E_{\delta_S,N}-E_{\delta_S,{N-1}}$ within a tower has a semiclassical interpretation as $\hbar \omega(n;\rho)$, where $\omega=\partial_n \mathcal{H}$ is the frequency of the classical orbit at the corresponding energy. This quantity varies smoothly as $N$ runs through a tower (at least away from isolated phase-space \textit{separatrices} where $\omega=0$, corresponding to singularities of the action-angle representation of $\mathcal{H}$ {--- cf. the related discussion in Sec.~\ref{sec_ESQPT} below}).
To make the local structure of the semiclassical spectrum manifest, we focus on a given energy shell and set $N = n \, L /2+ \delta_N$, with $\delta_N$ integer. Expanding in $1/L$, we obtain
\be
\label{eq_alpha0spectrum}
\begin{split}
E_{\delta_S,\delta_N}  \; \sim \; & \frac L 2  \, \mathcal{H}\bigg( n +  \frac {2 \delta_N} L ; 1- \frac{2\delta_S} L \bigg) \\
 \sim \; &   L \, \mathcal{E}(n)  %\\& 
 \; + \omega(n)  \, \delta_N + \bar\omega(n) \,  \delta_S %\\&
 \;  + \mathcal{O}\bigg(\frac 1 L \bigg) \, .
\end{split}
\ee
Thus, the energy spectrum around the energy density $\mathcal{E}(n) = \mathcal{H}(n;1)/2$ is built by combining integer multiples of two (positive) fundamental frequencies, $\omega \equiv \partial_n \mathcal{H}$ and $\bar\omega \equiv -\partial_\rho \mathcal{H}$, which vary smoothly with $n$. This is illustrated in the inset of Fig.~\ref{fig_alpha0spectrum}.\\

For later convenience, we note that the $\alpha=0$ eigenstates can be expressed in the language of
 \emph{spin-coherent states}. Here we synthetically report the necessary information on this formalism; a minimalistic self-contained review is in App.~\ref{app_spincoherent}.
Considering the $SU(2)$ tower identified by $(\delta_S,\kappa)$, we define the spin-coherent state with spherical angle $\boldsymbol{\Omega}\equiv(\theta,\phi)$ as the state with maximal spin projection in the direction $\vec n(\boldsymbol{\Omega}) = (\sin\theta\cos\phi,\sin\theta\sin\phi,\cos\theta)$ within the tower:
\be
\vec n(\boldsymbol{\Omega}) \cdot \hat{\vec S} \,  |\delta_S,\kappa;\boldsymbol{\Omega}\rangle = \bigg( \frac L 2 -\delta_S \bigg)  |\delta_S,\kappa;\boldsymbol{\Omega}\rangle.
\ee
The freedom in the choice of relative phases of spin-coherent states can be fixed by choosing a reference one --- usually in the $\vec z$ direction $(\theta=0)$, i.e. $|\delta_S,\kappa;\boldsymbol{\Uparrow}\rangle$ ---  and a particular rotation protocol --- usually the rotation by $\theta$ in the plane generated by $\vec z$ and $\vec n(\boldsymbol{\Omega})$: denoting $\vec{r}(\boldsymbol{\Omega}) \equiv \frac{\vec z \times \vec n(\boldsymbol{\Omega})}{|\vec z \times \vec n(\boldsymbol{\Omega})|}$, one has
\be
\label{eq_spincoherentdef}
|\delta_S,\kappa;\boldsymbol{\Omega}\rangle = \hat U(\boldsymbol{\Omega}) |\delta_S,\kappa;\boldsymbol{\Uparrow}\rangle, \quad \hat U(\boldsymbol{\Omega}) \equiv 
e^{-i \theta \, 
 \vec{r}(\boldsymbol{\Omega})
\cdot \hat{\vec S}} \, .
\ee
For every choice of $\boldsymbol{\Omega}$ we denote the ladder of eigenstates of the corresponding collective spin projection as
\be
\vec n(\boldsymbol{\Omega}) \cdot \hat{\vec S} \, |\delta_S,\kappa;M_{\boldsymbol{\Omega}}\rangle = \bigg( \frac L 2 -\delta_S - M \bigg)  |\delta_S,\kappa;M_{\boldsymbol{\Omega}}\rangle
\ee
for $M=0,1,\dots,L-2\delta_S$.
On varying $\vec n(\boldsymbol{\Omega})$ on the unit sphere, spin-coherent states span the full tower (``overcomplete basis''): one can show that
\be
\begin{split}
\mathbb{1}_{\delta_S,\kappa} = & \sum_{M=0}^{L-2\delta_S} 
|\delta_S,\kappa;M_{\boldsymbol{\Uparrow}}\rangle \langle\delta_S,\kappa;M_{\boldsymbol{\Uparrow}}| \\
= & \frac {L-2\delta_S+1}{4\pi}  \int d\boldsymbol{\Omega} \, |\delta_S,\kappa;\boldsymbol{\Omega}\rangle \langle\delta_S,\kappa;\boldsymbol{\Omega}|
\end{split}
\ee
where $\int d\boldsymbol{\Omega} \equiv \int_{0}^{2\pi}d\phi\int_0^\pi d\theta \sin\theta$.
Thus, every state in the tower can be written as a linear combination of spin-coherent states\footnote{This representation is not unique.}. In particular, for the eigenstates of $H_{\alpha=0}$ we can write
\be
|\delta_S,\kappa;N\rangle = \int d\boldsymbol{\Omega} \, \psi_{\delta_S,N}(\boldsymbol{\Omega}) |\delta_S,\kappa;\boldsymbol{\Omega}\rangle.
\ee
Here we introduced the \emph{coherent-state wave function} 
$\psi_{\delta_S,N}(\boldsymbol{\Omega}) \propto \langle\delta_S,\kappa;\boldsymbol{\Omega}|\delta_S,\kappa;N\rangle $ of the unperturbed eigenstate, which is efficiently accessible to both semiclassical and numerical computations (see App.~\ref{app_spincoherent}).
\\

In this work we will consider the  $\alpha=0$ spectrum and eigenbasis as known input data, which can be  computed either analytically (via a large-$L$ asymptotic expansion) or numerically (via exact diagonalization of the single-spin problem for $S \le  L/2 \approx 10^5$).
 We briefly mention in passing that the considered Lipkin-Meshkov-Glick Hamiltonian is actually Bethe-ansatz solvable~\cite{GGA}; however, this solution is not practically useful for large $L$, and semiclassical/numerical techniques give much easier access to the relevant information.

\subsection{Bosonic labelling of $\alpha=0$ eigenstates}
\label{subsec_bosonlabelling}

For $\alpha=0$ the many-body Hilbert space fragments into many single-spin Hilbert spaces. Each such $SU(2)$ irreducible representation is associated with a regular energy tower, identified by the collective spin size $S$ and by an additional quantum number $\kappa$ [cf. Eq.~\eqref{eq_eigenstatelabel}] that parametrizes the degeneracy of the towers with equal $S$.
As soon as interactions acquire a non-trivial spatial structure, e.g. for $\alpha\ne0$, different towers get coupled and the degeneracy gets lifted. To understand how the perturbation affects the eigenstates, we first need to make sense of $\kappa$ in terms of additional physical degrees of freedom, which %are ``inert''
{do not possess any dynamics}
for $\alpha=0$ and 
{acquire non-trivial dynamics} as $\alpha\ne0$.

In this Subsection we \emph{resolve the degeneracy of {the energy levels of}  $\hat H_{\alpha=0}$  in terms of bosonic excitations (physically interpreted as magnons) with non-vanishing momenta.}
This way, each $\alpha=0$ eigenstate can be identified 
by the collective spin state -- specified by the quantized classical action $N=0,1,\dots,L-2\delta_S$ -- and by 
the magnon content of the state -- specified e.g. by
$\delta_S$ non-vanishing momenta.
This allows us to replace the abstract labelling in Eq.~\eqref{eq_eigenstatelabel} with
\be
\label{eq_eigenstatelabelsw}
| \{ k_1,\dots, k_{\delta_S} \} ; N \rangle , \qquad k_1,\dots,k_{\delta_S} \neq0 \, ,
\ee
or, equivalently, the corresponding string of bosonic occupation numbers $ \{n_k\}_{k\neq0} $, 
with $\delta_S=\sum_{k\neq0}n_k$. 
In the following Subsections we will leverage this description of unperturbed eigenstates to reformulate the finite-range interacting spin Hamiltonian $\hat H_{\alpha\ne0}$ in terms of a large spin coupled to bosons.
\\

For simplicity of presentation we derive the bosonic labelling by working with the simplest collective Hamiltonian $\hat H_{\alpha=0} \mapsto \hat S^z$ {(i.e. with $J_0=0$)}. 
We construct a complete common eigenbasis of $\hat{\bold S}^2$ and $\hat S^z$, labelled by a magnon occupation string $\{n_k\}_{k\ne0}$ and the $z$-magnetization depletion $M$ (where $S=L/2-\delta_S=L/2-\sum_{k\ne0} n_k$ and  $S^z=S - M$).
The eigenstates of an arbitrary $\hat H_{\alpha=0}$ {[e.g. Eq.~\eqref{eq_lrxy0} with $J_0\neq0$]} are linear combinations of the eigenstates of $\hat S^z$ within individual $SU(2)$ towers, i.e. of states with different $M$ but fixed magnon content, as we describe at the end.\\

\begin{figure}
\centering
\includegraphics[width=0.44\textwidth]{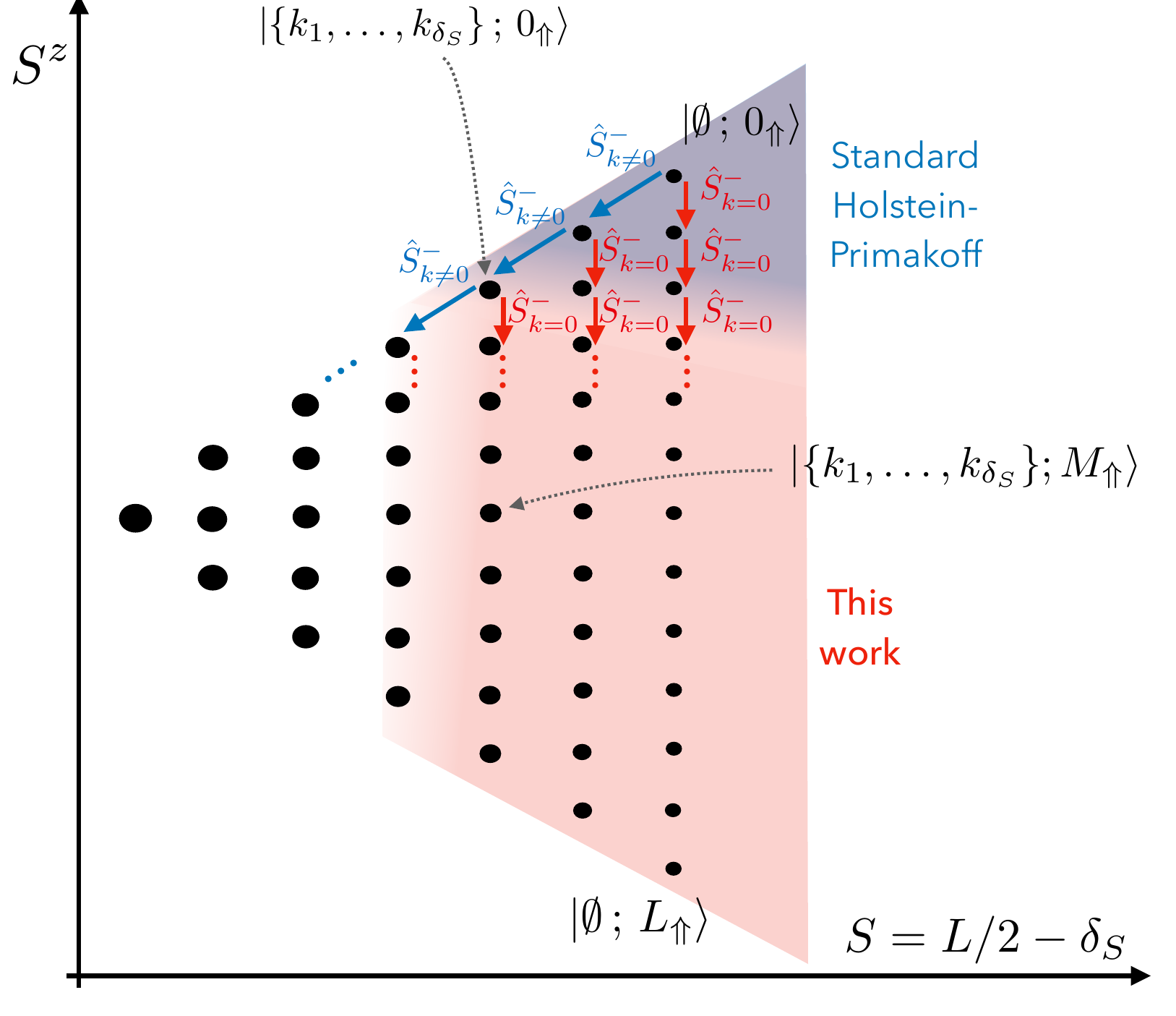}
\caption{
Illustration of the construction of a bosonic labeling.
Here, eigenspaces of $\hat{\bold{S}}^2$ correspond to vertical towers, with degeneracy increasing leftwards. Individual towers are internally spanned by collective operators (here $\tilde{S}^-_{k=0}$). We label the towers with $k\ne0$ magnon excitations created on top of the spin-coherent vacuum. The quasi-orthogonality of dilute magnon states (blue-shaded area) is then transferred through the entire towers, to large-spin states that cannot be easily described with standard Holstein-Primakoff bosons (red-shaded area). We use this bosonic labeling to resolve the degenerate eigenstates of an arbitrary $\hat H_{\alpha=0}$, and to express the matrix elements of an arbitrary $\hat H_{\alpha\ne0}$.
}
\label{fig_bosoniclabelling}
\end{figure}

We  start from the spin-coherent state $ | \boldsymbol{\Uparrow} \, \rangle  \equiv | \up_1\up_2\dots \up_L\, \rangle$  
and act with lowering operators $S^-_k$ [cf. Eq.~\eqref{eq_fourierspin}], generalizing the procedure leading to Clebsch-Gordan coefficients~\cite{landau3}. 
 To follow the construction it is helpful to make reference to the illustration in Fig.~\ref{fig_bosoniclabelling}: 
 \begin{itemize}
\item 
The parent state $ | \boldsymbol{\Uparrow} \, \rangle $ has collective quantum numbers $S=L/2$ (i.e. $\delta_S=0$) and $S^z=L/2$. 
By repeatedly applying the collective lowering operator $\tilde S^-_{k=0}$ to it, we generate the entire $\delta_S=0$ tower {(rightmost column in Fig.~\ref{fig_bosoniclabelling})}.
We normalize such states and denote them $|\emptyset ; M_{\boldsymbol{\Uparrow}} \rangle \propto (\tilde S^-_{k=0})^M {| \emptyset; \boldsymbol{\Uparrow} \, \rangle}$ for $M=0,1,\dots,L$.
(These permutationally-invariant states are known as \emph{Dicke states}~\cite{dicke1954coherence}.)
\item
 To construct the towers composing the $S=L/2-1$ subspace (i.e. $\delta_S=1$) we must act on $ | \boldsymbol{\Uparrow} \,  \rangle $ with spin lowering operators to generate $L-1$ parent states with maximal $S^z=L/2-1$ and orthogonal to the $\delta_S=0$ state $| \emptyset ; 1_{\boldsymbol{\Uparrow}} \, \rangle$ previously constructed {(with reference to Fig.~\ref{fig_bosoniclabelling}, the state $|0;1_\Uparrow\rangle$ corresponds to the second-from-top state in the rightmost column)}. A natural choice for this purpose is to act with $\tilde S^-_{k_1}$, $k_1\neq 0$. This directly yields orthogonal states $ | \{k_1\} ; \boldsymbol{\Uparrow} \rangle \propto \tilde S^-_{k_1} | \boldsymbol{\Uparrow} \, \rangle $ with $\delta_S=1$, which we properly normalize (no need to orthogonalize). By repeatedly applying the collective lowering operator $\tilde S^-_{k=0}$ to each of them, we generate the entire $\delta_S=1$ towers {(second-from-right column in Fig.~\ref{fig_bosoniclabelling})}.
 We normalize such states and denote them $|\{k_1\} ; M_{\boldsymbol{\Uparrow}} \rangle \propto (\tilde S^-_{k=0})^M | \{k_1\} ;\boldsymbol{\Uparrow} \, \rangle$ for $M=0,1,\dots,L-2$.
 \item
  We can then continue along this way: To construct the towers composing the $S=L/2-2$ subspace (i.e. $\delta_S=2$) we act with $\tilde S^-_{k_2}$ on $ | \{k_1\} ; \boldsymbol{\Uparrow} \rangle $ (with $k_1,k_2\neq0$) and orthogonalize with respect to the states with $\delta_S=0,1$, $\delta_S + M_{\boldsymbol{\Uparrow}} = 2$ previously constructed, to generate   $L(L-3)/2$ new parent states with maximal $S^z=L/2-2$. The states $\tilde S^-_{k_2}\tilde S^-_{k_1}  | \boldsymbol{\Uparrow} \, \rangle$ do not directly result in an orthogonal basis of the new subspace,  as they have a small $\mathcal{O}(1/L)$ overlap between themselves and with $\delta_S=0,1$ towers.  In this case, we define the normalized states
$
 |  \{k_1,k_2\} ; \boldsymbol{\Uparrow} \rangle \propto 
 \hat P_{\delta_S=2}  \tilde S^-_{k_2}\tilde S^-_{k_1}  | {\boldsymbol{\Uparrow}} \, \rangle 
 $,
where $k_1,k_2\neq0$ and $\hat P_{\delta_S}$ projects on the corresponding eigenspace of $\hat S^2$ (i.e., removes the components along the previously constructed states). 
These states span the full eigenspace with $\delta_S=2$ and $S^z=L/2-2$. 
They can be used to generate the entire $\delta_S=2$ towers by repeatedly applying the collective lowering operator $\tilde S^-_{k=0}$ to each of them {(third-from-right column in Fig.~\ref{fig_bosoniclabelling})}.
We normalize such states and denote them $|\{k_1,k_2\} ; M_{\boldsymbol{\Uparrow}} \rangle \propto (\tilde S^-_{k=0})^M | \{k_1,k_2\}; \boldsymbol{\Uparrow} \, \rangle$ for $M=0,1,\dots,L-4$.
\item
This procedure can be iterated to break down the full multi-spin Hilbert space into simultaneous eigenspaces of $\hat S^2$ and $\hat S^z$:
For arbitrary $\delta_S$ we define the normalized states
$
 |\{k_1,\dots,k_{\delta_S}\} ; \boldsymbol{\Uparrow} \rangle 
  \propto  \hat P_{\delta_S}  \tilde S^-_{k_{\delta_S}}\dots \tilde S^-_{k_1}  | \boldsymbol{\Uparrow} \, \rangle
$
where $k_1,\dots,k_{\delta_S}\neq0$. 
Such states span the full eigenspace with fixed ${\delta_S}$ and maximal $S^z=L/2-{\delta_S}$. 
By repeatedly applying the collective lowering operator $\tilde S^-_{k=0}$ to each of them we generate the entire towers that compose the ${\delta_S}$ eigenspace.
We normalize such states and denote them
\begin{multline}
\label{eq_defeigenstatelabelsw}
\quad\;\; | \{k_1,\dots,k_{\delta_S}\} ; M_{\boldsymbol{\Uparrow}}  \rangle \\
\propto   ( \tilde S^-_{k=0})^M |\{k_1,\dots,k_{\delta_S}\} ; \boldsymbol{\Uparrow} \rangle
\end{multline}
for $M=0,1,\dots,L-2\delta_S$.
(Note the slight redundancy of notation $| \{k_1,\dots,k_{\delta_S}\} ; 0_{\boldsymbol{\Uparrow}}  \rangle \equiv | \{k_1,\dots,k_{\delta_S}\} ; \boldsymbol{\Uparrow}  \rangle$.)
\end{itemize}

Using this procedure, we constructed states defined by a set of \emph{non-vanishing} momenta $k_1,\dots,k_{\delta_S}\neq0$ and an integer $M=0,1,\dots,L-2\delta_S$.
These states are exact simultaneous eigenstates of $\hat {\bold S}^2$ and $\hat S^z$, and they span the entire $2^L$-dimensional spin space.
This set of states is however massively redundant, as there is no restriction on the set of nonzero momenta.
Indeed, for $\delta_S = \mathcal{O} (L)$, two such states will typically have a finite overlap.
The states defined by Eq.~\eqref{eq_defeigenstatelabelsw} are only really useful when $\delta_S \ll L/2$. In this ``dilute'' regime one can check the \textit{quasi-orthonormality} property:
 \begin{widetext}
 \be
 \label{eq_quasiorthogonality}
\langle \{k'_1,\dots,k'_{\delta_S'}\} ; M'_{\boldsymbol{\Uparrow}} \,  | \, \{k_1,\dots,k_{\delta_S}\} ; M_{\boldsymbol{\Uparrow}} \rangle = \delta_{\delta_S,\delta_S'} \delta_{M,M'} \Big[
\delta_{\{k_1,\dots,k_{\delta_S}\},\{k'_1,\dots,k'_{\delta_S}\}} + \mathcal{O}\big(  \delta_S /L \big) \Big].
\ee
  \end{widetext}
 This property can be traced back to  
 the correspondence between \emph{nearly polarized} spin states with large~$S^z$ and dilute bosonic states, realized by the Holstein-Primakoff transformation.
 To see this, let us define the mapping
\be
\label{eq_HPlocal}
 \hat s^-_j %&
 \mapsto
      \hat b^\dagger_j (1- \hat b^\dagger_j \hat b_j ), \;
   % \\
   \hat s^+_j %&
   \mapsto   (1- \hat b^\dagger_j  \hat b_j )  \hat b_j, \;
    %\\
     \hat s^z_j %&
     \mapsto \frac 1 2  -   \hat b^\dagger_j \hat b_j,
\ee
where $\hat b_j,\hat b^\dagger_j$ are canonical bosonic annihilation and creation operators. {[Note that this is an equivalent simpler version of the standard exact Holstein-Primakoff transformation for $s=1/2$, see e.g. Ref.~\cite{vogl2020resummation}; our analysis does \textit{not} depend on this choice.]}
 The mapping~\eqref{eq_HPlocal} should be understood as an \emph{embedding} of the two-dimensional Hilbert space of a  spin-$1/2$ in the infinite-dimensional Hilbert space of a bosonic mode, with $|\up\,\rangle$ and $|\down\,\rangle$ mapped onto $|0\rangle\equiv |\emptyset\rangle$ and $|1\rangle \equiv b^\dagger|\emptyset\rangle$ respectively. 
 It is useful to define bosonic Fock states 
 \be
 \label{eq_bosonicfockstate}
 | \{n_k\} \rangle \equiv   \prod_{k} \frac {\big(\tilde b^\dagger_{k}\big)^{n_k}}{\sqrt{n_k !}} | \emptyset \rangle \, ,
 \ee
 where $\tilde b^\dagger_{k}=(1/\sqrt{L})\sum_j e^{ikj} b^\dagger_j$ creates a boson with momentum $k$ and $\{n_k\}$ is a string of occupation numbers.
 In App.~\ref{app_overlapspinboson} we show that, in the dilute regime, the \emph{maximally polarized} (i.e. $M=0$) spin states $| \{k_1,\dots,k_{\delta_S}\} ; \boldsymbol{\Uparrow}  \rangle$ defined above are close to the corresponding bosonic Fock states with unoccupied $k=0$ mode:
 \be
 \label{eq_overlapspinboson}
\langle  \{n_k\} | \{k_1,\dots,k_{\delta_S}\} ; \boldsymbol{\Uparrow} \rangle =
  \delta_{\{n_k\},\{k_1,\dots,k_{\delta_S}\}}
   + \mathcal{O}\big(  \delta_S /L \big) \, .
 \ee
 (Here, with a little abuse of notation, the Kronecker delta on the right-hand side forces the set of momenta counted with their multiplicities to be that identified by the string of occupation numbers.)
 Since Fock states are orthonormal, Eq.~\eqref{eq_overlapspinboson} immediately implies Eq.~\eqref{eq_quasiorthogonality} for $M=M'=0$. Hence, crucially, the quasi-orthonormality of the parent maximally polarized states is transferred down throughout the respective $SU(2)$ towers, to states $| \{k_1,\dots,k_{\delta_S}\} ; M_{\boldsymbol{\Uparrow}}  \rangle$ with arbitrary $M$.
 
 As illustrated in Fig.~\ref{fig_bosoniclabelling}, the construction above non-trivially extends  the standard Holstein-Primakoff mapping between spins and bosons. 
 Indeed, when not only $\delta_S\ll L/2$ but also  $M\ll L$, the  correspondence between spin states and bosonic states is one-to-one:
 \begin{multline}
 \label{eq_overlapspinbosonpolarized}
\langle  \{n_k\} | \{k_1,\dots,k_{\delta_S}\} ; M_{\boldsymbol{\Uparrow}} \rangle =
\\
  \delta_{n_{k=0},M} \, \delta_{\{n_k\}_{k\neq0},\{k_1,\dots,k_{\delta_S}\}} 
   + \mathcal{O}\big(  (M+\delta_S) /L \big).
 \end{multline}
 In other words, when the spin state is close to fully polarized ($L/2-S^z \ll L $), the quantum number $M$ can simply be identified with the occupation of the $k=0$ bosonic mode.
 In this corner of the multi-spin Hilbert space (blue-shaded area in Fig.~\ref{fig_bosoniclabelling}), the correspondence between nearly fully polarized spin states and dilute bosonic Fock states is complete.

Equation~\eqref{eq_quasiorthogonality} shows however that  spin states  $| \{k_1,\dots,k_{\delta_S}\} ; M_{\boldsymbol{\Uparrow}} \rangle$ with 
$\delta_S\ll L/2$ but \emph{arbitrarily large} $M$ can also be accurately labelled by their magnon content $\{n_k\}_{k\neq0}$ (red-shaded area in Fig.~\ref{fig_bosoniclabelling}). 
Loosely speaking, Eq.~\eqref{eq_quasiorthogonality} allows us to factorize the Hilbert space of $L$ spins as direct product of a large spin  and a Fock space of $L-1$ bosonic modes, which is accurate in the subspace $\delta_S \ll L/2$ with dilute bosons.
 It is tempting to picture the state $| \{k_1,\dots,k_{\delta_S}\} ; M_{\boldsymbol{\Uparrow}} \rangle$ as a macroscopic ``condensate'' of $M$ collective magnon excitations with $k=0$  
accompanied by dilute magnons with non-vanishing momenta $k_1,\dots,k_{\delta_S}\neq0$,
as suggested by the expression in Eq.~\eqref{eq_defeigenstatelabelsw}.  
One must keep in mind, however, that 
for extensive $M=\mathcal{O}(L)$ the total population $M+\delta_S =\mathcal{O}(L)$ of Holstein-Primakoff bosons is far from dilute, making the naive replacement of $\tilde S_{k=0}^-$ (the actual creation operator  of collective spin excitations) by $\tilde b^\dagger_{k=0}$ (the Holstein-Primakoff creation operator)  highly inaccurate. 
 The spin state $| \{k_1,\dots,k_{\delta_S}\} ; M_{\boldsymbol{\Uparrow}} \rangle$ 
 actually has exponentially-small-in-$M$ overlap with its candidate  bosonic partner $|n_0=M,\{n_k\}_{k\neq0}\rangle$. [In fact, it is straightforward to check that only an exponentially small fraction of the latter bosonic Fock state belongs to the physical spin space!]

 The set of {non-vanishing} momenta $\{n_k\}_{k\neq0}$ determines both $\delta_S = \sum_{k\neq0} n_k$ and the %silent 
 quantum number $\kappa$ appearing in Eq.~\eqref{eq_eigenstatelabel}, with accuracy $\mathcal{O}\big(  \delta_S /L \big)$.
In the construction above, states within each $SU(2)$ tower are identified by the eigenvalue of the collective operator $\hat S^z$. As anticipated above, the role of $\hat S^z$ can however be played as well by the quantized action operator $\hat N$ associated with an arbitrary collective Hamiltonian $\hat H_{\alpha=0}$. Correspondingly, the role of~$\tilde S_{k=0}^-$ is played by~$e^{i\hat\varphi}$.  
This way, states within each tower are identified by $N=0,1,\dots,L-2\delta_S$, 
 \be
 \label{eq_swlabellingH}
 |\{k_1,\dots,k_{\delta_S}\} ; N \rangle 
 \ee
 in place of the collective $z$-magnetization depletion $M=0,1,\dots,L-2\delta_S$.
 This labelling is equivalent to Eq.~\eqref{eq_eigenstatelabel} upon identifying $(\delta_S,\kappa) \leftrightarrow \{k_1,\dots,k_{\delta_S}\} \leftrightarrow \{ n_k\}_{k\neq0}$.
 This is the conclusion we aimed for.

\subsection{Intuition on eigenstate localization for $0<\alpha\ll  1$}

\label{subsec_intuition}

The fully connected Hamiltonian $\hat H_{\alpha=0}$ features an extensive tower of ``ground-state-like'' wavefunctions --- the $\delta_S=0$ states --- running throughout the many-body energy spectrum. From a thermodynamic viewpoint such states are anomalous highly-excited states: Differently from typical eigenstates in a given energy shell, they have large collective spin, low entanglement entropy, and large overlap with product states. Each such state finds exponentially many other states at equal energy density with normal properties: small collective spin, extensive entanglement entropy, and vanishing overlap with product states. As $\alpha>0$ states with different collective spin length get dynamically connected, and hybridization of the ``few'' anomalous states with small $\delta_S$ with the ``many'' normal states with $\delta_S = \mathcal{O}( L)$ becomes \textit{a priori} favorable. As discussed in Sec.~\ref{sec_numerics}, the numerical results indicate a different scenario, showing a weak or suppressed tendency towards hybridization.

We argue here --- and demonstrate below via explicit analytical arguments --- that  hybridization of non-thermal eigenstates is hindered for small $\alpha$ by a combination of two crucial ingredients: 
\begin{itemize}
\item[\textit{i)}] the suppression of matrix elements $\propto f_k(\alpha)$, ascribed to the long range of the interactions;
\item[\textit{ii)}] the {generic} absence of energy resonances in the subspaces $\delta_S \ll L/2$, ascribed to the classical integrability of the mean-field limit.
\end{itemize}

\begin{figure*}
\centering
\includegraphics[width=0.9\textwidth]{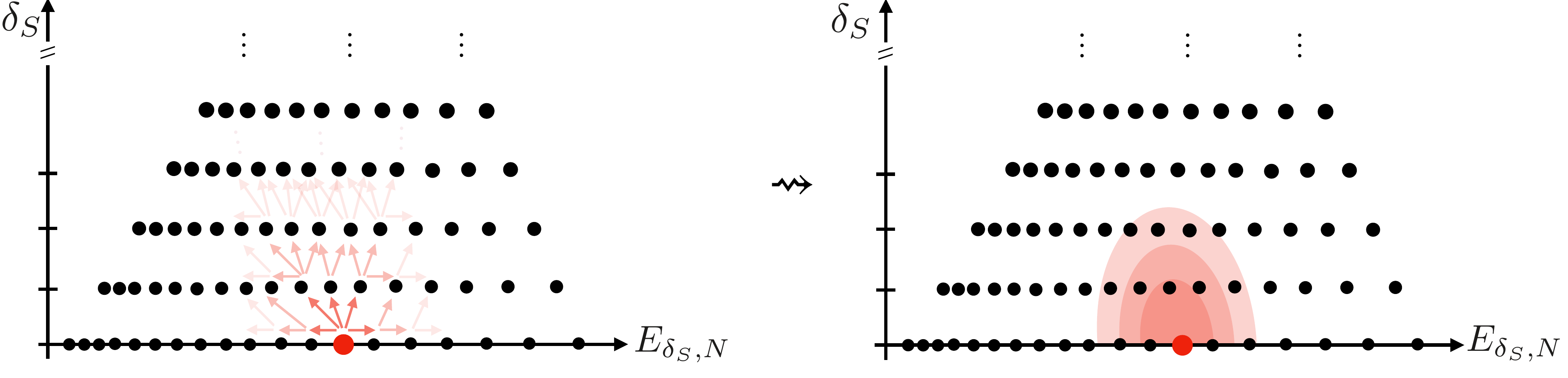}
\caption{Sketch of the intuitive argument for eigenstate localization in the main text.
Left panel: Transitions generated by $\hat V_\alpha$ are quasilocal on the graph associated with the unperturbed $\alpha=0$ spectrum. Thus, in stark contrast with conventional many-body integrability-breaking scenarios, perturbative transitions encounter finite energy mismatches. The shading of the arrows denotes the resulting amount of hybridization generated at the perturbative order for which the corresponding transition comes into play. 
Right panel: For $\alpha<1$, we demonstrate that an exact eigenstate originating from the $\delta_S \ll L/2$ subspace stays localized in a local neighbourhood, illustrated by the red-shaded area; the size of this neighborhood is sub-extensive. For more details we refer to the main text of Sec.~\ref{subsec_intuition}.
}
\label{fig_eigenstatedeloc}
\end{figure*}

This mechanism is pictorially illustrated in Fig.~\ref{fig_eigenstatedeloc}. The graph therein shows the energy levels of $\hat H_{\alpha=0}$ arranged in a smoothly varying 2d lattice spanned by the integers quantum numbers $(\delta_S,\delta_N)$, cf. Eq.~\eqref{eq_alpha0spectrum}, connected by shaded arrows representing matrix elements for $\alpha\neq0$. As the interaction operator flips at most two spins, each matrix element connects states with $\delta_S$ differing by at most $2$ --- and we will interpret the change in $\delta_S$ as creation and annihilation of magnons.\footnote{For a Hamiltonian with $p$-spin interactions, matrix elements connect states with $\delta_S$ differing at most by $p$.} At the same time, the dependency of matrix elements on the collective spin variables is rooted in semiclassics, resulting in analytic functions of $n=N/L$ and $\varphi$ (which will be rigorously shown below). These remarks show that transitions generated by the perturbation $\alpha\neq0$ are \emph{quasi-local} on the $(\delta_S,\delta_N)$-lattice. As the two fundamental frequencies $\omega,\bar\omega$ that locally define the lattice smoothly vary with $n$, for a generic choice of energy shell they are typically \emph{non-resonant} (i.e. incommensurate). This condition prevents energy denominators to become arbitrarily small at any given order in perturbation theory~\cite{giorgilli2022notes}.
The simultaneous suppression of matrix elements for small $\alpha$ thus prevents the occurrence of energy resonances to all orders, hindering the delocalization of perturbed eigenstates over the graph in Fig.~\ref{fig_alpha0spectrum}. This scenario is to be contrasted with conventional integrability-breaking scenarios in quantum many-body systems, where an unperturbed eigenstates is generically coupled to a continuum of other eigenstates at equal energy, resulting in a Fermi-golden-rule-type hybridization.

Remarkably, the basic mechanism illustrated above finds a natural counterpart in an emergent \emph{exact solvability} of the perturbed energy eigenstates originating from the subspaces $\delta_S\ll L/2$. The rest of this Section is devoted to this demonstration, which constitutes the main theoretical result of this paper. Combined with the observation of level repulsion
 at arbitrarily small $\alpha>0$ reported in Sec.~\ref{sec_numerics}, this result proves the coexistence of robust QMBS and a quantum-chaotic bulk many-body spectrum in long-range interacting systems, which is the main message of this paper.

\subsection{Matrix elements of $\hat H_{\alpha}$}

\label{subsec_matrixelements}

In Sec.~\ref{subsec_alphazero} we reviewed how $\hat H_{\alpha=0}$ {represents the dynamics of} a single collective degree of freedom, and in Sec.~\ref{subsec_bosonlabelling} we showed how the remaining $L-1$ dynamically frozen degrees of freedom can be conveniently expressed as bosonic modes --- physically associated with magnon excitations with non-vanishing momenta. 
It is natural to ask how the variable-range perturbation $\hat V_\alpha$ is expressed in terms of these degrees of freedom.
Considering the unperturbed spectrum as in Fig.~\ref{fig_eigenstatedeloc},
transitions between different towers --- generated by $\hat V_\alpha$ (shaded arrows) --- correspond to creation or destruction of few magnons 
simultaneously with jumps of the collective spin quantum number $N$. 
The goal of this and the next Subsection is to derive an explicit representation of the Hamiltonian $\hat H_\alpha$ as a large spin interacting with an ensemble of bosonic modes.
We will infer this representation by computing matrix elements.

Starting from the expression of $\hat H_\alpha$ in Eq.~\eqref{eq_lrxyfourier}, we insert on both sides resolutions of the identity with unperturbed eigenstates~\eqref{eq_eigenstatelabel}. The infinite-range Hamiltonian is by construction diagonal,
\be
\hat H_{\alpha=0}= \sum_{\delta_S,\kappa}\sum_N E_{\delta_S,N} |\delta_S,\kappa;N\rangle \langle \delta_S,\kappa ; N | ,
\ee
with $E_{\delta_S,N}$ given by Eq.~\eqref{eq_fullalpha0spectrum}.
The variable-range perturbation can be written as $\hat V_{\alpha} = \frac 1 L \sum_{k\neq0} f_k(\alpha) \hat V_k $, with
\be
\hat V_k = - {J_0}  \big[ (\tilde S^+_k \tilde S^-_{-k} + \tilde S^-_k \tilde S^+_{-k} )
+ \gamma (\tilde S^+_k \tilde S^+_{-k} + \tilde S^-_k \tilde S^-_{-k} ) \big] \, .
\ee
We need to compute the matrix elements
\be
\label{eq_Valphamatrixelement}
 \langle \delta_S',\kappa' ; N' | \hat V_k
|\delta_S,\kappa;N\rangle.
\ee
In principle we have an efficient recipe to apply $\hat V_k$ to maximally polarized states of the form $|\delta_S,\kappa;\boldsymbol{\Uparrow}\rangle$, by expressing both in terms of standard Holstein-Primakoff bosons. 
This strategy directly works for  low-energy states, e.g. $|\delta_S,\kappa;0\rangle$, which are nearly-polarized in the direction of the minimum of the classical Hamiltonian $\mathcal{H}$.
However, the ``condensate'' state $|\delta_S,\kappa;N\rangle$ is constructed as $e^{iN\hat \varphi}|\delta_S,\kappa;0\rangle$, where  $e^{i\hat\varphi}$ is a function of collective spin operators $\hat S^{x,y,z}$ (known from the $\alpha=0$ solution). We would then need to commute $N$ times the operators $\tilde S^\pm_k$ appearing in $\hat V_\alpha$ with $e^{i\hat\varphi}$ using the commutation relations $[\hat S^\alpha,\tilde S^\beta_k]=i\epsilon^{\alpha\beta\gamma} \tilde S^\gamma_k$. While a priori well defined, this procedure is impractical, as the expression of $\hat\varphi$ is generally cumbersome. 
We will thus follow a different route using the formalism of spin-coherent states, see Sec.~\ref{subsec_alphazero} and App.~\ref{app_spincoherent}.

The  advantage of decomposing unperturbed eigenstates as linear superpositions of spin-coherent states, is that we know how to apply $\hat V_k$ to the latter: Upon rotating the spin operators into a reference frame aligned with the coherent-state polarization, we can use the standard Holstein-Primakoff mapping to bosonize spin fluctuations. In other words, the complication of dealing with a macroscopically populated collective spin mode is transferred to computing integrals involving coherent-state wave functions. 

To see how this machinery works in practice, let us develop Eq.~\eqref{eq_Valphamatrixelement} in spin-coherent states [defined in Eq.~\eqref{eq_spincoherentdef} above]: 
\be
\int d\boldsymbol{\Omega'} \int d\boldsymbol{\Omega} \, \psi^*_{\delta_S',N'}(\boldsymbol{\Omega'}) \psi_{\delta_S,N}(\boldsymbol{\Omega})
 \langle \delta_S',\kappa' ; \boldsymbol{\Omega}' | \hat V_k
|\delta_S,\kappa;\boldsymbol{\Omega}\rangle.
\ee
As $\hat V_k$ flips at most two spins, the quantum numbers $S$ and $S^z$ can only change by at most $2$ units. Thus, its action on the maximally polarized  state $|\delta_S,\kappa;\boldsymbol{\Omega}\rangle$
results in a linear combination of states of the form $|\delta_S^{''},\kappa^{''};M_{\boldsymbol{\Omega}}\rangle$ with $|\delta_S^{''}-\delta_S|\le 2$ and $M_{\boldsymbol{\Omega}} \le 4$ (recall $\delta_S=L/2-S$ and $M=S-S^z$). 

To take the overlap with the bra state $\langle \delta_S',\kappa' ; \boldsymbol{\Omega}' |$, we need to  
decompose the latter on bra eigenstates 
of the collective spin projection along $ \vec n(\boldsymbol{\Omega})$, i.e. 
\be
\label{eq_omegaomegaprime}
\langle \delta_S',\kappa' ; \boldsymbol{\Omega}' | = \sum_{M=0}^{4}
g^*_{\delta_S',M}(\boldsymbol{\Omega}',\boldsymbol{\Omega}) \; \langle \delta_S',\kappa' ; M_{\boldsymbol{\Omega}}| \ ,
\ee
where we already implemented the restriction $M\le 4$ arising from applying $\hat V_k$ to the ket. 
The coefficients $g_{\delta_S,M}(\boldsymbol{\Omega}',\boldsymbol{\Omega}) $ are given by the standard single-spin coherent-state wave functions of spin-$S=L/2-\delta_S$ states with projection $S-M$ 
(see App.~\ref{app_spincoherent}). Explicitly,
\be
\label{eq_gfactor}
g_{\delta_S,M}(\boldsymbol{\Omega}',\boldsymbol{\Omega})= 
\binom{L-2\delta_S}{M}^{1/2}   
k(\boldsymbol{\Omega}',\boldsymbol{\Omega})^M
\langle \delta_S; \boldsymbol{\Omega} | \delta_S; \boldsymbol{\Omega}' \rangle;
\ee
where we set $k(\boldsymbol{\Omega}',\boldsymbol{\Omega})=\tan\big( \alpha(\boldsymbol{\Omega}',\boldsymbol{\Omega})/2 \big)e^{i\beta(\boldsymbol{\Omega}',\boldsymbol{\Omega})}$, and $\alpha$ and $\beta$ are the polar and azimuthal angle of $\vec{n}(\boldsymbol{\Omega}')$ in the rotated frame adapted to $\vec{n}(\boldsymbol{\Omega})$.

Now we are left with matrix elements of the form
\be
\langle \delta_S',\kappa' ; M_{\boldsymbol{\Omega}}|
\hat V_k 
|\delta_S,\kappa;\boldsymbol{\Omega}\rangle.
\ee
To evaluate this expression it is convenient to 
use Eq.~\eqref{eq_spincoherentdef} 
\be
\langle \delta_S',\kappa' ; M_{\boldsymbol{\Uparrow}}|
 \hat U^\dagger(\boldsymbol{\Omega}) \hat V_k \hat U(\boldsymbol{\Omega})
|\delta_S,\kappa;\boldsymbol{\Uparrow}\rangle \,;
\ee
and transfer the rotations to spin operators $  \hat U^\dagger(\boldsymbol{\Omega}) \tilde S^{\pm}_{\pm k} \hat U(\boldsymbol{\Omega}) \equiv \tilde S^\pm_{\pm k} (\boldsymbol{\Omega})$ appearing in $\hat V_k$, 
\be
\tilde S^\pm_{ k} (\boldsymbol{\Omega}) = %& 
R_{\pm, +}(\boldsymbol{\Omega})   \tilde S^+_{ k} 
%\\&
+ R_{\pm, -}(\boldsymbol{\Omega})   \tilde S^-_{ k}
%\\&
+ R_{\pm, z}(\boldsymbol{\Omega})  \tilde S_{ k}^z
\ee
where 
\be
\begin{split}
R_{+,+}(\boldsymbol{\Omega})= & R_{-,-}^*(\boldsymbol{\Omega})= -  \sin^2(\theta/2) e^{2i\phi}, \\
R_{+,-}(\boldsymbol{\Omega})= & R_{-,+}^*(\boldsymbol{\Omega})= +  \cos^2(\theta/2),  \\
R_{+,z}(\boldsymbol{\Omega})= & R_{-,z}^*(\boldsymbol{\Omega})= +  \sin(\theta) e^{i\phi}. \\
\end{split}
\ee
Plugging in these transformations we obtain
\be
\hat U^\dagger(\boldsymbol{\Omega}) \hat V_k \hat U(\boldsymbol{\Omega}) = \sum_{\mu,\nu=+,-,z} {J}^{\mu,\nu}(\boldsymbol{\Omega}) \, \tilde S^\mu_k \tilde S^\nu_{-k} \, ,
\ee
where we defined (omitting $\boldsymbol{\Omega}$-dependency for simplicity)
\begin{multline}
\label{eq_Jmunu}
{J}^{\mu,\nu} \equiv -J_0\Big[
\big(R_{+,\mu} R_{-,\nu} + R_{-,\mu} R_{+,\nu}\big) \\
+ \gamma \big(R_{+,\mu} R_{+,\nu} + R_{-,\mu} R_{-,\nu}\big) \Big] \, .
\end{multline}
Note the symmetry property ${J}^{\mu,\nu}\equiv {J}^{\nu,\mu}$.

Here comes the crucial point: We have now isolated the amplitudes of the collective spin's and the magnons' transitions in the expression of the matrix element. Putting everything together and performing the integrals over the spherical angles we obtain
\begin{widetext}
\be
\label{eq_exactmatrixelement}
 \langle \delta_S',\kappa' ; N' | \hat V_\alpha
|\delta_S,\kappa;N\rangle
=
 \sum_{k\neq0} f_k(\alpha)
 \sum_{\mu,\nu=+,-,z} \sum_{M=0}^{4} \; 
 \binom{L-2\delta_S}{M}^{1/2} \;
\mathcal{J}^{\mu,\nu}_{\delta_S',N';\delta_S,N;(M)} \;
 \langle \delta_S',\kappa' ; M_{\boldsymbol{\Uparrow}} | \frac {\tilde S^\mu_k \tilde S^\nu_{-k}} L 
|\delta_S,\kappa;\boldsymbol{\Uparrow}\rangle\ ,
\ee
where we defined the coefficients
\be
\label{eq_defJmunu}
\mathcal{J}^{\mu,\nu}_{\delta_S',N';\delta_S,N;(M)} \; \equiv \;
\int d\boldsymbol{\Omega'} \int d\boldsymbol{\Omega} \; \psi^*_{\delta_S',N'}(\boldsymbol{\Omega'}) \psi_{\delta_S,N}(\boldsymbol{\Omega}) \,
k(\boldsymbol{\Omega}',\boldsymbol{\Omega})^M \,
 {J}^{\mu,\nu}(\boldsymbol{\Omega})
 \, \langle \delta_S; \boldsymbol{\Omega} | \delta_S; \boldsymbol{\Omega}' \rangle\ .
\ee
\end{widetext}
This is an \emph{exact} expression for the matrix elements of the finite-range perturbation $\hat V_\alpha$ between \emph{arbitrary} eigenstates of an infinite-range Hamiltonian $\hat H_{\alpha=0}$ (no relation between the two has to be assumed). The computation of the matrix element is reduced to a finite sum of matrix elements of two-spin operators between maximally (ket) and a nearly-maximally (bra) polarized states in the corresponding $SU(2)$ towers, with coefficients computed from the single-spin $\alpha=0$ solution, i.e. by taking integrals of simple trigonometric polynomials and the two coherent-state wave functions.

\subsection{Interacting rotor-magnon effective Hamiltonian}

\label{subsec_rotorsw}

The expression~\eqref{eq_exactmatrixelement} is only really useful when $\delta_S \ll L/2$. In this case the remaining matrix element  can be efficiently accessed numerically by constructing the exact few-magnon subspace. 
Note that this procedure gives an alternative efficient projection method, based on resolving the unperturbed spectrum in terms of magnon occupations rather than the decorated trees exploited above in Sec.~\ref{sec_numerics} and reviewed in App.~\ref{app_approximethod}. A numerical comparison of the performances of the two methods is left to future work.

In this Subsection we show that the problem in this form is amenable to remarkable analytical progress.
The key observation is that the remaining matrix element states can be computed via the bosonic representation \emph{in the dilute approximation} [in stark contrast with the original matrix element in Eq.~\eqref{eq_Valphamatrixelement}!]. 
Concerning the operators, we take the Fourier transformation of the exact Holstein-Primakoff mapping~\eqref{eq_HPlocal},
 \be
 \label{eq_HPfourier}
\begin{split}
 \tilde S^-_k &\mapsto
      L^{1/2} \; \tilde b^\dagger_k - \frac 1 {L^{1/2}} \sum_{q_1,q_2} \tilde b^\dagger_{q_1} \tilde b^\dagger_{q_2} \tilde b_{q_1+q_2-k},
    \\
   \tilde S^+_k &\mapsto   L^{1/2} \; \tilde b_{-k} - \frac 1 {L^{1/2}} \sum_{q_1,q_2} \tilde b^\dagger_{q_1+q_2+k} \tilde b_{q_1} \tilde b_{q_2},
    \\
     \tilde S^z_k &\mapsto \frac L 2 \delta_{k,0}  -   \sum_{q} \tilde b^\dagger_{q+k} \tilde b_{q} \, .
\end{split}
 \ee
 Concerning the states, following Sec.~\ref{subsec_bosonlabelling}, for small $\delta_S/L$ we can identify $(\delta_S,\kappa)\mapsto \{n_k\}_{k\neq0}$ and 
make the following replacement in Eq.~\eqref{eq_exactmatrixelement}:
\be
\label{eq_bosonstate}
|\delta_S,\kappa;M_{\boldsymbol{\Uparrow}}\rangle \quad\mapsto \quad \frac {\big( b^\dagger_{k=0}\big)^M} {\sqrt{M!}} \prod_{k\neq0} \frac {\big(\tilde b^\dagger_k \big)^{n_k}}{\sqrt{n_k !}} |\emptyset\rangle  %\\ 
\ee
[Recall $0\le M\le 4 \ll L$ here, cf. Eq.~\eqref{eq_overlapspinbosonpolarized}.] 
Thus, the leading contributions to the matrix elements can be determined for each value of $M$ 
by straightforward substitution of Eqs.~\eqref{eq_HPfourier} and~\eqref{eq_bosonstate} into Eq.~\eqref{eq_exactmatrixelement}:
\begin{itemize}
\item 
For $M=0$, the leading contributions in $1/L$ are given by the terms
\be
\begin{split}
\frac 1 L  \tilde S^\mp_k \tilde S^\pm_{-k} &\; \mapsto \; \tilde b^\dagger_k \tilde b_k \, , \; \tilde b_{-k} \tilde b^\dagger_{-k }\, ; \\
\frac 1 L  \tilde S^\pm_k \tilde S^\pm_{-k} &\; \mapsto \; \tilde b_{-k} \tilde b_{k } \, , \; \tilde b^\dagger_k \tilde b^\dagger_{-k} \, .
\end{split}
\ee
\item
For $M=1$ contributions of the same leading order are given by cubic terms with $\tilde b_0^\dagger$ and a zero-momentum pair:
\be
\begin{split}
\frac 1 L  \tilde S^z_k \tilde S^-_{-k} &\; \mapsto \; 
- \frac 1 {L^{1/2}} \tilde b^\dagger_0 \, \tilde b_{-k} \tilde b^\dagger_{-k }\, ; \\
\frac 1 L  \tilde S^-_k \tilde S^z_{-k} &\; \mapsto \; 
- \frac 1 {L^{1/2}} \tilde b^\dagger_0 \, \tilde  b^\dagger_{k} \tilde b_{k}\, ;
\end{split}
\ee
indeed, in this case the extra factor $L^{1/2}$  in Eq.~\eqref{eq_exactmatrixelement}  compensates the $1/L^{1/2}$ in the right-hand sides.
\item Similarly, for $M=2,3,4$, contributions of the same leading order are given by terms with $(\tilde b_0^\dagger)^M$ and a zero-momentum pair.  
Here the factors $L^{-M/2}$ on the right-hand sides are exactly compensated by the extra factors $L^{M/2}$ in Eq.~\eqref{eq_exactmatrixelement}.
As it turns out, only the following term with $M=2$ 
contributes:\footnote{Note that here we used hermiticity $\langle \delta_S',\kappa' ; N' | \hat V_k
|\delta_S,\kappa;N\rangle=\langle \delta_S,\kappa ; N | \hat V_k
|\delta_S',\kappa';N'\rangle^*$ to rule out off-diagonal terms of the form $(\tilde b^\dagger_0)^M
 \, \tilde b_{-k} \tilde b_k$ for $M>0$.}
\be
\frac 1 L  \tilde S^-_k \tilde S^-_{-k} \; \mapsto \; - \frac 1 {L}
(\tilde b^\dagger_0)^2
\, \big(\tilde b^\dagger_{k} \tilde b_{k} + \tilde b_{-k} \tilde b^\dagger_{-k} \big)  \, .
\ee
\end{itemize}
We conclude that the transitions in magnon space are described by a \emph{quadratic} bosonic Hamiltonian (up to corrections of higher order in the magnon density $\delta_S/L$). 

What about the simultaneous transitions in collective-spin space?
Here the crucial observation is that the coefficients $\mathcal{J}^{\pm,\pm}_{\delta_S',N';\delta_S,N;(M)}$ can be identified with matrix elements of
a certain polynomial collective operator $P(\hat{\bold{S}}/S)$ between  eigenstates of the $\alpha=0$ problem,
as derived in App.~\ref{app_spincoherent}. 
It follows that the matrices $\mathcal{J}^{\mu,\nu}_{\delta_S',N';\delta_S,N;(M)}$  are independent of the spin size and only sensitive to $\delta_S-\delta_S'$ to leading order in $1/L$. 
Furthermore, these matrices have characteristic semiclassical structure, which is best appreciated upon expressing the spin components as analytic functions of the action and angle variables of the unperturbed problem: The analytic dependency on the angle $\varphi$ translates into an (at most) exponential decay of the matrix elements in $N'-N$ away from the diagonal, and the analytic dependency on the action $n$ translates into a smooth variation with $(N+N')/L$ along the diagonal. 
(These standard general properties can easily be checked numerically for an arbitrary polynomial collective operator $P(\hat{\bold{S}}/S)$ and  an arbitrary single-spin Hamiltonian $\hat H_{\alpha=0}$ eigenbasis.)

Putting everything together, we have found a remarkable result: for $\delta_S \ll L/2$, the matrix elements of the finite-range Hamiltonian $\hat H_\alpha$ coincide with those of an effective Hamiltonian describing a large spin coupled to bosons,
\begin{widetext}
\begin{multline}
\label{eq_spinoscillators}
\hat H_{\alpha,\text{eff}} =
\sum_N \bigg[
\frac L 2 \mathcal{H}\bigg( \frac {N}{L/2};1 \bigg) - \bar{\omega}\bigg( \frac {N}{L/2};1 \bigg) \sum_{k\neq0} \tilde b_k^\dagger \tilde b_k \bigg] |N\rangle\langle N|
\\ + \frac 1 2 
\sum_{N,N'}   \; \sum_{k\neq0} f_k(\alpha)
\Bigg[ \mathcal{J}_{N'-N}\bigg(\frac{N+N'}{L} \bigg) \, \Big( \tilde b^\dagger_k \tilde b_k + %\mathcal{J}^{(0)}_{N'-N}\bigg(\frac{N+N'}{L} \bigg)  \tilde b_{-k} \tilde b^\dagger_{-k } \\
\tilde b_{-k} \tilde b^\dagger_{-k } \Big)
+ \mathcal{K}_{N'-N}\bigg(\frac{N+N'}{L} \bigg)   \tilde b_{-k} \tilde b_{k } + \mathcal{K}^*_{N'-N}\bigg(\frac{N+N'}{L} \bigg) \tilde b^\dagger_k \tilde b^\dagger_{-k} \Bigg]
\, |N'\rangle\langle N|  ,
\end{multline}
where $\mathcal{J}_{N'-N}\big(\frac{N+N'}{L} \big) \equiv 2 \big( \mathcal{J}^{+,-}_{0,N';0,N;(0)}-\mathcal{J}^{z,-}_{0,N';0,N;(1)}- \mathcal{J}^{-,-}_{0,N';0,N;(2)}\big)$ and $\mathcal{K}_{N'-N}\big(\frac{N+N'}{L} \big) \equiv 2 \mathcal{J}^{+,+}_{0,N';2,N;(0)}$.
This equation is our first main result: \emph{We have rewritten our finite-range Hamiltonian as an effective model of a semiclassical particle (the collective spin), described by states $\{|N\rangle\}$, coupled to an ensemble of bosonic modes associated with magnon excitations with non-vanishing momentum ($k\neq0$). }

While the Hamiltonian~\eqref{eq_spinoscillators} is quadratic in the bosons, the coupling is highly non-linear as the coefficients of the quadratic bosonic Hamiltonian depend on the configuration of the collective spin. 
Since the semiclassical particle spans macroscopically different energies, we can simplify the expression by focusing on a given energy shell, and rewrite the effective Hamiltonian therein. Setting $N=nL/2+\delta_N$, $N'=nL/2+\delta_N+r$, and consistently neglecting $\mathcal{O}(1/L)$ terms, we obtain the following quantum rotor-magnon Hamiltonian:
\begin{multline}
\hat H_{\alpha,\text{eff}} (n)=
L \mathcal{E}(n)
+ \omega(n) \sum_{\delta_N} \delta_N |\delta_N\rangle\langle \delta_N|
+ \bar\omega(n) \sum_{k\neq0} \tilde b_k^\dagger \tilde b_k
\\ + \frac 1 2 
\sum_{\delta_N,r}   \; \sum_{k\neq0} f_k(\alpha)
\Bigg[ \mathcal{J}_r(n)\, \Big( \tilde b^\dagger_k \tilde b_k +   
\tilde b_{-k} \tilde b^\dagger_{-k }
\Big)
+ \mathcal{K}_r(n)\, \tilde b_{-k} \tilde b_{k }  + \mathcal{K}^*_r(n)\,  \tilde b^\dagger_k \tilde b^\dagger_{-k} \Bigg]
\; |\delta_N + r\rangle\langle \delta_N| \, .
\end{multline}
Alternatively, introducing the collective action displacement $\hat{\delta}_N = \sum_{\delta_N} \delta_N |\delta_N\rangle\langle \delta_N|$ and angle  $e^{\pm i\hat\varphi} = \sum_{\delta_N} |\delta_N\pm1\rangle\langle \delta_N|$ operators, we can rewrite the rotor-magnon Hamiltonian as
\begin{multline}
\label{eq_rotoroscillators}
\hat H_{\alpha,\text{eff}} (n)=
L \mathcal{E}(n)
+ \omega(n) \, \hat{\delta}_N
+ \bar\omega(n) \sum_{k\neq0} \tilde b_k^\dagger \tilde b_k
\\ +
\frac 1 2
 \sum_{k\neq0} f_k(\alpha)
\Big[ \mathcal{J}(n,\hat \varphi) 
\Big(
\tilde b^\dagger_k \tilde b_k + 
\tilde b_{-k} \tilde b^\dagger_{-k }
\Big)
+ 
\mathcal{K}(n,\hat \varphi) \;
\tilde b_{-k} \tilde b_{k } 
+ \mathcal{K}^*(n,\hat \varphi)  \; \tilde b^\dagger_k \tilde b^\dagger_{-k} \Big] \, .
\end{multline}
\end{widetext}
where $\mathcal{J}(n,\varphi) = \sum_r \mathcal{J}_r(n) \, e^{ir\varphi}$ and $\mathcal{K}(n,\varphi) = \sum_r \mathcal{K}_r(n) \, e^{ir\varphi}$ are $2\pi$-periodic analytic functions of $\varphi$, smoothly depending on the parameter~$n$. 
This equation accomplishes our goal. In the first line we recognize $\hat H_{\alpha=0}$, cf. Eq.~\eqref{eq_alpha0spectrum}, where $\delta_S$ is now resolved in terms of total occupation of magnon modes. The second line expresses  interactions between magnons and collective spin.

A neat physical picture  emerges:
\begin{itemize}
\item The non-linear  precession of the collective spin, with frequency $\omega$, constitutes a \emph{periodic drive} for magnon modes with mode-dependent driving  amplitude $f_k(\alpha)$; \item This drive induces a population of magnons, which in turn \emph{back-reacts} on the collective spin precession. 
\end{itemize}
The nature of these interacting dynamics is expected to crucially depend on the presence of \emph{resonances} between  the unperturbed magnon frequency $\bar\omega$ and the rotor frequency $\omega$.
In the next Subsection we will present an analytical solution of Eq.~\eqref{eq_rotoroscillators} which substantiates this physical intuition.

\subsection{Exact solution of the effective Hamiltonian}
\label{subsec_exactsol}

The structure of Eq.~\eqref{eq_rotoroscillators} is superficially reminiscent of the Bogolubov theory of a weakly interacting Bose gas~\cite{leggett2006quantum}.
However, such an analogy is only strictly valid 
at the isotropic point $\gamma=0$ of our spin Hamiltonian, where the total magnetization $\hat S^z$ is conserved. This constrains $\hat{\delta}_N$ to $\sum_{k\neq0} \tilde b^\dagger_k \tilde b_k$, making the former variable redundant, just like the conservation of the total number of particles in an interacting Bose gas enslaves the condensate population to the condensate depletion.
In this case, the condensate fluctuations (associated with the collective phase $\hat \varphi$) can be formally eliminated, and one is left with an effective quadratic theory for $k\neq0$ excitations (parametrically depending on the self-consistent condensate fraction)~\cite{leggett2006quantum}.
Crucially, in this case, the stability of the Bose-condensed long-range ordered ground state is \emph{a priori guaranteed} by the $U(1)$ symmetry. In our spin system with $\gamma=0$, the eigenstates with smallest $\delta_S$ are indeed the \emph{ground states} of the Hamiltonian in the various symmetry sectors.

The main goal of this paper is, however, to understand the fate of the QMBS when \emph{no symmetry protects them} --- e.g. in the Ising model ($\gamma=1$), see our numerical study in Sec.~\ref{sec_numerics}. There, the effective Hamiltonian~\eqref{eq_rotoroscillators} is \emph{not} equivalent to a standard interacting Bose gas and it incurs no further simplification: the dependency of the second line on $\hat\varphi$ cannot be trivially eliminated. The delocalization of large-spin eigenstates in Hilbert space for $\alpha\neq0$ is energetically possible and entropically favorable, making their stability far from obvious. 
\\

The crucial realization is that \emph{we can successfully diagonalize the effective Hamiltonian}~\eqref{eq_rotoroscillators}.
This will allow us to compute the properties of the presumed  eigenstates at finite $\alpha$ and to \textit{a posteriori} determine the regime of self-consistency of the approximation leading to Eq.~\eqref{eq_rotoroscillators}.
In the text below we  report a sketch of the diagonalization procedure, emphasizing its crucial points; we relegate a detailed derivation to App.~\ref{app_rotoroscillators}. We remark that, to the best of our knowledge, our exact solution of an interacting Hamiltonian  of the form~\eqref{eq_rotoroscillators} is novel and it may thus be of independent interest.

The diagonalization procedure involves a  sequence of two canonical transformations.
First, we introduce an ansatz
$e^{i\hat S}$ with a generator of the form 
\begin{multline}
\label{eq_generalizedbogolubov}
\hat S = 
\frac 1 2 \sum_{k\neq0} 
\bigg[  F^{(+)}_k(\hat \varphi) \, \Big( \tilde b_k \tilde b_{-k} +  \tilde b^\dagger_{-k} \tilde b^\dagger_k \Big)
\\ + i F^{(-)}_k(\hat \varphi) \, \Big( \tilde b_k \tilde b_{-k} -  \tilde b^\dagger_{-k} \tilde b^\dagger_k\Big) \bigg] \ .
\end{multline}
The ansatz~\eqref{eq_generalizedbogolubov} represents a kind of generalized Bogolubov transformation where the role of the unknown ``angles'' is taken by unknown \emph{functions} of the collective operator $\hat \varphi$.  Our goal is to determine functions ${F}^{(\pm)}_k$ in such a way that the transformed coefficients of $\tilde b_k \tilde b_{-k}$, $  \tilde b^\dagger_{-k} \tilde b^\dagger_k$ in the Hamiltonian get cancelled. 
  Parametrizing the hyperbolic rotation of modes $(k,-k)$ by a rapidity $\eta_k=\sqrt{\big(F^{(+)}_k\big)^2+\big(F^{(-)}_k\big)^2}$ and an angle $\xi_k$ with $\tan \xi_k = - F^{(+)}_k/F^{(-)}_k$, diagonalization imposes the following pair of \emph{ordinary differential equations} for $\eta_k(\varphi)$, $\xi_k(\varphi)$ (see App.~\ref{app_rotoroscillators}):
\begin{widetext}
\begin{multline}
\label{eq_ode1}
\omega \partial_\varphi \big( \eta_k \sin \xi_k\big) = 
+ \bar\omega \, \sinh \eta_k \cos \xi_k \\
+ f_k(\alpha) \Big[ 
\mathcal{J}^{(0)}(\varphi) \sinh \eta_k \cos \xi_k
+ \mathcal{J}^{(+)}(\varphi) \big( \cosh \eta_k \cos^2 \xi_k + \sin^2 \xi_k \big)
+ 2 \mathcal{J}^{(-)}(\varphi) \sinh^2 (\eta_k/2) \cos \xi_k \sin \xi_k
\Big] \, ,
\end{multline}
\begin{multline}
\label{eq_ode2}
\omega \partial_\varphi \big( \eta_k \cos \xi_k\big)  = 
- \bar\omega \, \sinh \eta_k \sin \xi_k \\
- f_k(\alpha) \Big[ 
\mathcal{J}^{(0)}(\varphi) \sinh \eta_k \sin \xi_k
+ 2 \mathcal{J}^{(+)}(\varphi) \sinh^2 (\eta_k/2) \cos \xi_k \sin \xi_k
+ \mathcal{J}^{(-)}(\varphi) \big( \cosh \eta_k \sin^2 \xi_k + \cos^2 \xi_k \big)
\Big] \; ,
\end{multline}
\end{widetext}
where $\mathcal{J}^{(0)} \equiv \mathcal{J} $, 
$ 
\mathcal{J}^{(+)} \equiv \Re \, \mathcal{K} $, $
\mathcal{J}^{(-)} \equiv  \Im \, \mathcal{K} $.
Note that we obtain {differential} equations rather than standard equations because of the {non-linear} rotor-magnon interactions. 

To understand the existence conditions of analytic $2\pi$-periodic solutions to Eqs.~\eqref{eq_ode1},~\eqref{eq_ode2} we first observe that for $\alpha\to0$ the second lines vanish, and hence $\eta_k\equiv 0$; linearization in $\eta_k \sim f_k(\alpha)$  yields the simplified equations
\be
\begin{split}
\omega \partial_\varphi \big( \eta_k \sin \xi_k\big) &= 
+ \bar\omega \, \eta_k \cos \xi_k 
+ f_k(\alpha)   \mathcal{J}^{(+)}(\varphi) \, , \\
\omega \partial_\varphi \big( \eta_k \cos \xi_k\big)  & = 
- \bar\omega \,  \eta_k \sin \xi_k 
- f_k(\alpha) \mathcal{J}^{(-)}(\varphi) 
 \; ,
\end{split}
\ee
which are equivalent (upon renaming  $P\equiv \eta_k \sin \xi_k$, $Q\equiv \eta_k \cos \xi_k$, and rescaling the ``time'' variable $\varphi \equiv \omega t$) to a fictitious harmonic oscillator with  natural frequency $\bar\omega$ externally driven at frequency $\omega$. As  is well known, there always exists a unique periodic trajectory of this system with the same frequency $\omega$ of the drive (i.e., a $2\pi$-periodic solution in $\varphi$), {provided}  $\bar\omega \neq r \omega$, for all integers $r$ appearing in the Fourier representation of $\mathcal{J}^{(\pm)}$. In correspondence of such \textit{resonances} $\{\bar\omega = r \omega\}$, all solutions are unbounded.
Remarkably,  the existence and uniqueness of a solution with the same periodicity as the drive persists under a general weak non-linearity~\cite{gallavotti2013elements}. The main qualitative effect of the non-linearity is to \emph{thicken} the resonances from discrete points to finite intervals  $\{ |\bar\omega - r \omega| \le \delta_{r}\}$ of width depending on the driving amplitude, thus upper bounded as $\delta_r \le \delta_0 \, e^{-\sigma |r|} \, \max_{k\neq0} |f_k(\alpha)|$ for some constant $\sigma>0$. This condition provides a theoretical guarantee of the existence of a finite range of parameter values for which the ansatz~\eqref{eq_generalizedbogolubov} successfully diagonalizes the magnon part of the effective Hamiltonian. From a practical point of view, the non-linear equations~\eqref{eq_ode1},~\eqref{eq_ode2} are known explicitly for a given model, and one can numerically determine its range of solvability throughout the parameter space.

Diagonalization of the magnon part of the Hamiltonian brings us within a stone's throw from the full solution of our problem. The transformed Hamiltonian reads
\be
\label{eq_diagonalizedswH}
e^{i\hat S} \, \hat H_{\alpha,\text{eff}} \, e^{-i\hat S} = 
L {\mathcal{E}} 
+ \omega \hat{\delta}_N' 
  +
 \sum_{k\neq0} \bigg[
 G_k(\hat \varphi)
  \bigg( \tilde \beta^\dagger_k \tilde \beta_k + \frac 1 2  \bigg) - \frac {\bar\omega} 2 \bigg]
  ,
\ee
where $\hat{\delta}_N'$ and  $\tilde \beta_k, \tilde \beta^\dagger_k$ denote the dressed (i.e. transformed) rotor and magnon operators, respectively, and
\begin{multline}
G_k(\hat \varphi) \equiv 
\cosh \eta_k(\hat \varphi) \big[ \bar\omega + f_k(\alpha) \mathcal{J}^{(0)}(\hat \varphi) \big]
 \\
  +  \sinh \eta_k(\hat \varphi) \cos \xi_k(\hat \varphi) \, f_k(\alpha) \mathcal{J}^{(+)}(\hat \varphi)
 \\
   +  \sinh \eta_k(\hat \varphi) \sin \xi_k(\hat \varphi) \, f_k(\alpha) \mathcal{J}^{(-)}(\hat \varphi) \, .
\end{multline}
 As the magnon part is now diagonal, we can
  diagonalize the rotor part separately in each sector defined by a Fock occupation $\{n_k\}_{k\neq0}$ of dressed magnon modes. 
In fact, it can be checked that a magnon-diagonal canonical transformation $e^{i\hat F}$ ansatz, with  
\be
\label{eq_hatF}
\hat F = \sum_{k\neq0} F^{(0)}_k(\hat \varphi) \bigg( \tilde \beta^\dagger_k \tilde \beta_k + \frac 1 2  \bigg)\, ,
\ee
successfully decouples the rotor from the magnons
if the generator satisfies
$
\omega \partial_\varphi F^{(0)}_k(\varphi) \equiv 
 {G_k(\varphi)}
$ for all $k$
(see App.~\ref{app_rotoroscillators} for more details).
In this case, we find that the sequential application of $e^{i\hat S}$ and $e^{i\hat F}$ fully diagonalizes our effective rotor-magnon Hamiltonian~\eqref{eq_rotoroscillators}:
\be
\label{eq_fullydiagonalizedHeff}
e^{i\hat F}e^{i\hat S} \, \hat H_{\alpha,\text{eff}} \, e^{-i\hat S}e^{-i\hat F} = 
L\mathcal{E}(\alpha) 
+ \omega \, \hat{\Delta}_N 
  +
 \sum_{k\neq0} 
 \bar\omega_k(\alpha)
   \tilde \beta^\dagger_k \tilde \beta_k 
 \, ,
\ee
with a reference energy $\mathcal{E}(\alpha)=\mathcal{E}+\frac 1 2 \sum_{k\neq0} (\bar\omega_k(\alpha)-\bar\omega)$ dressed by zero-point fluctuations and a nontrivial {dispersion relation for dressed magnon-like quasiparticles},
\be
\label{eq_dispersionrelation}
\bar\omega_k(\alpha) \equiv \overline{G_k(\varphi)}, 
\ee
where $\overline{\boldsymbol{\cdot}(\varphi)} = \int_0^{2\pi} \frac {d\varphi}{2\pi} \boldsymbol{\cdot}(\varphi)$. Here, $\bar\omega_k(\alpha)$  
replaces the flat band $\bar\omega$ of the limit $\alpha=0$, cf. Eq.~\eqref{eq_alpha0spectrum}; it is interpreted as spectrum of coherent excitations on top of a {highly excited} non-thermal eigenstate.

The many-body spectrum thus takes the form
$
E_{\{n_k\},  \delta_N }(\alpha) = L\mathcal{E}(\alpha) + \omega \, \delta_N + \sum_{k\neq0} \bar\omega_k(\alpha) n_k
$. {Eigenstates are labelled \emph{continuously} from $\alpha=0$ to $\alpha$ (away from resonances!) by the integer quantum number $\delta_N$, associated with the choice of reference non-thermal vacuum, and $\{n_k\}_{k\neq0}$, associated with the quasiparticle content.}
  Importantly,
  however, in the original unperturbed basis the eigenstates feature non-trivial quantum entanglement between rotor and magnons degrees of freedom, encoded in the inverse canonical transformation $e^{-i\hat S}e^{-i\hat F}$, as is manifest in the expressions~\eqref{eq_generalizedbogolubov} and~\eqref{eq_hatF}.
  \\

\subsection{Self-consistency of localized eigenstates for $0<\alpha<1$}

\label{subsec_selfconsistency}

In the previous Subsection we determined the exact eigenstates of the effective rotor-magnon Hamiltonian~\eqref{eq_rotoroscillators} for arbitrary $\alpha$.
An important property of these eigenstates is the extent of delocalization --- i.e. the amount of bare rotor and bare magnon excitations they contain --- as a function of $\alpha$.
These quantities can be evaluated explicitly and 
play a crucial role to establish the regime of self-consistency of the theory, i.e. consistency with the assumptions leading to the effective Hamiltonian~\eqref{eq_rotoroscillators}:
An eigenstate of the effective Hamiltonian can represent a legitimate eigenstate of the original long-range XY spin chain only if its wavefunction remains localized in the subspace $\delta_S\ll L/2$, $|\delta_N| \ll L$ (associated with the ``vertical'' and ``horizontal'' delocalizations in Fig.~\ref{fig_eigenstatedeloc}, respectively). 

Both these quantities can be computed within the exact solution above. 
Considering for simplicity an eigenstate in the magnon vacuum tower $\{ n_k \}_{k\neq0} = \emptyset$
(calculation is analogous for other eigenstates), $\Big\langle \hat{\delta}_S \Big\rangle$ is readily evaluated  as (see App.~\ref{app_rotoroscillators})
\be
\Big\langle \hat{\delta}_S \Big\rangle = \Bigg\langle \sum_{k\neq0} \tilde b^\dagger_k \tilde b_k \Bigg\rangle = \sum_{k\neq0} \overline{ \sinh^2 \eta_k(\varphi)}
\ee
On the other hand, the amount of rotor fluctuations in an eigenstate  can be estimated by evaluating, e.g., $\Big\langle\hat{\delta}_N^2\Big\rangle$.
Calculation yields an explicit expression, reported in App.~\ref{app_rotoroscillators}.

The case $0<\alpha<1$ is particularly interesting, as the  values $f_{k_\ell}(\alpha)\equiv f_\ell(\alpha)$ are discrete and decay as $\ell^{-(1-\alpha)}$ (see App.~\ref{app_fkalpha}). Hence, for sufficiently large $\ell$ we are always in the linear regime where  $\eta_\ell \propto f_\ell(\alpha)$. This directly leads to the   estimates:%obtained for the ground state in Sec.~\ref{subs_groundstate} 
\be  
\label{eq_depletionestimate}
\Big\langle \hat{\delta}_S \Big\rangle \sim \sum_{\ell=-L/2}^{L/2} |f_\ell(\alpha)|^2 \sim
\left\{
\begin{split}
\mathcal{O}(1) & \quad \text{for } 0<\alpha<1/2 \, ,\\
\log L & \quad \text{for } \alpha=1/2 \, ,\\
L^{2\alpha-1} & \quad \text{for } 1/2<\alpha<1 \, . \\
\end{split}
\right.
\ee
In this regime, one finds the same qualitative behavior for
$\sqrt{\Big\langle\hat{\delta}_N^2\Big\rangle}$ (see App.~\ref{app_rotoroscillators}). %, leading to similar estimates as for $\Big\langle \hat{\delta}_S \Big\rangle$ above.
This prediction for $\langle \hat \delta_S\rangle$ has been shown in Fig.~\ref{fig_proj_obs} above and Fig.~\ref{fig_kick_conv} below, along with the numerical data. The plot shows complete compatibility even for finite system size.
The joint occurrence 
\be
\Big\langle \hat{\delta}_S \Big\rangle\ll L/2 \, , \quad  \sqrt{\Big\langle\hat{\delta}_N^2\Big\rangle} \ll L\, , \quad \text{for } 0<\alpha<1
\ee
graphically illustrated in the right panel of  Fig.~\ref{fig_eigenstatedeloc}, constitutes a \emph{solid analytical argument for our description of QMBS  
being self-consistent for $0<\alpha<1$.}

On the other hand, for $\alpha\gtrsim1$, the delocalization of eigenstates  extends to sectors with extensively depleted collective spin $ \delta_S = \mathcal{O}(L)$, as
$\Big\langle \hat{\delta}_S \Big\rangle \sim L \int_{-\pi}^\pi \frac{dk}{2\pi} |f_k(\alpha)|^2$. The small but finite density of magnons breaks the eigenstate self-consistency, and full hybridization of the QMBS with the thermal bulk may be expected for large $L$. The dynamical counterpart of this expectation is that non-linear magnon-magnon interactions come into play and trigger the eventual thermalization of initial states with large collective spin. 
As such processes are suppressed with the magnon density, thermalization is (at least) parametrically slow as~$\alpha$ is decreased toward~$1$, in agreement with previous numerical observations~\cite{ZunkovicPRL18_Merging,HalimehPRB17_PersistentOrder,GorshkovConfinement}. We stress, however, that we are not in a position to fully rule out other possible subtler effects which may further delay (or even suppress) thermalization for $\alpha\gtrsim 1$.

\subsection{Generality of the theory}

\label{subsec_generality}

Here we comment on the generalization of the results of this Section, covering the full range of parameters of model~\eqref{eq_xxzlrd}.

\subsubsection{Higher dimension}
Our stability result becomes stronger for higher dimensional spin lattices. The only change we need is from $f_k(\alpha)$ to the analogous function $f_{\mathbf{k}}(\alpha)$ defined in the $d$-dimensional Brillouin zone. Throughout our derivation, we just replace $\sum_k f_k(\alpha) \dots$ with $\sum_{\mathbf{k}} f_{\mathbf{k}}(\alpha) \dots$ everywhere.  
As shown in App.~\ref{app_fkalpha}, the properties of $f_k(\alpha)$ for $\alpha<1$ directly generalize to $f_{\mathbf{k}}(\alpha)$ for $\alpha<d$; in particular,
$f$ squeezes toward $\mathbf{k}=\mathbf{0}$ as $L\to\infty$, and becomes a function of the discrete wavenumbers $\mathbf{k}L = 2\pi \boldsymbol{\ell}$, decaying algebraically as $|\boldsymbol{\ell}|^{-(d-\alpha)}$. By analogy we denote $f_{\mathbf{k}_{\boldsymbol{\ell}}}\equiv f_{\boldsymbol{\ell}} $.

It follows that the collective spin depletion $\langle \hat{\delta}_S\rangle$  of  solvable eigenstates is 
\be  
\label{eq_depletionestimatehigherd}
\Big\langle \hat{\delta}_S \Big\rangle \sim \sum_{\boldsymbol{\ell}} |f_{\boldsymbol{\ell}}(\alpha)|^2 \sim
\left\{
\begin{split}
\mathcal{O}(1) & \quad \text{for } 0<\alpha<d/2 \, ,\\
\log L & \quad \text{for } \alpha=d/2 \, ,\\
L^{2\alpha-d} & \quad \text{for } d/2<\alpha<d  \\
\end{split}
\right.
\ee
Similar estimates apply to $\sqrt{\Big\langle\hat{\delta}_N^2\Big\rangle}$. This shows self-consistency of QMBS for $0<\alpha<d$.

\subsubsection{General spin-spin interactions}
Our stability result applies equally well to arbitrary type of (long-range) spin interactions.
For example, concerning the XYZ spin interactions with $\Delta\neq0$ in Eq.~\eqref{eq_xxzlrd}, our derivation can be straightforwardly generalized.
The only addition we need is for terms arising from $\tilde S^z_{k}\tilde S^z_{-k}$ in $\hat V_k$: The additional rotated operators read
\be
%\begin{split}
\tilde S^z_{\pm k} (\boldsymbol{\Omega}) = %& 
R_{z, +}(\boldsymbol{\Omega})   \tilde S^+_{\pm k} 
%\\&
+ R_{z, -}(\boldsymbol{\Omega})   \tilde S^-_{\pm k} 
%\\&
+ R_{z, z}(\boldsymbol{\Omega})  \tilde S_{\pm k}^z 
%\end{split}
\ee
with coefficients computed analogously:
\be
\begin{split}
R_{z,+}(\boldsymbol{\Omega})= & R_{z,-}^*(\boldsymbol{\Omega}) = -\sin(\theta) e^{i\phi}, \\
R_{z,z}(\boldsymbol{\Omega})= & +\cos(\theta). \\
\end{split}
\ee
These extra terms just result in the addition $+2\Delta R_{z,\mu}R_{z,\nu}$ inside the square brackets in Eq.~\eqref{eq_Jmunu}.
The rest of the discussion is unchanged.

\subsubsection{Higher spins}
Our stability result becomes stronger when individual spins are larger.
In fact, an equivalent analysis can be performed when spins-$1/2$ are replaced by spins-$s$ with arbitrary $s$. This replacement gives more room to the accuracy of the bosonic description of spin states, which requires $\delta_S \ll Ls$. 

The only important caveat here is the absence of self-interactions $J_{\mathbf{r},\mathbf{r}}=0$ (which model e.g. single-ion anisotropy in magnetic materials with higher local spins). We will comment on their effect in Sec.~\ref{sec_predictions} below.\\

To summarize, our analysis straightforwardly carries over to arbitrary lattices, multi-spin interactions, and spin sizes, resulting in the same qualitative conclusion in all cases.

\section{Predictions: eigenstate delocalization by semiclassical chaos}
\label{sec_predictions}

The theory of Sec.~\ref{sec_theory} predicts the stability of QMBS %the collective scars 
with \textit{i)} sufficiently long range of interactions and \textit{ii)} absence of spectral resonances for $\alpha=0$. % -- associated with classical integrability of the mean-field Hamiltonian.
This theory rationalizes the numerical findings of Sec.~\ref{sec_numerics} for the long-range quantum Ising chain with  $\alpha<1$  and  $h\ge J_0$, which generically satisfies both conditions. 

 What if conditions \textit{i)} or \textit{ii)} fail?
In Sec.~\ref{sec_numerics} we already presented evidence that violation of the condition \textit{i)} (i.e. $\alpha>1$) results in asymptotic hybridization of the seemingly anomalous finite-size eigenstates.
Concerning condition \textit{ii)},
the representation of the Hamiltonian in Sec.~\ref{subsec_rotorsw} in terms of a collective spin coupled to magnon modes  gives us a clear and predictive understanding of QMBS stability.
In this Section we explore situations where, according to our theory, stability is challenged by  resonances in the mean-field energy spectrum.

We begin by discussing proximity to rare accidental resonances in Sec.\ref{sec_accidenti}, which generically exist for large system size. 
Afterwards, we demonstrate how  
exponentially diverging classical trajectories of the mean-field Hamiltonian result in instability of QMBS.  
This happens in correspondence of isolated phase-space separatrices (often associated with ESQPT) discussed in Sec.~\ref{sec_ESQPT}, as well as of fully developed classical chaos, e.g. generated by strong periodic drives, which we analyze in Sec.~\ref{sec_kicked}. 
In all these cases numerical data indicate instability of QMBS in agreement with our theory predictions.

\begin{figure}[t]
\centering
\hspace{-.5cm}
\includegraphics[width=.5\textwidth]{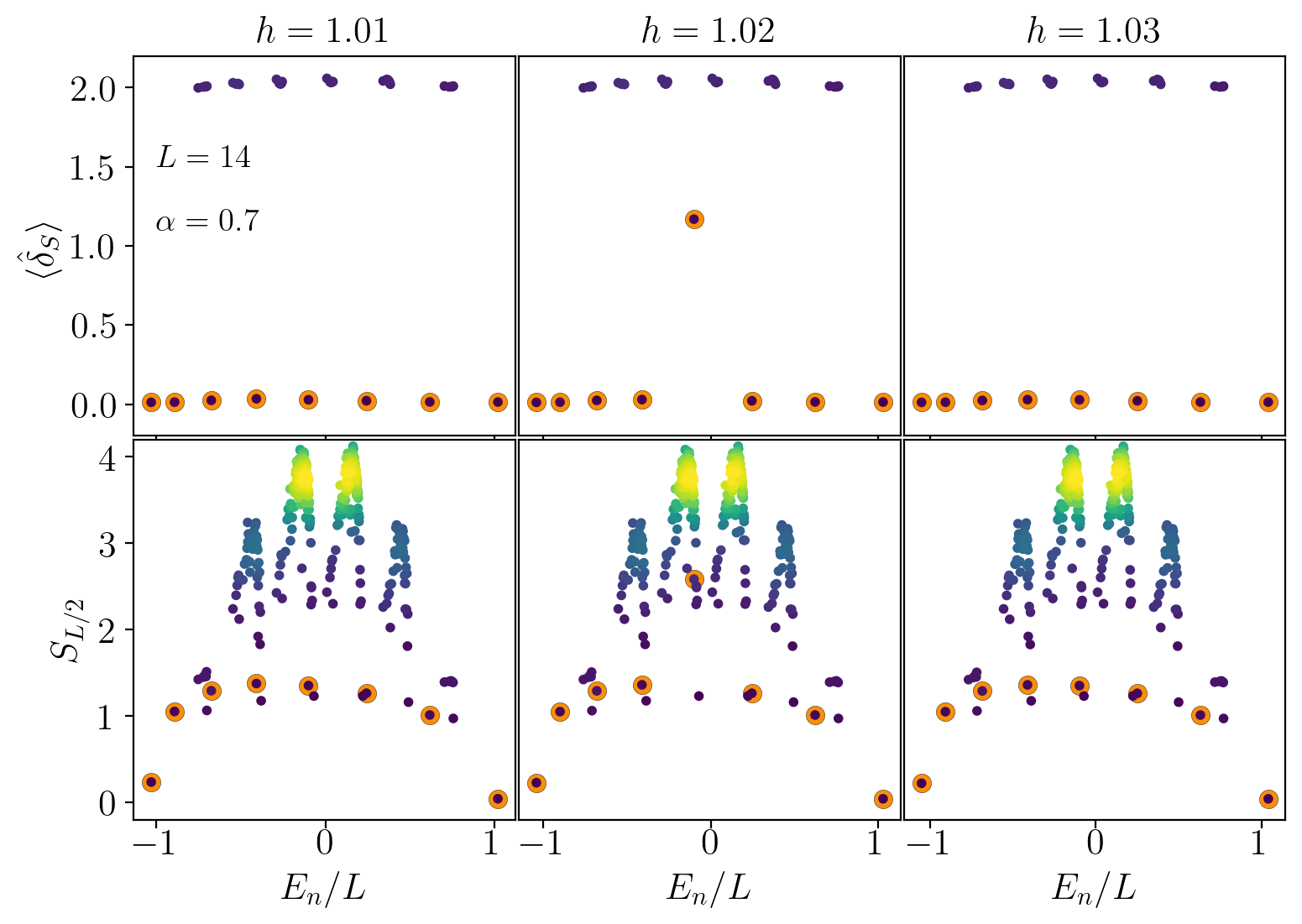}
\caption{QMBS hybridization due to an accidental spectral resonance. Data are shown for the quantum Ising chain with $L=14$, $\alpha=0.7$, $J_0=0.5$ and $h=1.01$, $1.02$ and $1.03$.  Scatter plot of the observables $\langle \delta_S\rangle_{E_n}$ (top row) and entanglement entropy $S_{L/2}(E_n)$ (bottom row) of each eigenstate $|E_n\rangle$ versus energy density $E_n/L$.
{The heatmap color illustrates the density of eigenstates, from darkest (isolated eigenstates) to lightest (maximum density).}
}
\label{fig_acci}
\end{figure}

\subsection{Accidental resonances}
\label{sec_accidenti}

The derivation of a self-consistent exact solution for QMBS in Sec.~\ref{subsec_exactsol} required the absence of resonances $\bar \omega \neq r \omega$ between the two fundamental rotor and magnon frequencies $\omega$ and $\bar \omega$.
This is a generic feature of energy shells of Hamiltonians with collective spin-spin interactions. However, accidental resonances may occur for specific values of model parameters and energy density. In this case the QMBS are expected to  hybridize with the rest of the spectrum. Observing such rare instances  in finite systems requires a careful search in parameter space.

An example of a fine-tuned resonance is shown in the middle panel of Fig.~\ref{fig_acci}, where the spin depletion and the entanglement entropy of each eigenstate are plotted for the quantum Ising chain Hamiltonian with $L=14$ and $\alpha=0.7$. The accidental resonance takes place at $h=1.02 J_0$ (central panel), causing significant hybridization of one of the QMBS (large orange dots). The spikes in the spin depletion $\braket{\delta_S}\simeq 1.2$ and entanglement entropy $S_{L/2}$ for the central scar eigenstate are well visible. 
This effect disappears upon an slight change in any parameter. This is illustrated in the two lateral panels of the same Figure, where we report analogous plots with $\approx1\%$ variations of the transverse field, i.e., $h/J_0=1.01$ and $1.03$.

\subsection{Excited-state quantum phase transitions}
\label{sec_ESQPT}

In this Section, we discuss another effect leading to instability of QMBS: 
mean-field criticality at finite energy density. 
This happens whenever the classical mean-field Hamiltonian $\mathcal{H}$ has saddle points, which are accompanied by isolated unstable trajectories terminating on them (\textit{phase-space separatrices}) with diverging period.
These systems are thus characterized by singularities of the density of states at some finite energy density $E_{\text{cr}}/L$.
Such mean-field criticality can (but does not necessarily) result from the spontaneous breaking of a discrete symmetry.
 In this case the critical energy separates ordered eigenstates (below) from disordered ones (above).
 
Such excited-state quantum phase transitions (ESQPTs)~\cite{CAPRIO20081106,ceynar2008impact,cejnar2021excited} have been discussed at mean-field level, see also Refs.~\cite{STRANSKY201473,stransky2015excited,STRANSKY20162637,santos2016excited,kloc2017quantum,nader2021avoided,gamito2022excited,khalouf2023excited}.
The accumulation of energy eigenvalues at a given energy density is necessarily associated with resonances -- in the language of Sec.~\ref{subsec_rotorsw} one has $\omega(n)=0$. This leads to the failure of the eigenstate localization theory. We are thus led to conjecture that {a finite region of} quantum many-body chaos develops for $\alpha>0$ near the location of a mean-field ESQPT.

\begin{figure}[H]
\centering
%\hspace{-.5cm}
\includegraphics[width=.41\textwidth]{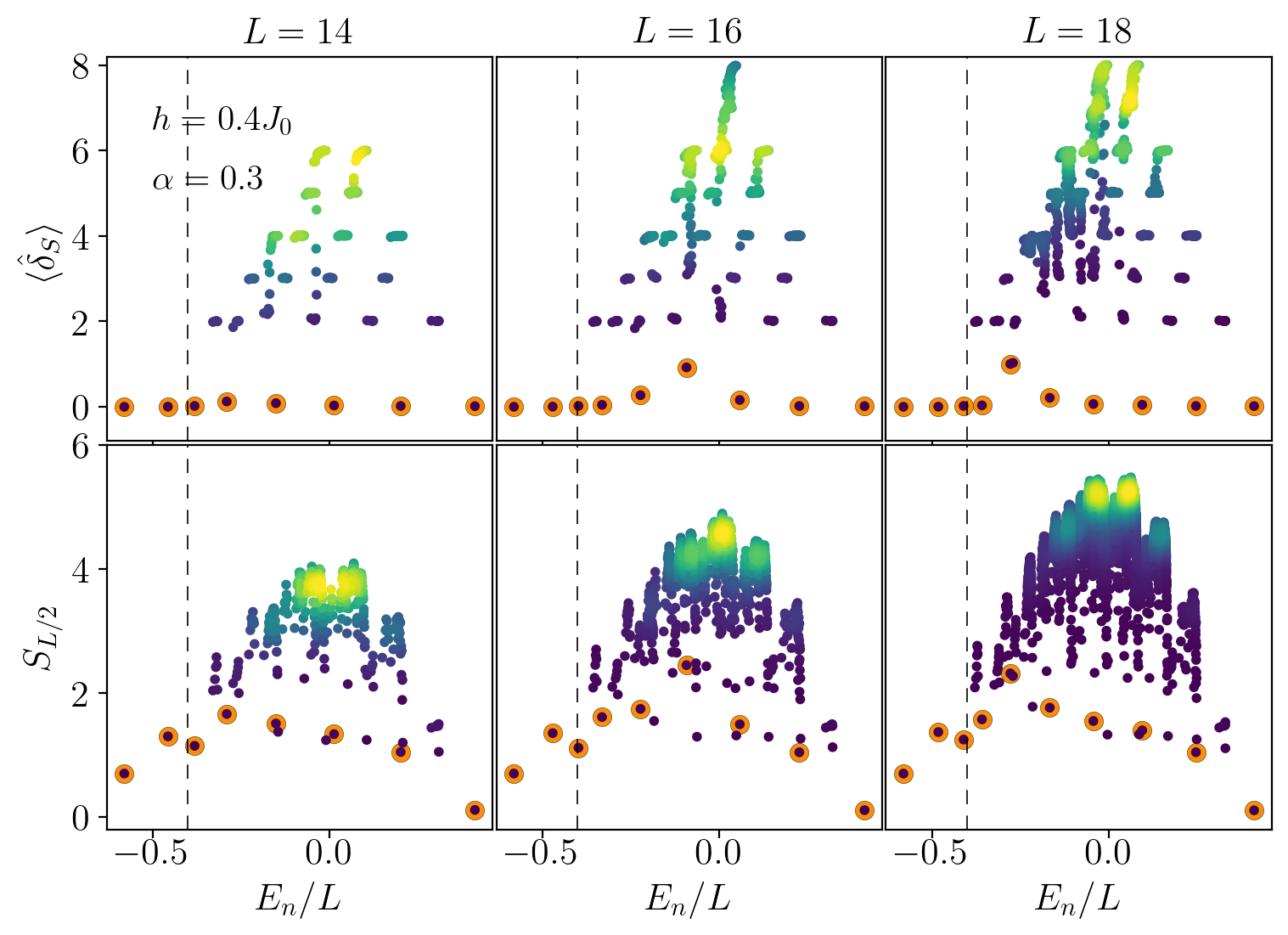}
\includegraphics[width=.41\textwidth]{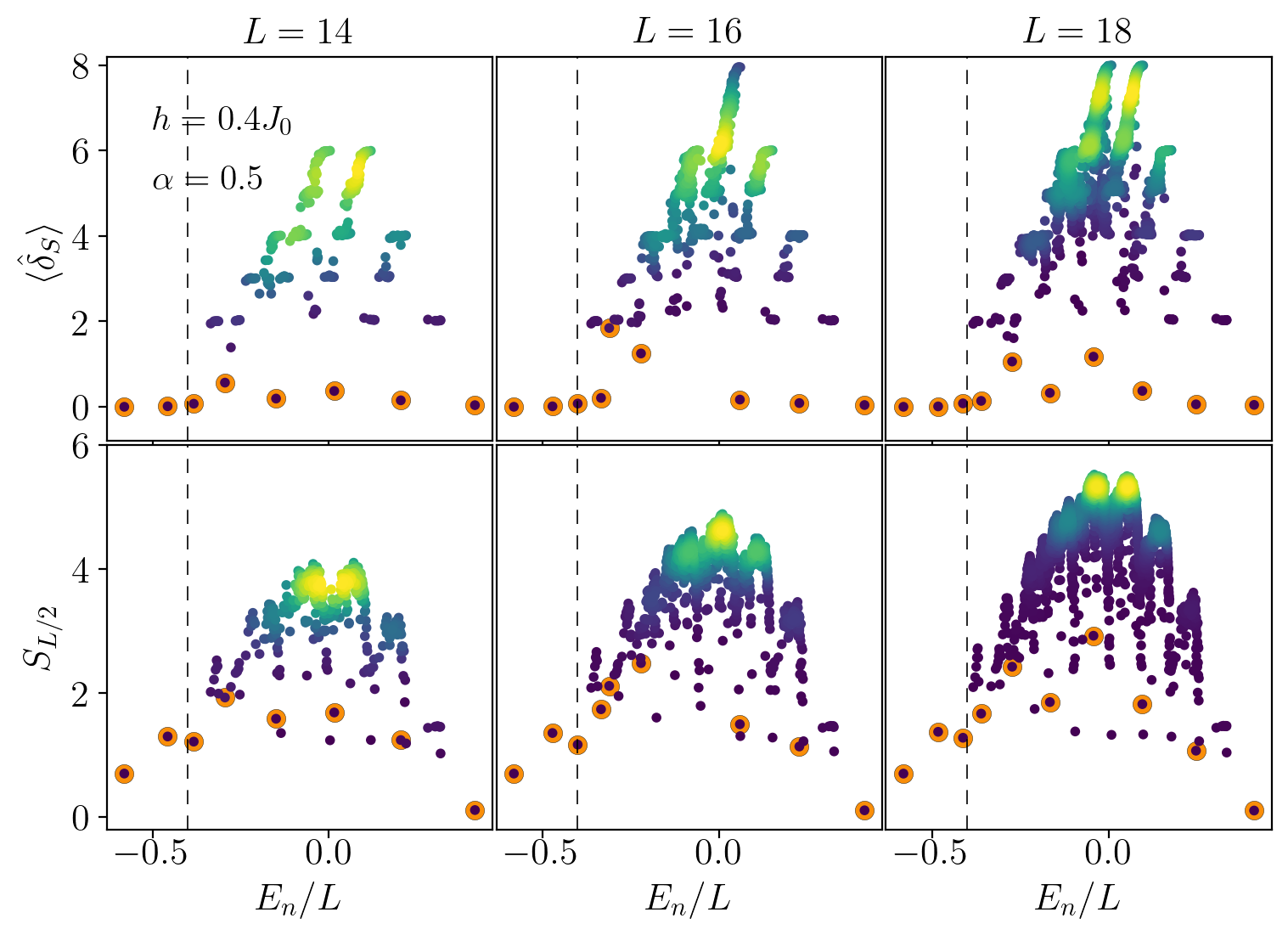}
\includegraphics[width=.41\textwidth]{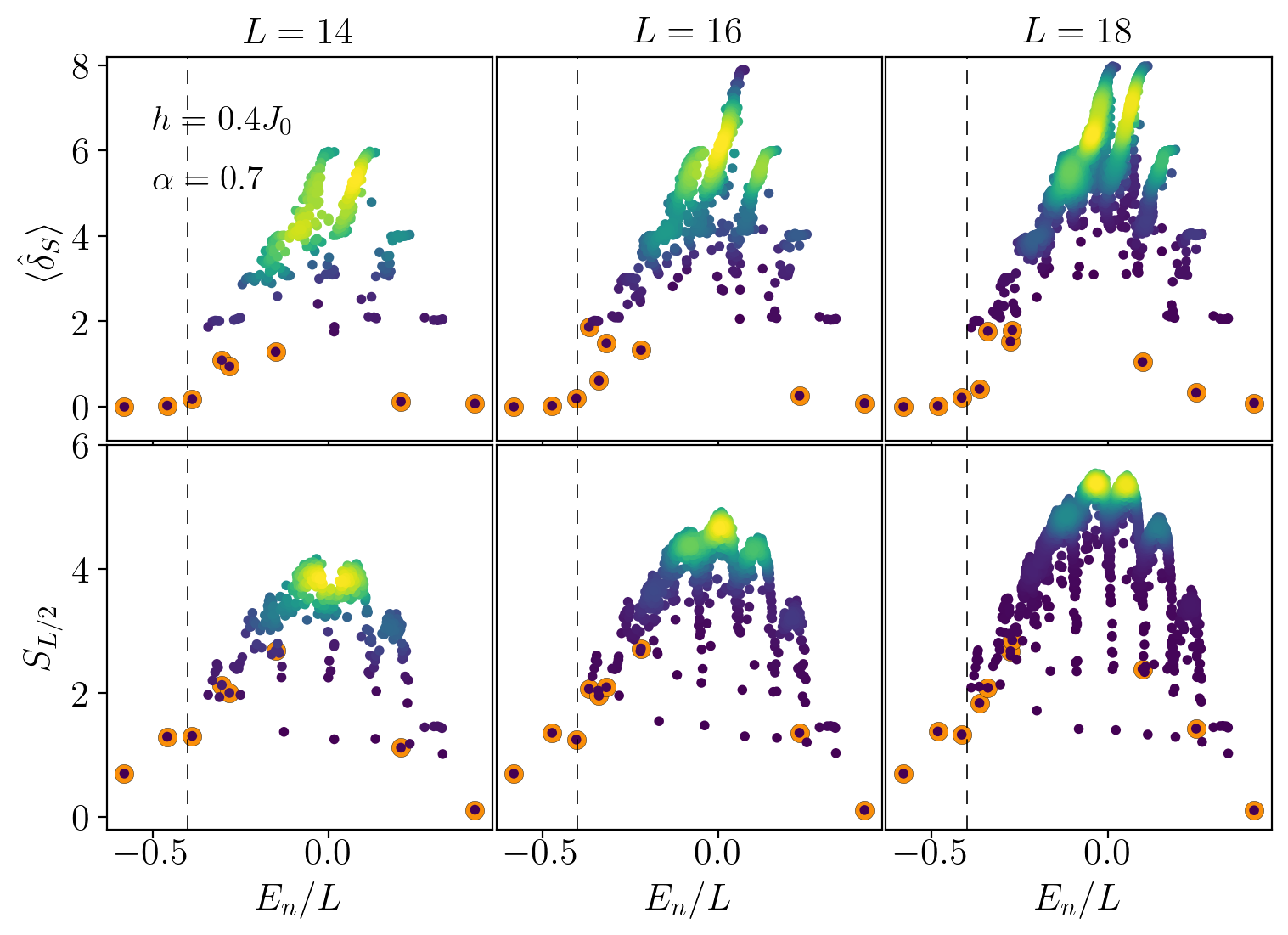}
\caption{Fate of the ESQPT for finite-range interactions $\alpha>0$. Scatter plot of the observables $\langle \delta_S\rangle_{E_n}$ (first row) and entanglement entropy $S_{L/2}(E_n)$ (second row) of each eigenstate $|E_n\rangle$ versus energy density $E_n/L$, for {$h=0.4 J_0$}, {$J_0=0.5$},   {$\alpha=0.3$ (top panel), $\alpha=0.5$ (central panel) and $\alpha=0.7$ (bottom panel)} with increasing system sizes $L=14$, $16$, $18$ from left to right. The dashed vertical lines indicate {the energy density $E_{\text{cr}}/L=-0.4$} of the mean-field ESQPT.
{The heatmap color illustrates the density of eigenstates, from darkest (isolated eigenstates) to lightest (maximum density).}
}
\label{fig_dpt}
\end{figure}

We numerically assess the impact of finite-range interactions on an ESQPT. In Fig.~\ref{fig_dpt} we consider the quantum Ising chain Hamiltonian in Eq.~\eqref{eq_lrxy} with $\gamma=1$ and in the ordered phase  {$h<2 J_0$}. In this range of parameters, the level spacing distribution agrees with Wigner-Dyson statistics for $\alpha>0$, in agreement with the findings of Refs.~\cite{SredinickiETHinLR,russomanno2021quantum}. We take $h=0.4 J_0$ and {$\alpha=0.3$, $0.5$, $0.7$ in the three panels, respectively, inspecting stability of QMBS for $L=14$, $16$, $18$ in each case}. 
{In an energy density window just above the mean-field ESQPT ($E_{\rm{cr}}/L=-h$)} a small number of QMBS show signs of instability and hybridization with the rest of the spectrum, while states above or below $E_{\rm{cr}}/L$ are arguably unaffected. 
Our theory predicts that, for large $L$, \emph{the ESQPT instability will only affect a narrow energy shell} at  energy density $\mathcal{E}_{\rm{cr}}\equiv E_{\rm{cr}}/L$, where $\omega=0$. Thus, the relative size of the instability window is expected to shrink as $L\to\infty$. Our numerical data in Fig.~\ref{fig_dpt} is compatible with this expectation, even though $L$ is manifestly not large enough to provide a conclusive numerical confirmation.
While signs of QMBS hybridization are already visible for small $\alpha>0$ (top panel), the surmised instability is significantly more dramatic and pronounced for larger $\alpha$ (bottom panel).

Interestingly, we observe that the energy window of instability of QMBS shows a marked asymmetry, lying entirely \emph{above} the mean-field energy density value $E_{\text{cr}}/L$  associated with the ESQPT (dashed vertical lines in Fig.~\ref{fig_dpt}). This global displacement  towards higher energies can be attributed to finite-size effects, as the window slowly drifts downwards in energy (i.e. leftwards in the Figure) upon increasing $L$. 
In the weak coupling regime (small $\alpha$), one can however give an intuitive physical argument for the persistence of the asymmetry for large $L$  based on energy conservation as follows. 
 As the collective spin is weakly coupled to empty bosonic modes, the latter \emph{subtract} energy from the former, pushing it downward in the mean-field energy landscape. Thus, for energy {density} slightly below $E_{\text{cr}}/L$, interactions push the collective spin \emph{away} from the resonance, leading to stability; instead, for energy {density} above $E_{\text{cr}}/L$, they push it \emph{towards} the resonance, leading to instability~\cite{LeroseShort,LeroseLong}.  

 \begin{figure*}
\centering
\includegraphics[width=0.96\textwidth]{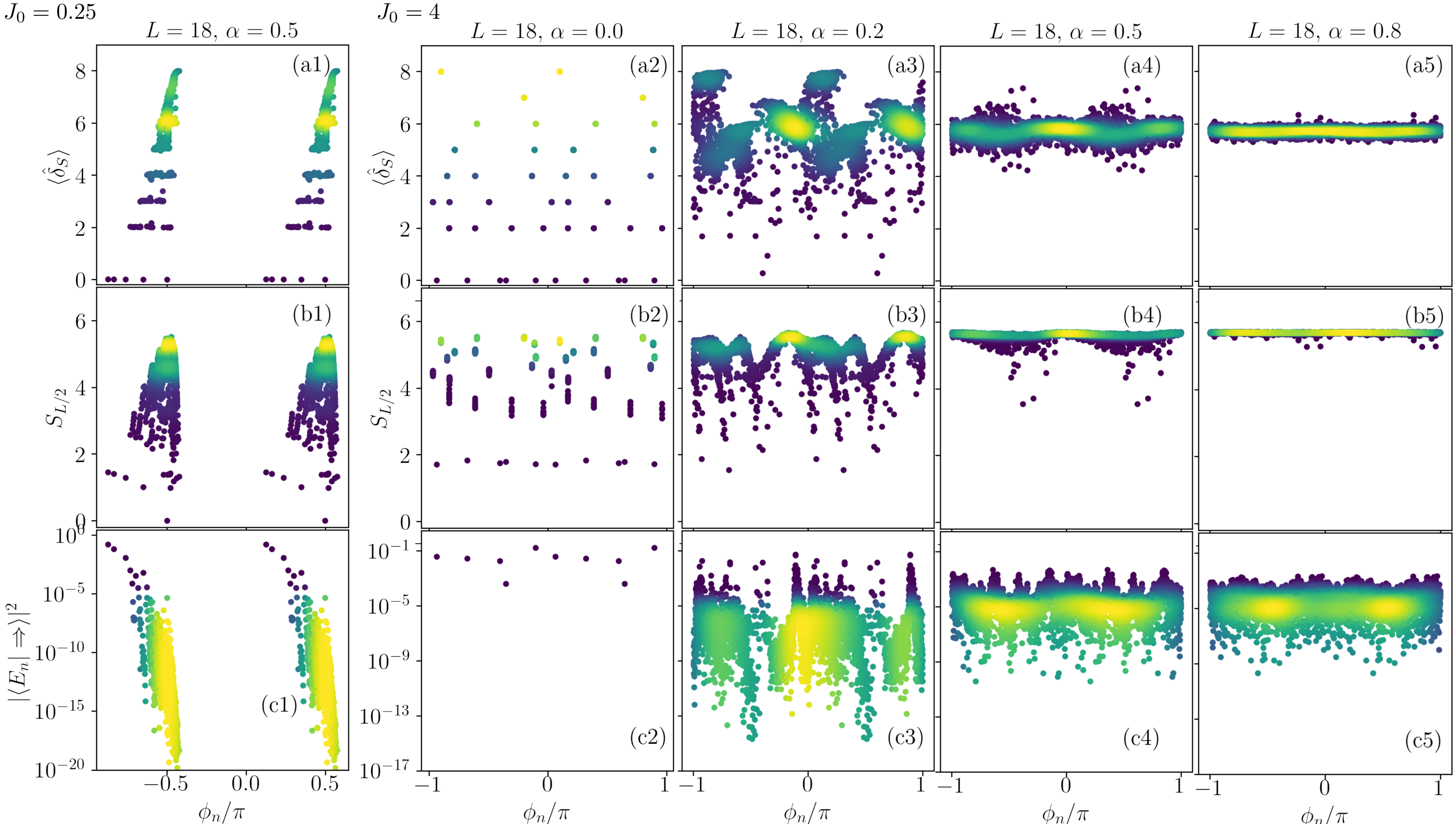}
\caption{Fate of QMBS in the kicked variable-range quantum Ising model. Here $L=18$. Scatter plot of the observables $\langle \hat\delta_S\rangle_{E_n}$ (top row), entanglement entropy $S_{L/2}(E_n)$ (middle row), and overlap with the fully polarized state {along $x$ $|\langle E_n|\Rightarrow\rangle|^2$} (bottom row), for each eigenstate $|E_n\rangle$, versus energy density $E_n/L$.   Column 1 has {$h=\pi/2$ and} \textcolor{blue}{ $J_0=0.25$} [weak-integrability breaking (KAM) regime]   and {$\alpha=0.5$}. Columns 2-5 have {$h=\pi/2$ and} \textcolor{blue}{ $J_0=4$} [fully chaotic regime of the mean-field dynamics]  and increasing $\alpha=0$, $0.2$, $0.5$, $0.8$, respectively.
{The heatmap color illustrates the density of eigenstates, from darkest (isolated eigenstates) to lightest (maximum density).}
}
\label{fig_kick_L18}
\end{figure*}

\subsection{Semiclassical chaos: Periodic kicking}
\label{sec_kicked}

A cornerstone of our stability theory in Sec.~\ref{sec_theory} is classical integrability of the mean-field Hamiltonian, which implies generic absence of resonances in the unperturbed spectrum.  
{Here, we discuss the effects of classical breaking of integrability at mean-field level at $\alpha=0$ on QMBS for $\alpha>0$.}  
The {close} relationship between classical trajectories and semiclassical eigenstates is well established~\cite{Percival_1973}.
When the mean-field integrability-breaking perturbation is small, Kolmogorov-Arnold-Moser (KAM) theory~\cite{giorgilli2022notes} guarantees that most classical trajectories are deformed tori.
In this case the semiclassical spectrum is regular. Our theory thus predicts generic stability of QMBS for finite $0<\alpha<d$. Conversely, for stronger integrability breaking the classical phase space gradually crosses over to fully developed chaos. In this case the semiclassical spectrum for $\alpha=0$ is well-known to display statistical properties compatible with random-matrix universality. 
As energy eigenvalues are randomly {distributed} according to Wigner-Dyson statistics and uncorrelated between adjacent towers in $\delta_S$, low-order resonances become generic, and we expect full eigenstate delocalization for small $\alpha>0$.

We numerically verify the above prediction by studying the kicked version of the quantum Ising model,
with a time-periodic Hamiltonian defined by the following two-step protocol, 
\begin{equation}
    \hat H_{\alpha} (t)= 
    \begin{dcases}
        - \frac{J_0}{\mathcal{N}_{\alpha,L}}  \sum_{j=1}^L \sum_{r=1}^{\floor{L/2}}
    {\vphantom{\sum}}'
\; \frac{\hat\sigma^x_{j} \hat\sigma^x_{j+r}}{ r^\alpha} \\ \qquad\qquad\qquad\qquad\, t\in \left[-\frac T4 , \frac T4 \right ) \mod T \\
         - h  \sum_{j=1}^L  \hat\sigma^z_{j} \qquad \qquad t\in \left [\frac T4 , \frac 34 T \right ) \mod T\ ,
    \end{dcases} 
\end{equation}
 where $T$ is the period of the drive. The evolution {operator} over one period reads
\begin{align}
\begin{split}
\label{Ukick}
      \hat U  &= T_t \exp \left ( - i \int_0^T H(t) dt \right ) 
      %\\ & 
      = e^{ i \hat  W_\alpha \frac T4} e^{ i h \hat S^z  T }  e^{ i  \hat W_\alpha \frac T4}  \ ,  
\end{split}
\end{align}
where $ \hat  W_\alpha = \frac{J_0}{\mathcal{N}_{\alpha,L}}  \sum_{j=1}^L \sum_{r=1}^{L/2} \frac{\hat\sigma^x_{j} \hat\sigma^x_{j+r}}{ r^\alpha}$. In our numerical simulations, we fix $T=1$ and $h=\pi/2$. 
For $\alpha=0$ this is a textbook model of quantum chaos known as \emph{quantum kicked top}. Depending on the value of the interaction strength $J_0$, the model exhibits a transition between a regular phase-space -- described by KAM theory -- and a chaotic one \cite{haake1987classical, haake1991quantum}. Its dynamical properties and their relations to quantum information dynamics have been intensively investigated~\cite{zarum_quantum-classical_1998, miller_signatures_1999, chaudhury_quantum_2009,  piga_quantum_2019, wang_entanglement_2004, trail_entanglement_2008, ghose_chaos_2008, lombardi_entanglement_2011, stamatiou_quantum_2007, JessenKickedTop, Madhok2014,  pappalardi2018scrambling, sieberer_digital_2019, PilatowskyCameo2020, pappalardi2023quantum}.

We perform exact diagonalization of the unitary operator in Eq.~\eqref{Ukick} and study the Floquet spectrum
\begin{equation}
    \hat U \ket{\phi_n} = e^{i \phi_n} \ket{\phi_n} \ ,
\end{equation}
with $\phi_n\in (-\pi,\pi]$. We examine the behavior of Floquet eigenstates as a function of the chaoticity parameter $J_0$ and of $\alpha$.
 This is illustrated in Fig.~\ref{fig_kick_L18}, where -- paralleling the analysis of the time-independent Hamiltonian version in Fig.~\ref{fig_stable_L18} -- we report a scatter plot the collective spin depletion, the entanglement entropy, and the overlap with a fully polarized state, for each Floquet eigenstate, versus the Floquet quasi-energy $\phi_n/\pi$, with fixed $L=18$. 
 
 Results are remarkably sharp even for the small system sizes we can simulate.

For small {$J_0=0.25$} the mean-field classical phase-space is regular. Correspondingly, the spectral properties in this regime (column 1 of Fig.~\ref{fig_kick_L18}) show clear signatures of stability of QMBS for all $0<\alpha\lesssim 1$. 
   The conditions identified by our theory in Sec.~\ref{sec_theory} thus appear necessary and sufficient for robust QMBS to exist.
By contrast, for \textcolor{blue}{ $J_0=4$} and $\alpha=0$, the classical phase-space is fully chaotic. The quantized single-spin spectra within sectors labelled by $\delta_S$ thus exhibit random-matrix features (while retaining exact degeneracy among equal-weight $SU(2)$ irreducible representation spaces). This is shown in column~2 of Fig.~\ref{fig_kick_L18}.
   As soon as permutational symmetry is broken by finite $\alpha>0$ (columns~3-5 of Fig.~\ref{fig_kick_L18}) we observe that the eigenstates quickly hybridize, in a way compatible with the onset of strong ETH. In particular, \emph{no signatures of QMBS phenomena appear}.  %showing the absence of collective quantum scars in the presence of 
   We conclude that classical chaos at the mean-field level completely destroys the mechanism of eigenstate localization and suppresses all signatures of QMBS, for arbitrarily slow decay of the interactions. 

\begin{figure}[t]
\centering
\includegraphics[width=.46\textwidth]{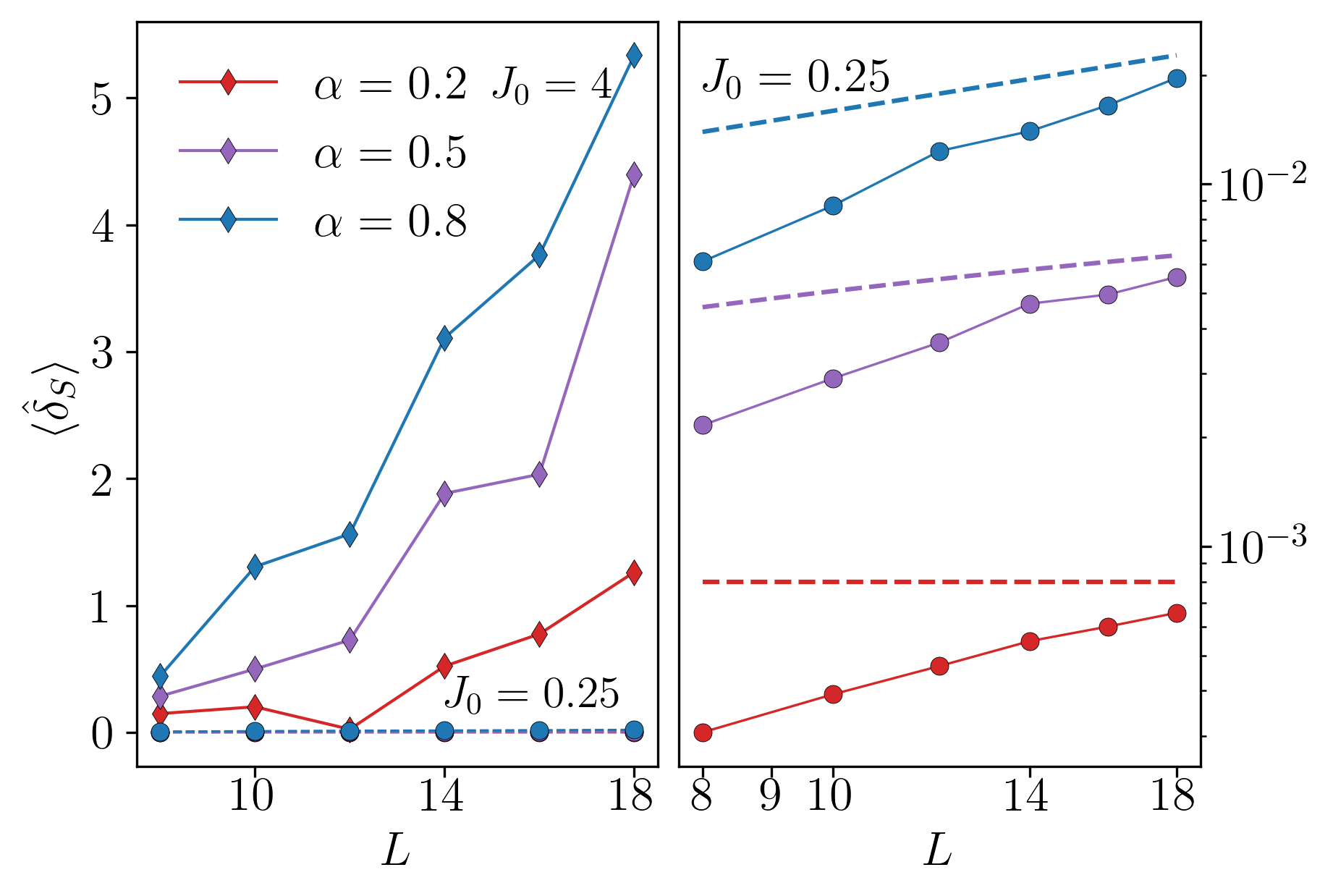}
\caption{Collective spin depletion averaged over the $L/4+1$ most polarized Floquet eigenstates of the kicked quantum Ising chain vs system size. Left: Comparison between the behavior in the KAM (\textcolor{blue}{ $J_0=0.25$}, dashed) and chaotic  (\textcolor{blue}{ $J_0=4$}, solid) regimes, for three values of $\alpha$ (see legend). In the right panel we zoom on the former, and compare them with the theory prediction in Eq.~\eqref{eq_depletionestimate}
(dashed lines).}
\label{fig_kick_conv}
\end{figure}

We further analyze the behavior upon increasing the system size $L$. In Fig.~\ref{fig_kick_conv} we plot the collective spin depletion $\braket{\hat\delta_S} = L/2-\langle |\hat{\bold{S}}|\rangle$ averaged over the $L/4+1$ eigenstates with minimum $\braket{\hat\delta_S}$ (i.e., the candidate QMBS). As shown in the left panel, in the strongly chaotic regime $\braket{\hat\delta_S}$ increases rapidly with $L$, while the growth is strongly suppressed in the regular regime. In this case the scaling of $\braket{\hat\delta_S}$ with $L$ looks roughly compatible with the behavior predicted by the theory of eigenstate localization, Eq.~\eqref{eq_depletionestimate}, highlighted by the dashed lines in the right panel.
We note that this kind of instability might also occur in resonantly driven non-KAM systems~\cite{varikuti2024probing}.

As an important byproduct of our analysis, we have demonstrated that \emph{in periodically driven quantum many-body systems with slowly decaying interactions and non-chaotic mean-field dynamics, energy absorption from the drive and entropy growth (``heating'') is suppressed for arbitrarily long times}. Conversely, in presence of chaotic mean-field dynamics,  heating is fast even for small $\alpha>0$. This demonstration underlies the findings of Ref.~\cite{LeroseKapitza}. Suppression of heating opens the door to hosting  long-range ordered non-equilibrium phases of matter such as discrete time crystals~\cite{pizzi2021discrete,giachetti2022highorder} protected by QMBS stability, similarly to Refs.~\cite{maskara21dtc,bluvstein2021controlling}. 
See also Ref.~\cite{daniel2023bridging}.

\subsection{Higher spin models}

Higher spin Hamiltonians with $s>1/2$ can exhibit chaotic mean-field dynamics even without an external drive. 
This happens when the  
Hamiltonian features additional “self-interaction” terms with $\mathbf{i}=\mathbf{j}$ in Eq.~\eqref{eq_xxzlrd}, for example $+D\sum_{{\mathbf j}}(\hat \sigma_{\mathbf j}^z)^2$. Such terms break the conservation of the
collective spin length $\hat{\bold{S}}^2$.  
In this case the $\alpha=0$ Hamiltonian involves $n=2s>1$ collective degrees of freedom~\cite{SciollaBiroliMF}, and its classical limit may display chaotic behavior depending on the parameters~\cite{MoriLRETH,evrard2024quantum}. 

Our theory thus suggests that spin-$s$ Hamiltonians with slowly decaying interactions $0<\alpha<d$ \emph{and} self-interactions exhibit a similar phenomenology as the periodically kicked model discussed in Sec.~\ref{sec_kicked}: QMBS arising from $\alpha=0$ KAM regimes and strong ETH arising from $\alpha=0$ fully chaotic regimes.  
Note that mean-field integrability in this setup would lead to a generalization of our effective Hamiltonian in Sec.~\ref{subsec_rotorsw}, consisting of multiple rotors coupled to bosons. We envisage that this generalized effective Hamiltonian can be approached with the same technique and leads to self-consistent eigenstate localization under the assumption of strong incommensurability of all the unperturbed fundamental frequencies.

\section{Conclusions and perspectives}
\label{sec_conclusions}

In this paper we investigated the existence of \emph{robust} quantum many-body scars (QMBS) in quantum spin lattices.  
We identified a stability mechanism of QMBS based on
two conditions:  slowly decaying spin-spin interactions  and classical  integrability of the underlying few-body mean-field Hamiltonian.
We provided extensive numerical evidence supporting our analytical predictions {on stability as well as verifying its predicted instabilities}.

The results of this work have broader consequences:
\begin{itemize}
    \item 
    Dynamics in mean-field models ($\alpha=0$) are well understood in terms of the classical limit of few collective degrees of freedom~\cite{SciollaBiroliMF}. Mean-field integrability-breaking  gives rise to the established Kolmogorov-Arnold-Moser (KAM) scenario~\cite{giorgilli2022notes}, whereby  non-ergodic motion is largely stable for small perturbations, and chaos gradually invades phase space as the perturbation gets stronger.
    However, no general KAM-like scenario has been established for systems with many interacting degrees of freedom. On the contrary, it is largely believed that generic infinitesimal perturbations destroy integrability.
    The question remains open on whether specific classes of interacting (classical or quantum) many-body systems allow for a KAM-like scenario.
Long-range interacting systems are natural candidates, as the parameter $\alpha$ allows to interpolate between few-body classical dynamics ($\alpha=0$) and conventional many-body quantum dynamics ($\alpha=\infty$).
The findings of this paper imply \emph{a sense of structural stability of non-ergodic dynamics arising from mean-field integrability, reminiscent of KAM scenario}.
    To the best of our knowledge this might be the first controlled extension of KAM-like phenomena to many-body physics.
 
    \item As a byproduct, we have shown when and how \emph{heating can be suppressed in periodically driven long-range interacting spin systems}. 
         We found that regularity of classical discrete-time orbits for $\alpha=0$ determines stability of large-spin Floquet eigenstates to finite-range interactions $0<\alpha<d$. Robustness of such QMBS implies bounded energy absorption and bounded  entropy growth. This finding extends existing Floquet-prethermalization theory, enabling stable non-equilibrium phases of matter.

  \item  Excited state quantum phase transitions (ESQPTs)  are usually discussed only in the context of  mean-field models ($\alpha=0$).
    Our results  elucidate the \emph{fate of ESQPTs upon decreasing the interaction range}:  For finite $0<\alpha<d$ the unavoidable spectral resonances at the ESQPT   cause hybridization of large-spin eigenstates in a finite spectral window right above the critical energy.
     This work thus provides a step toward a many-body theory of ESQPTs, {which remains to be investigated}. 
     
     \item 
     The class of  QMBS wavefunctions discovered here 
     represents an extension of 
    the standard collectively entangled states arising from all-to-all interactions, e.g. Dicke states (which are actually akin to the known exact QMBS wavefunctions~\cite{moudgalya2022quantum}).
    Thus, bounds on multipartite entanglement of these states {as well as on the associated spin squeezing dynamics} will inform extensions of the quantum metrology and sensing toolbox \cite{braunstein1994statistical, giovannetti2011advances, toth2012multipartite, hyllys2012fisher, pezze2018quantum} beyond known instances of QMBS{, so far restricted to models of Rydberg-blockaded atom arrays or with exact symmetries}~\cite{dooley2021robust,desaules2022extensive, dooley2023entanglement}.
    Experimental applications of these ideas can be envisaged in the atomic-molecular-optical platforms featuring slowly decaying interactions, most notably ion traps~\cite{blatt2012quantum,britton2012engineered,richerme2014nonlocal}. 
    These ideas, {particularly in the scenario discussed at the end of Sec.~\ref{subsec_selfconsistency}, may} also apply  to dipolar lattices in two but not in three dimensions, as the strong anisotropy of interactions in the latter case has been theoretically and experimentally shown to yield rapid thermalization~\cite{lepoutre2019out,alaoui2022measuring}.
    
\end{itemize}

This work furthermore opens new avenues of research: 
\begin{itemize}

    \item[-] Our approach --- more specifically Eqs.~\eqref{eq_exactmatrixelement} and~\eqref{eq_spinoscillators} --- may form the basis of \emph{a new class of computational methods} to study non-equilibrium quantum spin dynamics for arbitrary $\alpha$. At variance with other non-equilibrium spin-wave approaches~\cite{LeroseShort,LeroseLong,SacredLog},  which inevitably lose accuracy over the Ehrenfest time scale  (usually $T_{\rm Eh} \sim \sqrt{L^d}$) in the best case, the new method promises to remain accurate at  long times, its accuracy being only controlled by~$\alpha$. As it fully takes into account the quantum nature of the collective spin wavefunction, 
    subtle long-time non-classicality effects are expected to be accurately captured  (see also the recent Refs.~\cite{roscilde2023entangling,roscilde2023rotor}, which explore a related idea in the low-energy sector of $U(1)$-symmetric spin interactions). Implementation and systematic investigation of the accuracy and of the efficiency of such numerical method, as well as comparison with that of App.~\ref{app_approximethod}, will be the subject of future work.

 \item[-]  
    Our construction of a rotor-magnon effective Hamiltonian could be of help to \emph{elucidate the  origin of QMBS in the ``PXP'' Hamiltonian describing Rydberg-blockaded arrays}~\cite{BernienRydberg}. We envisage that a related rotor-magnon description may arise in this context: On the one hand these QMBS possess an approximate description as ($k=\pi$) magnon condensates, similarly to the large-spin eigenstates discussed here; on the other hand, the existence of quasiparticle excitation bands on top of them, with a well-defined dispersion relation [cf. Eq.~\eqref{eq_dispersionrelation}], was shown in Ref.~\cite{surace2020lattice} (see also Ref.~\cite{ljubotina2023superdiffusive}).  As a crucial difference, perfect decoupling of the rotor (i.e., ``perfect scars''~\cite{khemani2018signatures,choi2019emergent}) requires now to fine-tune Hamiltonian parameters. 

    \item[-] Finally, the approach developed here may be extended to open quantum systems, including in particular systems with \emph{long-range dissipation}~\cite{seetharam2022correlation,seetharam2022dynamical}, opening the door to new technical possibilities, e.g. describing non-stationary quantum states (limit cycles) away from the mean-field limit.  
    
\end{itemize}

\section*{Data Availability}
The data that support the findings of this study are openly available at Zenodo, at the link in Ref. \cite{zenodo}.

\begin{acknowledgments}
We are grateful to G. Giudici and I. Protopopov for useful suggestions on the numerical methods. We also thank N. Defenu, L. Mazza, S. Moudgalya, M. M\"uller, N. Regnault, A. M. Rey, T. Roscilde, R. Senese, and F. M. Surace for discussions on the subject of this work and/or comments on the manuscript.

This work was supported by the Swiss National Science Foundation (SNSF) (A.L., D.A.) under project 501100001711-188532, as well as by
the European Research Council via the grant agreements TANQ, 864597 (A.L., D.A.) and RAVE, 101053159 (R.F.);
the Italian PNRR MUR project PE0000023- NQSTI (R.F.);
the Deutsche Forschungsgemeinschaft (DFG, German Research Foundation) under Germany’s Excellence Strategy - Cluster of Excellence Matter and Light for Quantum Computing (ML4Q) EXC 2004/1 -390534769 and CRC 183
(project number 277101999, B02)
(S.P.).
Views and opinions expressed are however those of the authors only and do not necessarily reflect those of the European Union or the European Research Council. Neither the European  Union nor the granting authority can be held responsible for them.
\end{acknowledgments}

\appendix

\section{Properties of $f_{\mathbf k}(\alpha)$}
\label{app_fkalpha}

Here we review the properties of the Fourier transform of $1/{\lvert \lvert \mathbf{r} \rvert\rvert^\alpha}$ on a periodic $d$-dimensional lattice of $V=L^d$ sites [same conventions as for Eq.~\eqref{eq_xxzlrd}], which we denote $f_{\mathbf{k}}(\alpha)$:
\be
\label{eq_fourier}
{f}_{\mathbf{k}}(\alpha) = 
\sum_{\mathbf{r}\neq\mathbf{0}} \frac{e^{-i\mathbf{k}\cdot \mathbf{r}}}{\lvert\lvert\mathbf{r}\rvert\rvert^{\alpha}}
\Bigg/
\sum_{\mathbf{r}\neq\mathbf{0}} \frac{1}{\lvert\lvert\mathbf{r}\rvert\rvert^{\alpha}}
 .
\ee
The function $f_k(\alpha)$ defined in Eq.~\eqref{eq_fkalpha} corresponds to the case $d=1$.
The properties derived below only rely on the asymptotic decay of interactions $J_{\mathbf{r},\mathbf{r'}} \sim 1/{\lvert \lvert \mathbf{r} - \mathbf{r'}\rvert\rvert^\alpha}$ --- neither on the details of $J_{\mathbf{r},\mathbf{r'}}$ at short distances nor on the specific lattice.

\subsubsection{Case $0<\alpha < d$}

For $0<\alpha < d$ the leading behavior is captured by approximating sums with integrals in Eq.~\eqref{eq_fourier}. As we are interested in the scaling with $L$ only, we do not keep track of prefactors. 
Following the standard procedure for Fourier transforming a radial function, we switch to spherical coordinates and integrate over all the angles:
\be
\label{eq_fourierradial}
{f}_{\mathbf{k}\ne\mathbf{0}} (\alpha) \thicksim \frac{1}{L^{d-\alpha}} \int_1^{L} d\rho \, \rho^{d-1-\alpha} \, 
\frac{\mathcal{J}_{d/2-1} (|\mathbf{k}|\rho)}{(|\mathbf{k}| \rho)^{d/2-1}},
\ee
where  $\mathcal{J}_\nu(x)$ is the standard Bessel function of order $\nu$.

For finite $|\mathbf k |$ the right-hand side always vanishes in the limit $L\to\infty$. 
A finite value of ${f}_{\mathbf{k}\ne\mathbf{0}}$ is only obtained when $|\mathbf k | \propto 1/L$.
Recalling the definition $\mathbf{k}\equiv\mathbf{k}_{\boldsymbol{\ell}}\equiv 2\pi \boldsymbol{\ell} /L$,
we make the substitution $\rho=Ls$ and take $L\to\infty$:
\be
{f}_{\mathbf{k}_{\boldsymbol{\ell}}\ne\mathbf{0}} (\alpha) \equiv {f}_{{\boldsymbol{\ell}}\ne\mathbf{0}} (\alpha) \thicksim  \int_0^{1} ds \, s^{d-1-\alpha} \, 
\frac{\mathcal{J}_{d/2-1} (2\pi |\boldsymbol{\ell} | s)}{(2\pi |\boldsymbol{\ell} | s)^{d/2-1}}.
\ee
Thus, for $0<\alpha<d$, ${f}_{\mathbf{k}_{\boldsymbol{\ell}}\ne\mathbf{0}}$ is actually a function of the discrete index $ \boldsymbol{\ell} $.
For large $|\boldsymbol{\ell}|$ we obtain the asymptotic estimate
\be
\label{eq_estimateftildealpha<d}
{f}_{\boldsymbol{\ell}\ne\mathbf{0}} (\alpha) \thicksim   \frac {A(\alpha)} {|\boldsymbol{\ell} |^{d-\alpha} }+ \frac { B(\alpha)} { |\boldsymbol{\ell} |^{(d+1)/2} }\, .
\ee
The first  [second] term governs the asymptotic decay of the discrete coefficients ${f}_{\boldsymbol{\ell}\ne\mathbf{0}} (\alpha)$ for $(d-1)/2 < \alpha < d$ [respectively $0 < \alpha < (d-1)/2$].

\subsubsection{Case $\alpha > d$}

For $\alpha>d$ the function ${f}_{\mathbf{k}}(\alpha)$ attains a finite limit for all $\mathbf{k}$ as $L\to\infty$.
For small $\mathbf{k}$, this function has a singular behavior.
In this ``large-scale'' limit it is again legitimate to replace the sum by the corresponding integral. Proceeding similarly to Eqs.~\eqref{eq_fourierradial}, we find
\be
{f}_{\mathbf{k}\ne\mathbf{0}} (\alpha) \thicksim 
\frac{
\int_1^{\infty} d\rho \, \rho^{d-1-\alpha} \,
\frac{\mathcal{J}_{d/2-1} (|\mathbf{k}|\rho)}{(|\mathbf{k}| \rho)^{d/2-1}}
}{
\int_1^{\infty} d\rho \, \rho^{d-1-\alpha} C_d
}
\ee
where $C_d=2^{-(d/2-1)}/\Gamma(d/2)$.
Here the short-distance part gives a regular contribution $\mathcal{O}(|\mathbf k |^2)$ and the long-distance part gives a singular contribution $\mathcal{O}(|\mathbf k |^{\alpha-d})$:
\be
\label{eq_estimateftildealpha>d}
{f}_{\mathbf{k}\ne\mathbf{0}} (\alpha) \thicksim 1 -  \bar A(\alpha) |\mathbf{k}|^{\alpha-d} - \bar B(\alpha)  |\mathbf{k}|^2.
\ee
The first  [second] term governs the asymptotic low-momentum behavior of  ${f}_{\mathbf{k}\ne\mathbf{0}} (\alpha)$ for $d < \alpha < d+2$ [respectively $ \alpha > d+2$].
\\

By numerically evaluating $f$ one can check all these properties, cf. Fig.~\ref{fig_fkalpha}.

\section{Projecting a spin Hamiltonian onto large-spin subspaces}
\label{app_approximethod}

\begin{figure}[t]
\centering
\includegraphics[width=.46\textwidth]{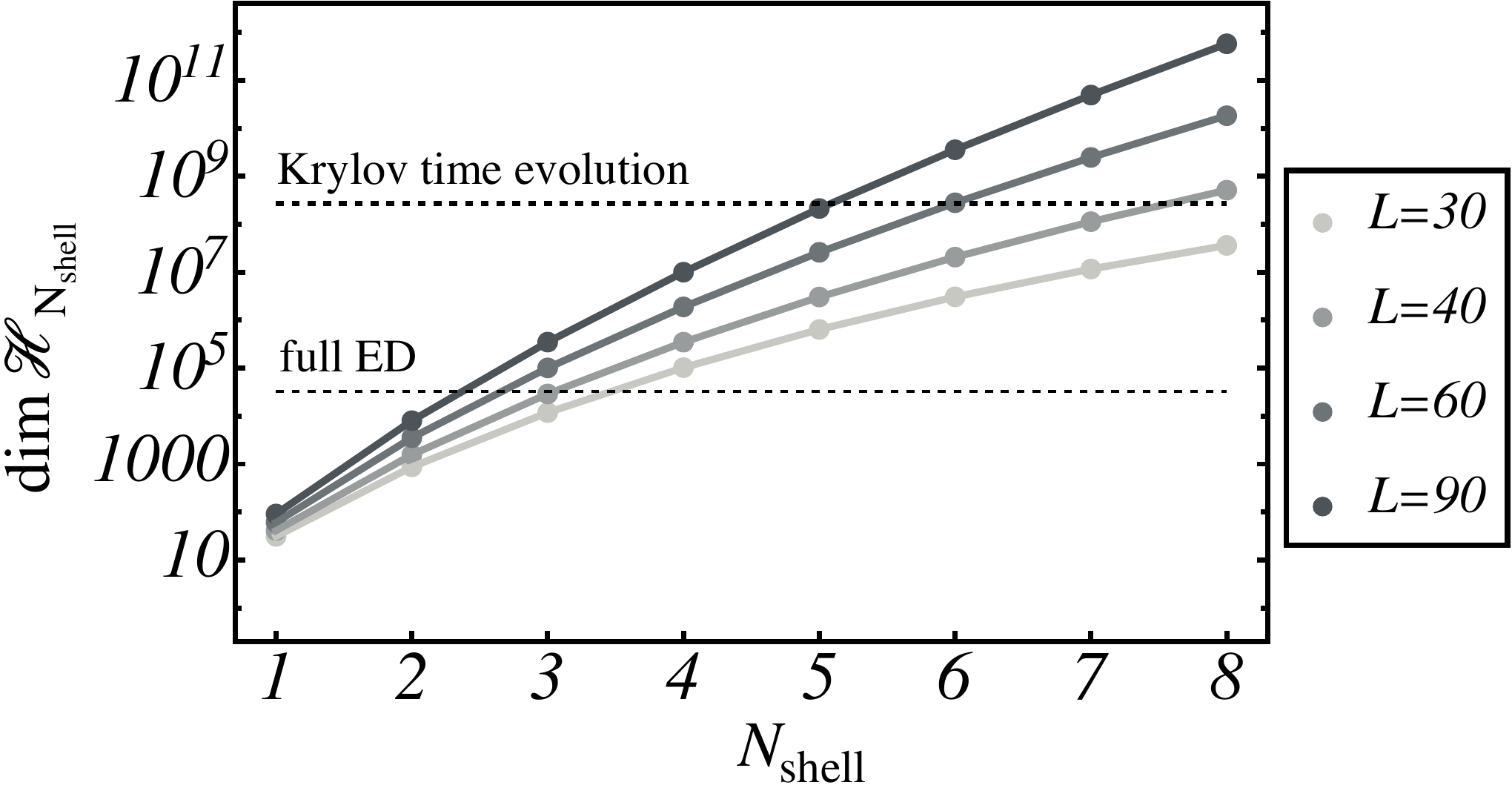}
\caption{
Illustration of the reach of the projection method developed here, suited to long-range interacting spin lattices. For systems of $L=30,40,60,90$ spins we plot the size of the projected Hilbert space vs the number $N_{\text{shell}}$ of kept eigenspaces of $\hat{\bold S}^2$ relative to the largest eigenvalues.
In practise, for a given system, the parameter $N_{\text{shell}}$ is increased until convergence of observables of interest is reached. 
Dashed horizontal lines denote the Hilbert-space sizes commonly at reach for full ED ($\approx 2^{15}$) and Krylov time evolution ($\approx 2^{28}$) with modest computational resources, illustrating the potential gain of the method. 
}
\label{fig_projectionmethod}
\end{figure}

Here, we review the method used in Sec.~\ref{sec_numerics} 
to project the Hamiltonian onto the eigenspaces of the collective spin size operator $\hat{\bold S}^2$.
This approach builds on Refs.~\cite{protopopov2017effect, protopopov2020nonabelian}, which developed related ideas in the context of $SU(2)$-symmetric disordered spin chains. The version of the method  presented here may be of independent interest.

We introduce the restricted Hilbert space $\mathcal H_{N_\text{{shell}}}$, which is defined as the direct sum of the eigenspaces relative to the ${N_\text{{shell}}}$ largest  eigenvalues, namely
\begin{equation}
\label{eq:Hnflip}
\mathcal H_{{N_\text{{shell}}}} = \bigotimes_{\delta_S=0}^{{N_\text{{shell}}-1}} \mathcal H(S=L/2-\delta_S) \ ,
\end{equation}
where $\mathcal H(S)$ is the eigenspace $\hat{\bold S}^2$ associated with the eigenvalue $S(S+1)$. 
Hence, ${N_\text{{shell}}}=1$ corresponds to the maximal spin sector $\mathcal H(S=L/2)$ (i.e. $\delta_S=0$), also referred to as \emph{Dicke manifold}~\cite{dicke1954coherence},
while the full Hilbert space is retrieved for ${N_\text{{shell}}=L/2+1}$.
The size of the reduced Hilbert space is 
\begin{equation}
    \label{eq_dimH}
    \dim \mathcal H_{{N_\text{{shell}}}} \sim \frac{L^{N_\text{{shell}}}}{(N_\text{{shell}}-1)!} \qquad \text{for } L\gg N_\text{{shell}}  \ .
\end{equation}
Despite being only polynomial in the system size, for moderate $L$ and $N_\text{{shell}}$  the resulting $\dim \mathcal H_{{N_\text{{shell}}}}$ can easily become comparable with the maximum sizes allowed by full ED, e.g. for $L=32$ with $N_\text{{shell}}=3$, one has $\mathcal H_{{N_\text{{shell}}}=3} \simeq 1.5 \cdot 10^4 \simeq 2^{14}$. The reach of the method is graphically illustrated in Fig.~\ref{fig_projectionmethod}.

\begin{figure}[t]
\centering
\includegraphics[width=.46\textwidth]{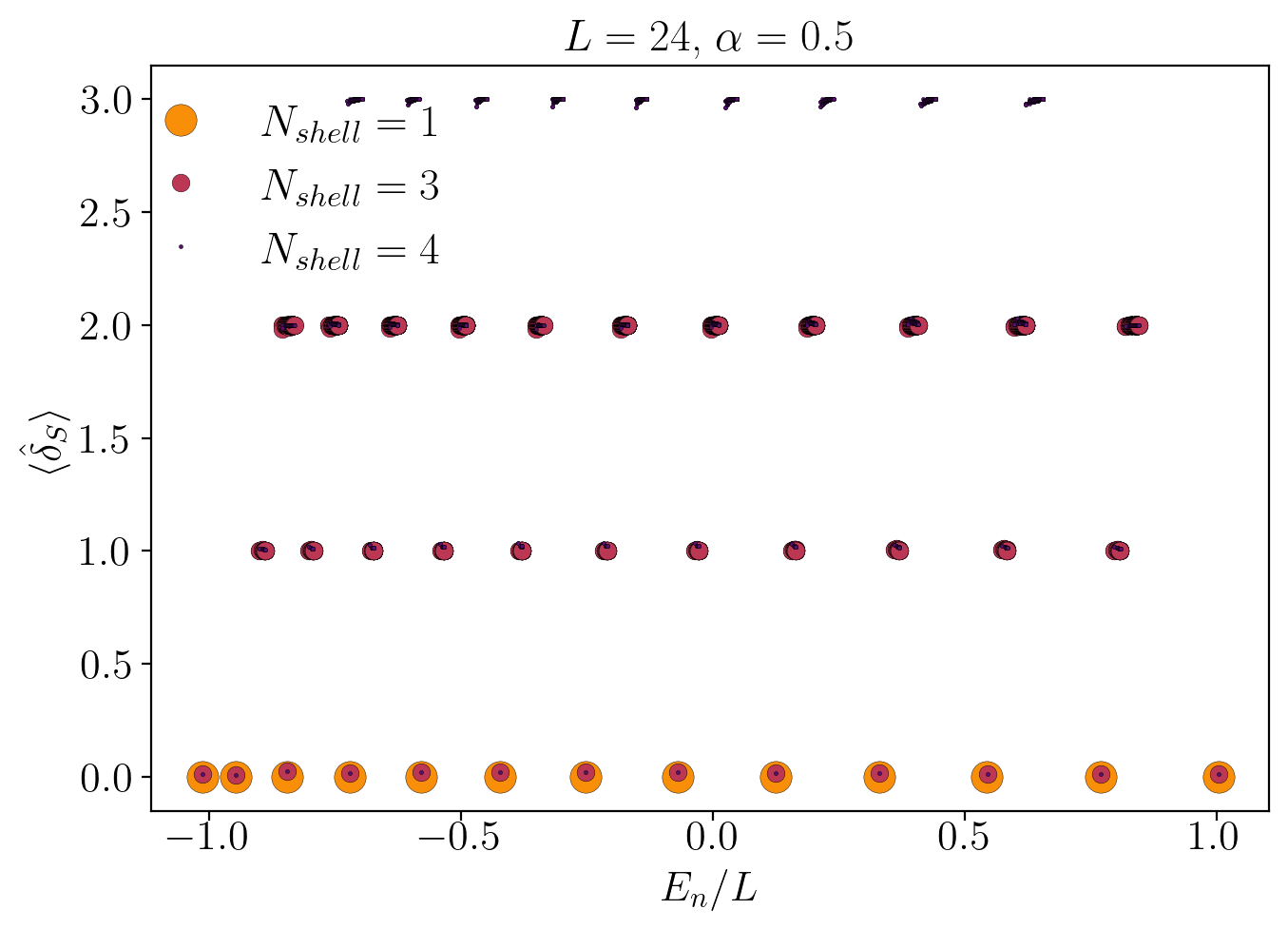}
\caption{
Numerical convergence of the eigenstates of the long-range quantum Ising chain projected onto $\mathcal H_{{N_\text{{shell}}}}$. We plot the expectation value of the collective spin depletion from the maximum value $\langle \delta_S\rangle$ for each eigenstate of the projected Hamiltonian for increasing cutoff $N_\text{{shell}}=1, 3, 4$. Parameters: $L=24$, $\alpha=0.5$ and ${h=2J_0=1}$.
}
\label{fig_proj}
\end{figure}

\begin{figure}[t]
\centering
\includegraphics[width=.43\textwidth]{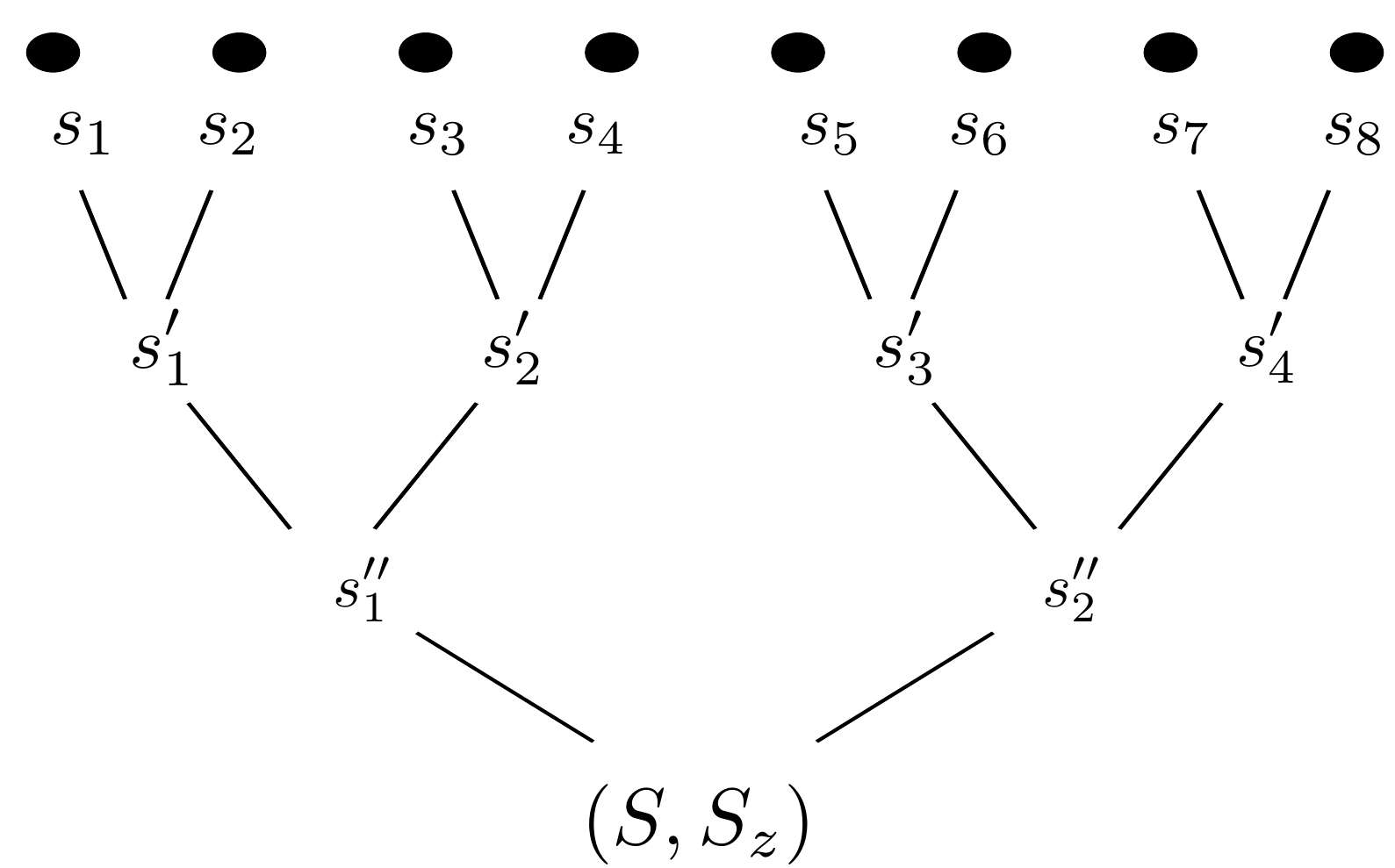}
\caption{
Decorated tree representation of basis states of a system of 8 spins. The quantum numbers $ s_1'', \dots, s_4'$ describe the spin values of the partial fusions, compatible with the total spin $S$ and the total magnetization $S^z$ of the state. 
}
\label{fig_tree}
\end{figure}

We construct $\hat{H}$ in the Hilbert space $\mathcal H_{N_\text{{shell}}}$ by generalizing the technique of Refs.~\cite{protopopov2017effect, protopopov2020nonabelian}. As first step of the method, eigenstates of the collective spin size operator $\hat{\bold S}^2$ are represented by decorated trees as follows. 
The state is built by successively fusing pairs of spins (i.e. decomposing the composite space in irreducible $SU(2)$-representations of the collective spin of the pair). Assigning a fusion ordering is equivalent to assigning a tree, where the leaves correspond to the physical spins of size~$s$ (usually $s=1/2$) and the root of the tree corresponds to the collective spin size~$S$ of the system, as illustrated in Fig.~\ref{fig_tree}. Each internal node of the tree is associated with a spin of size $s'$, obtained by the fusion of higher nodes, in a way compatible with $S$ at the root. Finally, each multiplet of $2S+1$ states associated with such a fully decorated tree is resolved by e.g. the collective magnetization label $S^z=S-M$, $M=0,1,\dots,2S$ (or, equivalently, by the eigenvalue of any other non-degenerate collective operator), which we also assign to the root. 
Such decorated trees are in one-to-one correspondence with many-body eigenstates of $\hat{\bold S}^2$, $\hat S^z$. 
Thus, the implicit quantum number $\kappa$ in Eq.~\eqref{eq_eigenstatelabel} can be made explicit through the set of compatible decorations of a tree with root $S$.

As second step, the matrix elements of basic spin operators $\hat \sigma^\mu_j$, with $j=1,\dots,L$, $\mu=x,y,z$, between a pair of decorated tree states, can be computed analytically. 
The formula has been worked out in Ref.~\cite{protopopov2017effect}, to which we refer for details. %\al{Formula for the matrix elements of $\hat\sigma^\mu_j$?}
Projection to $\mathcal{H}_{N_{\text{shell}}}$ is imposed by restricting the basis to decorated trees with root label $S\ge L/2- N_{\text{shell}}+1$. 
The resulting sparse matrices can thus be constructed efficiently for system sizes much larger than those at the reach of full ED (see Fig.~\ref{fig_projectionmethod}).

The validity of Hilbert space truncation \eqref{eq:Hnflip} is justified \emph{a posteriori}, by studying the convergence of the observables upon increasing the number of kept sectors ${N_\text{shell}}$. 
While the method is applicable to arbitrary systems of interacting quantum spins, its usefulness strongly depends  on the nature of the problem.
Concerning the applications of interest in this paper,
if an exact eigenstate possesses large collective spin, then cutting off states with $\delta_S \ge N_\text{shell}$ from the Hilbert space amounts to only neglecting a small tail of the wavefunction, and observables will be weakly affected by this approximation. On the other hand, for exact eigenstates with small average collective spin size, convergence will only be achieved for $N_{\text{shell}}\sim L/2$, and the approach is in no way useful.

The central theoretical result of this paper guarantees that the approximation above is accurate for long but finite range of interactions.
Specifically, we showed that in $d$-dimensional systems of $L^d$ quantum spins with interactions decaying with the distance as $1/r^\alpha$, a class of highly-excited energy eigenstates (the QMBS) can be described with a finite $L$-independent $N_{\text{shell}}$ for $0<\alpha<d/2$, and with $N_{\text{shell}}\sim L^{2\alpha-d}$ for $d/2<\alpha<d$, see Eq.~\eqref{eq_depletionestimatehigherd}.

In the main text, we used this method as a numerical tool to check the stability of the tower of QMBS with small $\langle \hat\delta_S\rangle$ in the variable-range quantum Ising chain. 
In this case, we also incorporated the $\mathbb Z_2$ symmetry of the model in the construction of the projected Hamiltonian and fixed $P_z=1$.  This allowed us to reduce the Hilbert space size in Eq.~\eqref{eq_dimH} by a further factor of $2$. However, this method does not allow us to easily exploit translational invariance; for this reason we compute eigenstates in all momentum sectors, cf. those with $\langle\hat\delta_S\rangle\approx 1$ in Fig.~\ref{fig_proj} (see however Ref.~\cite{heitmann19combining}).
We diagonalized the projected Hamiltonian and computed the observables in Eqs.~\eqref{eq:totalS}-\eqref{eq:overlap} for each eigenstate. For an illustration, see Fig.~\ref{fig_proj}, where we plot the 
collective spin depletion $\langle \hat\delta_S\rangle$ for different $N_\text{{shell}}=1, 3, 4$ for $L=24$ and $\alpha=0.5$.  The data are consistent with rapid convergence of the eigenstates upon increasing $N_\text{{shell}}$. Indeed, the values of the observables for the QMBS remain nearly constant upon increasing the cutoff.

\section{Spin-coherent states}
\label{app_spincoherent}

We provide a minimalistic self-contained review of the formalism of spin-coherent states. For a more complete treatment we refer the reader to specialized textbooks, e.g., Ref.~\cite{robert2021coherent}.

\subsubsection{Definition}

Let us consider a single quantum spin of size $S$, associated with a $(2S+1)$-dimensional Hilbert space and spin operators $\hat S^{x,y,z}$.

The spin-coherent state $|\boldsymbol{\Omega}\rangle$ has maximal spin projection in the direction $\vec n(\boldsymbol{\Omega}) = (\sin\theta\cos\phi,\sin\theta\sin\phi,\cos\theta)$:
\be
\vec n(\boldsymbol{\Omega}) \cdot \hat{\vec S} \,  |\boldsymbol{\Omega}\rangle =  S  |\boldsymbol{\Omega}\rangle.
\ee
The relative phases of spin-coherent states can be fixed by choosing a reference one --- usually in the $\vec z$ direction $(\theta=0)$, i.e. $|\boldsymbol{\Uparrow}\rangle$ ---  and a rotation protocol --- usually the rotation by $\theta$ 
in the plane generated by $\vec z$ and $\vec n(\boldsymbol{\Omega})$, i.e. around the axis $\vec{r}(\boldsymbol{\Omega}) \equiv \frac{\vec z \times \vec n(\boldsymbol{\Omega})}{|\vec z \times \vec n(\boldsymbol{\Omega})|}$:
\be
\label{eq_spincoherentdefapp}
|\boldsymbol{\Omega}\rangle = \hat U(\boldsymbol{\Omega}) |\boldsymbol{\Uparrow}\rangle, \qquad \hat U(\boldsymbol{\Omega}) \equiv 
e^{-i \theta \, 
 \vec{r}(\boldsymbol{\Omega})
\cdot \hat{\vec S}} \, .
\ee
The spin-coherent state can also be expressed through Euler angles,
\be
|\boldsymbol{\Omega}\rangle = e^{i\phi \hat S^z} e^{i\theta \hat S^y} e^{-i\phi \hat S^z}  |\boldsymbol{\Uparrow}\rangle \, .
\ee

For every choice of $\boldsymbol{\Omega}$ we denote the ladder of eigenstates of the corresponding collective spin projection as
\be
\vec n(\boldsymbol{\Omega}) \cdot \hat{\vec S} \, |M_{\boldsymbol{\Omega}}\rangle = (S - M )  |M_{\boldsymbol{\Omega}}\rangle
\ee
for $M=0,1,\dots,2S$.
They can be generated by applying $M$ times the rotated lowering operator \begin{multline}
U(\boldsymbol{\Omega}) \hat S^- U^\dagger(\boldsymbol{\Omega})
= \\
\cos^2(\theta/2) \, \hat S^+
-  \sin^2(\theta/2) e^{-2i\phi} \, \hat S^-
- \sin(\theta)e^{i\phi} \, \hat S^z 
\end{multline}
to the vector $|\boldsymbol{\Omega}\rangle\equiv |0_{\boldsymbol{\Omega}}\rangle $.

The overlap between two spin-coherent states is
\be
\langle\boldsymbol{\Omega}'|\boldsymbol{\Omega}\rangle =
[\cos(\alpha/2)]^{2S} \, e^{i S \Phi}
\ee
where $\alpha$ is the angle between the two directions [$\cos\alpha\equiv\vec n(\boldsymbol{\Omega})\cdot \vec n(\boldsymbol{\Omega}')$] and $\Phi$ is the area enclosed in the spherical triangle with vertices $\vec n(\boldsymbol{\Omega}),\vec n(\boldsymbol{\Omega}'),\vec z$.
In particular,
\be
\langle\boldsymbol{\Uparrow}|\boldsymbol{\Omega}\rangle = [\cos(\theta/2)]^{2S} \, .
\ee
  For arbitrary $S$ this can be proved by writing $S=L/2$ and realizing each spin-coherent state as a product state of $L$ spin-$1/2$ coherent states: The overlap reduces to the $L$-th power of the spin-$1/2$ overlap, which can be readily computed with elementary $SU(2)$ algebra.

\subsubsection{Overcompleteness}

Spin-coherent states span the full Hilbert space: 
\be
\mathbb{1} =  \sum_{M=0}^{2S} 
|M_{\boldsymbol{\Uparrow}}\rangle \langle M_{\boldsymbol{\Uparrow}}| 
=  \frac {2S+1}{4\pi}  \int d\boldsymbol{\Omega} \, |\boldsymbol{\Omega}\rangle \langle\boldsymbol{\Omega}| \, ,
\ee
where $\int d\boldsymbol{\Omega} \equiv \int_{0}^{2\pi}d\phi\int_0^\pi d\theta \sin\theta$.
To prove this equation, first observe that the operator $\int d\boldsymbol{\Omega} \, |\boldsymbol{\Omega}\rangle \langle\boldsymbol{\Omega}|$ commutes with all rotations, hence it is proportional to the identity (Schur's lemma). The proportionality factor can be found e.g. by taking the expectation value on $|\Uparrow\rangle$:
\be
\int_{0}^{2\pi}d\phi\int_0^\pi d\theta \sin\theta \bigg( \frac {1+\cos\theta}{2} \bigg)^{2S}
=
\frac{4\pi}{2S+1} \, .
\ee

\subsubsection{Coherent-state wave functions}

The overcomplete basis property implies that every state in Hilbert space can be written as a linear combination of spin-coherent states:
\be
\label{eq_functionrepresentation}
|\psi\rangle = \int d\boldsymbol{\Omega} \, \psi(\boldsymbol{\Omega}) |\boldsymbol{\Omega}\rangle \, .
\ee
Here we have defined the \emph{coherent-state wave function} of $|\psi\rangle$:
\be
\psi(\boldsymbol{\Omega}) \equiv \frac{2S+1}{4\pi} \langle \boldsymbol{\Omega} | \psi \rangle \, .
\ee
The overlap between two arbitrary states can  thus be recast in terms of coherent-state wave functions:
\be
\langle \psi' | \psi \rangle = \int \int d\boldsymbol{\Omega}' d\boldsymbol{\Omega} \, 
\psi^*(\boldsymbol{\Omega}') \psi(\boldsymbol{\Omega}) \langle\boldsymbol{\Omega}'|\boldsymbol{\Omega}\rangle \, .
\ee
The representation~\eqref{eq_functionrepresentation} is non-unique for every finite $S$: the null vector $| \psi \rangle=0$ can be represented with non-vanishing functions $\psi(\boldsymbol{\Omega})$ (the vector space of complex functions on the sphere is infinite-dimensional, and the map $\psi(\boldsymbol{\Omega}) \mapsto |\psi\rangle$ is a projection.)

In Eq.~\eqref{eq_omegaomegaprime} of the main text the coherent-state wave functions of the eigenstates $|M_{\boldsymbol{\Uparrow}}\rangle$ of $\hat S^z$ appear. To compute them, we can proceed as follows. First, we can write
\be
\label{eq_Sminusrepresentation}
|\boldsymbol{\Omega}\rangle = 
\frac{1}{(1+|\mu|^2)^S} e^{\mu \hat S^-} |\boldsymbol{\Uparrow}\rangle, \quad \mu \equiv  \tan(\theta/2) e^{i\phi} \, .
\ee
(To prove this equation, one can e.g. decompose $\hat U(\boldsymbol{\Omega})$ in terms of exponentials of $\hat S^+$ and $\hat S^-$, which can be worked out for $S=1/2$ using standard $SU(2)$ algebra.) 
Hence, in Eq.~\eqref{eq_Sminusrepresentation} we can expand the exponential
\be
|\boldsymbol{\Omega}\rangle = 
\frac{1}{(1+|\mu|^2)^S} \sum_{M=0}^{2S} \frac{\mu^M}{M!} (\hat S^- )^M|\boldsymbol{\Uparrow}\rangle \, ,
\ee
whence we read off
\begin{multline}
\label{eq_cswfSzeigenstates}
\langle M_{\boldsymbol{\Uparrow}} |\boldsymbol{\Omega}\rangle =
\frac{1}{(1+|\mu|^2)^S}
\frac{\mu^M}{M!}
\prod_{p=0}^{M-1}
(p+1)^{1/2} (2S-p)^{1/2}
\\
= 
\binom{2S}{M}^{1/2}
[\cos(\theta/2)]^{2S-M}
[\sin(\theta/2)]^M   e^{i M\phi} \, .
\end{multline}
Taking large $S$ with fixed $M/S\equiv m$, the absolute value of the coherent-state wave function concentrates on the parallel identified by $\theta=\arccos(1-m)$ exponentially fast in $S$, 
consistent with the WKB semiclassical picture of the eigenstate of $\hat S^z$ as a classical trajectory on the sphere.

\subsubsection{Semiclassical limit of spin-coherent kernels}

In Eq.~\eqref{eq_defJmunu} we encounter  an expression of the form
\be
\int d\boldsymbol{\Omega'} \int d\boldsymbol{\Omega} \; \psi^{'*}_{\delta}(\boldsymbol{\Omega'}) \psi_0(\boldsymbol{\Omega}) \,
K(\boldsymbol{\Omega}',\boldsymbol{\Omega}) \,
 {J}(\boldsymbol{\Omega})
 \, \langle \boldsymbol{\Omega} |  \boldsymbol{\Omega}' \rangle \, ,
\ee
where the braket is between spin-$S$ coherent states, and $\psi_0$, $\psi'_\delta$ are the coherent-state wavefunctions of two eigenstates of a Hamiltonian $\mathcal{H}(\hat{\vec{S}}/S)$ of a single spin of size $S$ and $S\mapsto S+\delta$, respectively, and $K$ and $J$ are analytic functions on the sphere.
We want to show for fixed $\delta$ this expression is equivalent to
\be
\label{eq_scmatrixelementform}
\langle \psi'_\delta | \, P(\hat {\vec S}/S) \, | \psi_0 \rangle + \mathcal{O}(1/S)
\ee
where $P$ is an analytic function defined by the kernels $K$ and $J$, and $|\psi'_\delta\rangle$ is the spin-$S$ (not $S+\delta$ !) state with coherent-state wavefunction $\psi'_\delta$.

The claim follows  from the well-known fact that for large $S$ every polynomial operator can be approximated by its classical symbol~\cite{lieb1973classical},
\be
\langle \boldsymbol{\Omega} | J(\hat {\vec S}/S) |\psi_0\rangle = J(\boldsymbol{\Omega}) \psi_0(\boldsymbol{\Omega})
+ \mathcal{O}(1/S) \, .
\ee
 Hence, if $\hat K$ is the operator with kernel $K(\boldsymbol{\Omega}',\boldsymbol{\Omega}) \, \langle \boldsymbol{\Omega}' | \boldsymbol{\Omega}\rangle
 \equiv
 \langle \boldsymbol{\Omega}' | \hat K| \boldsymbol{\Omega}\rangle$, then $P(\hat {\vec S}/S)$ is given by $ \hat K \hat J$.
Note that in Eq.~\eqref{eq_defJmunu} $\hat K$ can be identified with the $M$-th power of a spin component  (the normalized spin-lowering operator relative to the axis $\boldsymbol{\Omega}'$).
 
Concerning the resulting semiclassical properties of the matrix elements~\eqref{eq_scmatrixelementform} for large $S$, exploited in the main text, it suffices to observe that upon decomposing 
$|\psi'_\delta\rangle$ onto the spin-$S$  Hamiltonian eigenbasis,  overlaps are sensibly non-vanishing only with eigenstates with nearby eigenvalue. This property can easily be checked, e.g. analytically for the eigenstates of $\hat S^z$ using Eq.~\eqref{eq_cswfSzeigenstates}, or numerically for an arbitrary collective Hamiltonian $\hat H_{\alpha=0}$.

\section{Derivation of Eq.~\eqref{eq_overlapspinboson}}
\label{app_overlapspinboson}

We will prove Eq.~\eqref{eq_overlapspinboson} in two steps:
\begin{itemize}
\item First we show that, for any $\{n_k\}$, the bosonic Fock state is close to a corresponding spin state;
\item Second, we take $n_{k=0}=0$ and show that the bosonic state is close to the target spin state --- in other words we show Eq.~\eqref{eq_overlapspinboson}, where $M=0$. As a byproduct, we also prove Eq.~\eqref{eq_overlapspinbosonpolarized} for $M\neq0$.
\end{itemize}

\subsubsection{First step}
As a first step, we make the connection between a bosonic Fock state and the corresponding spin state manifest. Let us expand the bosonic state $|\{n_k\}\rangle$ [cf. Eq.~\eqref{eq_bosonicfockstate}] in real space:
\be
| \{n_k\} \rangle = \frac{1}{L^{\delta_S/2}}\sum_{j_1,\dots,j_{\delta_S}} e^{i (k_1 j_1+\dots+k_{\delta_S}j_{\delta_S})}\hat b^\dagger_{j_1} \dots \hat b^\dagger_{j_{\delta_S}} |\emptyset\rangle.
\ee
Let us split this sum as follows:
\be
| \{n_k\} \rangle = | \Psi \rangle + | \delta\Psi \rangle
\ee
where
\begin{multline}
| \Psi \rangle = 
 \frac{1}{L^{\delta_S/2}}\sum_{j_1\neq\dots\neq j_{\delta_S}} e^{i (k_1 j_1+\dots+k_{\delta_S}j_{\delta_S})}\hat b^\dagger_{j_1} \dots \hat b^\dagger_{j_{\delta_S}} |\emptyset\rangle
\\ \leftrightarrow \; \frac{1}{L^{\delta_S/2}} \tilde S^-_{k_1}\cdots \tilde S^-_{k_{\delta_S}} | \boldsymbol{\Uparrow} \rangle
\end{multline}
and 
\be
 | \delta\Psi \rangle = \frac{1}{L^{\delta_S/2}}\sum_{\substack{j_a= j_b \\ \text{ for some }a,b}} e^{i (k_1 j_1+\dots+k_{\delta_S}j_{\delta_S})}\hat b^\dagger_{j_1} \dots \hat b^\dagger_{j_{\delta_S}} |\emptyset\rangle
\ee
In this splitting, we identified the first term as a  spin state. (For $k_1\neq0$, \dots, $k_{\delta_S}\neq0$,  this spin state differs from our target spin state $|\{k_1,\dots,k_{\delta_S}\};\boldsymbol{\Uparrow} \rangle$ only by the lack of projection to the $\delta_S$ eigenspace of $\hat S^2$ and of normalization.)
The ``remainder'' $|\delta\Psi\rangle$ is orthogonal to the physical spin subspace, thus $\langle \delta\Psi | \Psi \rangle =0$.

The crucial point is the evaluation of the square norm $\langle\delta\Psi|\delta\Psi\rangle$ of the deviation. One can organize the sum in terms of the number of bosons at equal sites. For example, terms with only one pair of equal indices $j_a=j_b$ and all other indices distinct contribute with weight
\be
\binom{\delta_S}{2}
(\delta_S-2)! 
\binom{L}{\delta_S-1}
\bigg(\frac{1}{L^{\delta_S/2}}\bigg)^2 \; \sim \;\frac{\delta_S}{L} \, .
\ee
Likewise, terms with a higher number of common indices contribute higher powers of the density $\delta_S/L$. Thus
\be
\langle \delta\Psi|\delta\Psi\rangle=\mathcal{O}\big(\delta_S/L\big) \, .
\ee
As the bosonic Fock state is normalized, we must have
\be
\langle \Psi |\Psi\rangle = 1 - \mathcal{O}\big(\delta_S/L\big) \, .
\ee
Thus, the target overlap is close to $1$:
\be
\label{eq_firststep}
\frac{\langle \Psi | \{n_k\} \rangle}{\sqrt{\langle \Psi |\Psi\rangle}}= 1-
\mathcal{O}\big(  \delta_S /L \big) \, .
\ee

\subsubsection{Second step}
For $n_{k=0}=0$, we aim to show that the overlap of the spin state above with our target spin state is high:
\be
\langle \Psi | \{k_1,\dots,k_{\delta_S}\};\boldsymbol{\Uparrow} \rangle = 1-
\mathcal{O}\big(  \delta_S /L \big) \, .
\ee
Since the former has high overlap with the bosonic state, cf. Eq.~\eqref{eq_firststep}, this will imply that our target state has high overlap with its bosonic partner.

Our target spin state is defined as
\be
| \{k_1,\dots,k_{\delta_S}\};\boldsymbol{\Uparrow} \rangle = \frac{\hat P_{\delta_S} |\Psi\rangle}{\sqrt{\langle\Psi|\hat P_{\delta_S}|\Psi\rangle}}
\ee
[cf. Eq.~\eqref{eq_defeigenstatelabelsw} with $M=0$].
To see why the application of the projector $\hat P_{\delta_S}$ is harmless,  
let us embed the operator $\hat S^2$ in the bosonic Fock space using the exact Holstein-Primakoff transformation~\eqref{eq_HPlocal}:
\begin{multline}
%\begin{split}
\hat S^2  = \hat S^z_{k=0} \hat S^z_{k=0} + \frac 1 2 (\hat S^+_{k=0} \hat S^-_{k=0} + \hat S^-_{k=0} \hat S^+_{k=0}) \\
 = \bigg(\frac{L}{2} - \sum_k \tilde b^\dagger_k \tilde b_k\bigg)^2 + \frac L 2 (\tilde b^\dagger_{k=0} \tilde b_{k=0} + \tilde b_{k=0} \tilde b^\dagger_{k=0})+\hat R
%\end{split}
\end{multline}
The remainder $\hat R$ is given by multi-boson corrections to the truncated Holstein-Primakoff, which we analyze below. In the truncated part the $k=0$ terms cancel out to leading order in $L$: %can be rewritten as
\begin{multline}
\label{eq_S2}
\hat S^2 = \bigg( \frac L 2 - \sum_{k\neq0} \tilde b^\dagger_k \tilde b_k \bigg)\bigg( \frac L 2 - \sum_{k\neq0} \tilde b^\dagger_k \tilde b_k + 1\bigg)
\\ + \sum_{k\neq0} \tilde b^\dagger_k \tilde b_k  + 2 \tilde b^\dagger_{k=0} \tilde b_{k=0} \sum_{k\neq0} \tilde b^\dagger_k \tilde b_k + \hat R \, .
\end{multline}
From the first line we recognize the leading-order identification of the collective spin size eigenvalue
\be
S = \frac L 2 - \sum_{k\neq0} \tilde b^\dagger_k \tilde b_k , \qquad \text{i.e. } \delta_S = \sum_{k\neq0} \tilde b^\dagger_k \tilde b_k \, .
\ee
The bosonic state $|\{n_k\}\rangle$  is an eigenstate of this truncated operator with $\delta_S=\sum_{k\neq0} n_k$. If we were to replace the full expression of $\hat S^2$ by this truncated version, our target spin state would thus be $\mathcal{O}(\delta_S/L)$-close to the $\delta_S$ eigenspace [Eq.~\eqref{eq_firststep}], which is our claim. 
Therefore, we just need to show that the multi-boson terms beyond quadratic order in the second line of Eq.~\eqref{eq_S2} modify the projector by no more than $\mathcal{O}(\delta_S/L)$.

\textit{Proof:}\\
Let us examine the action of terms in the second line in Eq.~\eqref{eq_S2} on a bosonic Fock state $|\{n_k\}\rangle$ with $n_{k=0}=0$. The first term is diagonal, hence it does not affect the projection space.
The second term vanishes. Let us finally turn to $\hat R$: From Eq.~\eqref{eq_HPfourier} we have
\begin{multline}
\hat R = - \frac 1 2 \sum_{q_1,q_2}
\Big( \tilde b_0 \tilde b^\dagger_{q_1} \tilde b^\dagger_{q_2} \tilde b_{q_1+q_2}+\tilde b^\dagger_{q_1} \tilde b^\dagger_{q_2} \tilde b_{q_1+q_2} \tilde b_0 + \text{H.c.} \Big) \\
+ \frac1{2L} \sum_{q_1,q_2,q_3,q_4}
\Big(  \tilde b^\dagger_{q_1+q_2} \tilde b_{q_1} \tilde b_{q_2} \tilde b^\dagger_{q_3} \tilde b^\dagger_{q_4}  \tilde b_{q_3+q_4}  \\ + \tilde b^\dagger_{q_1} \tilde b^\dagger_{q_2} \tilde b_{q_1+q_2} \tilde b^\dagger_{q_3+q_4} \tilde b_{q_3} \tilde b_{q_4}   \Big) \, .
\end{multline}
Now, let us evaluate the action of these operators on the state $|\{n_k\}\rangle$ with $\delta_S$ bosons with non-vanishing momenta. It is straightforward to see that terms in the first line generate $\delta_S^2$ block-off-diagonal contributions (they scatter two $k\neq0$ magnons to one $k\neq0$ magnon plus one $k=0$ magnon).
The perturbative dressing of the state $|\{n_k\}\rangle$, i.e. $\sum (\text{matrix element/eigenvalue mismatch})^2$, is thus of order $\mathcal{O}(\delta_S^2/L^2)$.
The remaining terms give both block-diagonal contributions (which do not affect the projection space) and $\delta_S^3$ block-off-diagonal contributions. By the same line of argument, the latter yield a perturbative dressing of the state $|\{n_k\}\rangle$ of order $\mathcal{O}(\delta^3_S/L^4)$. $\square$

We finally note that the exact projector is also harmless for states with $M>0$, as $\tilde S^-_{k=0}$ commutes with the projector. This immediately implies Eq.~\eqref{eq_overlapspinbosonpolarized}.

\section{Diagonalization of the rotor-magnon Hamiltonian}
\label{app_rotoroscillators}

In this Appendix we report the full procedure to diagonalize the Hamiltonian of a relativistic  quantum rotor parametrically coupled to quadratic bosons, such as  Eq.~\eqref{eq_rotoroscillators}.
Diagonalization of the latter effective Hamiltonian is a crucial step in our argument for eigenstate localization in long-range interacting quantum spin systems.
To the best of our knowledge the analytical approach presented here is new, and it may thus be of independent interest. For this reason we explain the steps in detail and generality. \\

We work with a Hamiltonian of the form in Eq.~\eqref{eq_rotoroscillators}, reported here for convenience,
 \begin{widetext}
 \begin{multline}
\label{eq_rotoroscillatorsapp}
\hat H =
 \omega \, \hat{\delta}_N
+ \bar\omega \sum_{k\neq0} \tilde b_k^\dagger \tilde b_k
 + \frac 1 2 
 \sum_{k\neq0} f_k
\Big[ \mathcal{J}(\hat \varphi) \, \tilde b^\dagger_k \tilde b_k + \mathcal{J}(\hat \varphi) \, \tilde b_{-k} \tilde b^\dagger_{-k }
+ \mathcal{K}(\hat \varphi) \, \tilde b_{-k} \tilde b_{k } + \mathcal{K}^*(\hat \varphi)  \, \tilde b^\dagger_k \tilde b^\dagger_{-k} \Big] \, ,
\end{multline}
describing a quantum rotor with quantized angular momentum $\hat {\delta}_N$ and conjugate phase $\hat \varphi$, interacting with an ensemble of bosonic modes labelled by $k$, described by canonical annihilation and creation operators $\tilde b_k, \tilde b^\dagger_k$.

 Such a Hamiltonian can be  diagonalized using a sequence of two canonical transformation.
 First, we introduce a canonical transformation $e^{i\hat S}$ with a generator of the form 
\be
\label{eq_generalizedbogolubovapp}
\hat S = \sum_{k\neq0} 
\bigg[ F^{(0)}_k(\hat \varphi) \, \hat A^{(0)}_k + F^{(+)}_k(\hat \varphi) \, \hat A^{(+)}_k + F^{(-)}_k(\hat \varphi) \, \hat A^{(-)}_k \bigg] 
\equiv \sum_{k\neq0} \vec{F}_k(\hat \varphi) \cdot \hat{\vec{A}}_k
\ee
where we introduced the notation
\be
\hat A^{(0)}_k 
%& 
= \frac 1 {2 } \Big( \tilde b^\dagger_k  \tilde b_k + \tilde b_{-k} \tilde b^\dagger_{-k}  \Big) \, , 
\qquad
%\\
\hat A^{(+)}_k 
%&
=   \frac 1 {2 } \Big( \tilde b_k \tilde b_{-k} +  \tilde b^\dagger_{-k} \tilde b^\dagger_k \Big) \, , 
\qquad
%\\
\hat A^{(-)}_k 
%& 
=    \frac i {2 } \Big( \tilde b_k \tilde b_{-k} -  \tilde b^\dagger_{-k} \tilde b^\dagger_k\Big) \, . 
\ee
\end{widetext}
These operators are hermitian generators of the two-boson quadratic algebra for each pair $(k,-k)$:
\be
[ \hat A^{(0)}_k , \hat A^{(\pm)}_q ]  
%&
= \pm i \hat A^{(\mp)}_k \, \delta_{k,q} , 
\quad
%\\
[ \hat A^{(-)}_k , \hat A^{(+)}_q ]  
%& 
= + i \hat A^{(0)}_k \, \delta_{k,q}  
\ee
(here we recognize $SO(2,1)$ commutation relations).
The generator~\eqref{eq_generalizedbogolubovapp} has the following properties: \textit{i)} it is independent of $\hat{\delta}_N$;  \textit{ii)} it is quadratic in the bosonic operators;  \textit{iii)} it does not mix pairs of bosonic modes with different momentum.
Basically, the ansatz~\eqref{eq_generalizedbogolubovapp} represents a kind of generalized Bogolubov transformation where the ``angles'' depend on the collective operator $\hat \varphi$, similarly to polaron-type transformations. Our goal is to determine functions $\vec{F}_k(\hat \varphi)$ which cancel the last two terms in Eq.~\eqref{eq_rotoroscillatorsapp}.

To this aim, let us compute the transformed effective Hamiltonian $\hat H \mapsto e^{i\hat S} \hat H e^{-i\hat S}$. 
First of all, we rewrite $\hat H$ in the compact form
\begin{multline}
\label{eq_Hcompactapp}
\hat H =
 \omega\, \hat{\delta}_N 
+ \bar\omega \sum_{k\neq0} \hat A^{(0)}_k
 +
 \sum_{k\neq0} f_k
 \vec{\mathcal{J}}(\hat \varphi) \cdot \hat {\vec{A}}_k
 \, ,
\end{multline}
where we omitted the constant $-L\bar\omega/2$ for simplicity, and
\be
\begin{split}
\mathcal{J}^{(0)}( \varphi) & \equiv 
  \mathcal{J} ( \varphi) ,\\
\mathcal{J}^{(+)}( \varphi) & \equiv 
 \Re \, \mathcal{K} ( \varphi) ,\\
\mathcal{J}^{(-)}( \varphi) & \equiv 
 \Im \, \mathcal{K} ( \varphi) . \\
\end{split}
\ee 
Hence, we need the following ingredients:
\begin{align}
\label{eq_transformNapp}
e^{i\hat S} \hat{\delta}_N e^{-i\hat S} & = \hat{\delta}_N - \partial_\varphi \hat S \; ; \\
e^{i\hat S} \vec{\mathcal{J}}(\hat \varphi) e^{-i\hat S} & = \vec{\mathcal{J}}(\hat \varphi) \; ; \\
e^{i\hat S} \hat{\vec{A}}_k e^{-i\hat S} &= \exp\Big( \mathcal{F}_k(\hat \varphi) \Big) \cdot \hat{\vec{A}}_k \; ;
\end{align}
here we defined the matrix
\be
\mathcal{F}_k(\hat \varphi) \equiv 
\begin{pmatrix}
0 & - F^{(-)}_k & + F^{(+)}_k \\
- F^{(-)}_k & 0 & - F^{(0)}_k \\
  + F^{(+)}_k & + F^{(0)}_k & 0 \\
\end{pmatrix}
\ee
implementing the adjoint representation of the $SO(2,1)$ algebra.

With these ingredients at hand we can compute the transformed Hamiltonian:
\begin{multline}
e^{i\hat S}\, \hat H \, e^{-i\hat S} = 
- \omega \, \hat{\delta}_N \\
 +
 \sum_{k\neq0} 
 \bigg[
 - \omega \partial_\varphi \vec{F}_k(\hat \varphi) + 
 \Big( \bar\omega \vec{e}_1 + f_k \vec{\mathcal{J}}(\hat \varphi) \Big)
 e^{ \mathcal{F}_k(\hat \varphi) }
 \bigg] \cdot \hat {\vec{A}}_k
 \, ,
\end{multline}
where $\vec{e}_1 \equiv \begin{pmatrix} 1 & 0 & 0 \end{pmatrix}^T $.
We see that, compared to Eq.~\eqref{eq_Hcompactapp}, the effect of the transformation is to mix the coefficients of the operator vector $\hat {\vec{A}}_k$ via the matrix $e^{ \mathcal{F}_k }$ transposed and to add $-\omega \partial_\varphi \vec{F}_k$. Our ansatz is successful if we can choose functions $\vec{F}_k(\varphi)$ such that the transformed coefficients of $\hat A^{(\pm)}_k$ become zero, i.e. the magnon Hamiltonian becomes diagonal.

As we have two equations and three variables per momentum to play with  (i.e. ${F}^{(0)}_k,{F}^{(\pm)}_k$), we can restrict the ansatz. A convenient choice is to set ${F}^{(0)}_k\equiv 0$. In this case the matrix $e^{ \mathcal{F}_k }$ becomes a hyperbolic rotation (or Lorentz boost) parametrized by a rapidity $\eta_k=\sqrt{\big(F^{(+)}_k\big)^2+\big(F^{(-)}_k\big)^2}$ and an angle $\xi_k$ with $\tan \xi_k = - F^{(+)}_k/F^{(-)}_k$.
Imposing the vanishing of the second and third coefficients yields the following pair of ordinary differential equations for the unknown functions $\eta_k(\varphi)$, $\xi_k(\varphi)$:
\begin{widetext}
\begin{multline}
\label{eq_ode1app}
\omega \partial_\varphi \big( \eta_k \sin \xi_k\big) = 
+ \bar\omega \, \sinh \eta_k \cos \xi_k \\
+ f_k \Big[ 
\mathcal{J}^{(0)}(\varphi) \sinh \eta_k \cos \xi_k
+ \mathcal{J}^{(+)}(\varphi) \big( \cosh \eta_k \cos^2 \xi_k + \sin^2 \xi_k \big)
+ 2 \mathcal{J}^{(-)}(\varphi) \sinh^2 (\eta_k/2) \cos \xi_k \sin \xi_k
\Big] \, ;
\end{multline}
\begin{multline}
\label{eq_ode2app}
\omega \partial_\varphi \big( \eta_k \cos \xi_k\big)  = 
- \bar\omega \, \sinh \eta_k \sin \xi_k \\
- f_k \Big[ 
\mathcal{J}^{(0)}(\varphi) \sinh \eta_k \sin \xi_k
+ 2 \mathcal{J}^{(+)}(\varphi) \sinh^2 (\eta_k/2) \cos \xi_k \sin \xi_k
+ \mathcal{J}^{(-)}(\varphi) \big( \cosh \eta_k \sin^2 \xi_k + \cos^2 \xi_k \big)
\Big] \; .
\end{multline}
\end{widetext}

We look for analytic $2\pi$-periodic solutions. To convince ourselves that such solutions exist, we can proceed as follows. First, a quick inspection of the equations reveals that for $f_k\to0$ the second lines vanish, and hence $\eta_k \equiv 0$ solves the equations. (Indeed, in this limit the effective Hamiltonian is already in diagonal form, as we know.)
Thus, for small $f_k$ we have $\eta_k \sim f_k$, and it makes sense to linearize the equations in $f_k$. This gives us the simplified pair of equations
\be
\begin{split}
\omega \partial_\varphi \big( \eta_k \sin \xi_k\big) &= 
+ \bar\omega \, \eta_k \cos \xi_k 
+ f_k   \mathcal{J}^{(+)}(\varphi) \, , \\
\omega \partial_\varphi \big( \eta_k \cos \xi_k\big)  & = 
- \bar\omega \,  \eta_k \sin \xi_k 
- f_k \mathcal{J}^{(-)}(\varphi) 
 \; .
\end{split}
\ee
Now, renaming the variables as $P\equiv \eta_k \sin \xi_k$, $Q\equiv \eta_k \cos \xi_k$, and rescaling the ``time'' variable $\varphi \equiv \omega t$, we recognize equivalence with a fictitious harmonic oscillator with coordinate $Q$, momentum $P$, and natural frequency $\bar\omega$, externally driven at frequency $\omega$. As it is well known, there always exists a unique periodic trajectory of this system with the same frequency $\omega$ of the drive (i.e., a $2\pi$-periodic solution in the original variable $\varphi$), \textit{provided} the natural frequency $\bar\omega$ does not equal any frequency component of the drive, i.e. provided $\bar\omega \neq r \omega$, for all integers $r$ appearing in the Fourier representation of $\mathcal{J}^{(\pm)}$. In correspondence of such \textit{resonances} $\{\bar\omega = r \omega\}$, all solutions are unbounded.

Remarkably, the analysis of the linearized regime enlightens the general scenario, as the existence and uniqueness of a solution with the same periodicity as the drive persists under a general weak non-linearity: see, e.g., Chapter 2 of the book in Ref.~\cite{gallavotti2013elements}. The main qualitative effect of the non-linearity is to \emph{thicken} the resonances from discrete points to finite intervals  $\{ |\bar\omega - r \omega| \le \delta_{r}\}$ of width depending on the driving amplitude, thus upper bounded as $\delta_r \le \delta_0 \, e^{-\sigma |r|} \, \max_{k\neq0} |f_k|$. This condition provides a theoretical guarantee of the existence of a finite range of parameter values for which our generalized Bogolubov transformation successfully diagonalizes the bosonic part of the Hamiltonian. From a practical point of view, the non-linear equations~\eqref{eq_ode1app},~\eqref{eq_ode2app} are known explicitly for a given model, and one can numerically determine its range of solvability throughout the parameter space.

Diagonalization of the bosonic Hamiltonian brings us within a stone's throw from the full solution of our problem. The transformed Hamiltonian reads
\be
\label{eq_diagonalizedswHapp}
e^{i\hat S} \, \hat H \, e^{-i\hat S} = 
 \omega \, \hat{\delta}'_N 
  +
 \sum_{k\neq0} 
 G_k(\hat \varphi)
  \bigg( \tilde \beta^\dagger_k \tilde \beta_k + \frac 1 2  \bigg)
 \, ,
\ee
where $\hat{\delta}'_N$ and $\tilde \beta_k, \tilde \beta^\dagger_k$ denote here the dressed (i.e. transformed) rotor and bosons, and
\begin{multline}
G_k(\hat \varphi) \equiv 
\cosh \eta_k(\hat \varphi) \big[ \bar\omega + f_k \mathcal{J}^{(0)}(\hat \varphi) \big]
 \\
  + f_k \, \sinh \eta_k(\hat \varphi) \cos \xi_k(\hat \varphi) \,  \mathcal{J}^{(+)}(\hat \varphi)
 \\
   + f_k \, \sinh \eta_k(\hat \varphi) \sin \xi_k(\hat \varphi) \,  \mathcal{J}^{(-)}(\hat \varphi) \, .
\end{multline}
Importantly, the functions $G_k$ are analytic in $\varphi$ (and smoothly vary with $n$).

From Eq.~\eqref{eq_diagonalizedswHapp} we can read off the eigenvalues and eigenstates of the Hamiltonian. As the bosonic part is now diagonal, we can choose a factorized  ansatz for the eigenstate wavefunction
\be
|\Psi_{\{n_k\}_{k\neq0} }\rangle \equiv |\psi\rangle \otimes |\{n_k\}_{k\neq0} \rangle.
\ee
where the bosonic part is a Fock state in the Bogolubov-rotated basis.
The rotor wavefunction $|\psi\rangle$ is subject to a reduced Hamiltonian with a potential that parametrically depends on the Fock state:
\be
\langle {\{n_k\}_{k\neq0} } |\hat H|{\{n_k\}_{k\neq0} } \rangle =  
 \omega \, \hat{\delta}'_N   + G_{\{n_k\}}(\hat \varphi)  
\ee
where we defined
\be
 G_{\{n_k\}}(\hat \varphi) \equiv
\sum_{k\neq0} 
 G_k(\hat \varphi)
  \bigg( n_k + \frac 1 2  \bigg) \, .
\ee
This Hamiltonian is equivalent to a Wannier-Stark ladder, i.e. a single particle hopping on a chain and subject to a linear potential. Here the integer $\delta_N$ plays the role of the chain sites, $\bar\omega$ plays the role of the potential slope, and the Fourier components of $G_{\{n_k\}}(\varphi)$ play the role of hopping amplitudes; analyticity guarantees that the hopping amplitudes decay exponentially for all choices of $\{n_k\}$.
This problem can be easily diagonalized by a unitary transformation of the form
$
e^{i F_{\{n_k\}}(\hat \varphi)}
$,
upon choosing the function $F_{\{n_k\}}(\varphi)$ so as to remove the oscillating part of $G_{\{n_k\}}$.
To this aim, let us define the averaging symbol $\overline{\boldsymbol{\cdot}(\varphi)} = \int_0^{2\pi} \frac {d\varphi}{2\pi} \boldsymbol{\cdot}(\varphi)$, and split
\be
G_{\{n_k\}}(\varphi) = E_{\{n_k\}} + T_{\{n_k\}}(\varphi)
\ee
where 
\be
E_{\{n_k\}}\equiv \overline{G_{\{n_k\}}(\varphi)} = \sum_{k\neq0} 
   \overline{G_k(\varphi)} 
   \bigg( n_k + \frac 1 2  \bigg)
\ee 
is a constant and $T_{\{n_k\}}(\varphi)$ is an oscillating function with zero average, $\overline{T_{\{n_k\}}(\varphi)}=0$.
Now, we can pick a generating function $F_{\{n_k\}}$ such that $\omega\partial_\varphi F_{\{n_k\}}= T_{\{n_k\}}(\varphi)$ [cf. Eq.~\eqref{eq_transformNapp}].
After this transformation, the reduced Hamiltonian becomes diagonal.
The spectrum is still an equi-spaced ladder with spacing $\omega$, rigidly shifted by the quantity $E_{\{n_k\}}$ depending on the magnon state. 

We can thus assemble a unitary transformation $e^{i\hat F}$ in the full many-body Hilbert space by taking the direct product of the unitaries $e^{iF_{\{n_k\}}(\hat\varphi)}$ in each sector with fixed bosonic state $|\{n_k\}\rangle$.
Explicitly, the rotor part of $e^{i\hat S} \, \hat H \, e^{-i\hat S}$ in Eq.~\eqref{eq_diagonalizedswHapp} is diagonalized by the canonical transformation $e^{i\hat F}$ with generator
\be
\hat F = \sum_{k\neq0} F_k(\hat \varphi) \bigg( \tilde \beta^\dagger_k \tilde \beta_k + \frac 1 2  \bigg)
\ee
where 
\be
F_k(\varphi) \equiv \frac 1 \omega \int_0^\varphi d\varphi'
\Big( {G_k(\varphi')} - \overline{G_k}
\Big) \, .
\ee

Thus, the sequential application of $e^{i\hat S}$ and $e^{i\hat F}$ fully diagonalizes the original Hamiltonian~\eqref{eq_rotoroscillatorsapp}.
The many-body spectrum acquires the form
\be
E_{\{n_k\},  \delta_N } = \Delta\mathcal{E} + \omega \, \delta_N + \sum_{k\neq0} \bar\omega_k n_k
\ee
with a reference energy $\Delta\mathcal{E}=\frac 1 2 \sum_{k\neq0} (\bar\omega_k-\bar\omega)$ dressed by zero-point fluctuations and a nontrivial dispersion relation for dressed bosons,
\be
\bar\omega_k \equiv \overline{G_k(\varphi)}, 
\ee
which replaces the flat band $\bar\omega_k = \bar\omega$ of the unperturbed limit $f_k=0$.

  Importantly, despite the many-body eigenstates are labelled by the quantum numbers $\delta_N$ and $\{n_k\}_{k\neq0}$ continuously in the perturbation (provided resonances are absent), in the original unperturbed basis they feature non-trivial quantum entanglement between the rotor and the bare bosons.
An important property of perturbed eigenstates is the extent of delocalization --- i.e. the amount of rotor excitations and of bare bosonic excitations they contain --- as a function of the perturbation (here expressed by $\{f_k\}$).
These quantities can be evaluated explicitly.

Firstly, the population of bare bosonic excitations in an eigenstate  is measured by $\sum_{k\neq0} \langle  {\tilde b}^\dagger_k \tilde b_k\rangle$.
Considering for simplicity an eigenstate in the vacuum tower $\{ n_k \}_{k\neq0} = \emptyset$, the bare bosonic population  can be readily evaluated as
\be
\sum_{k\neq0} \langle  \tilde b^\dagger_k \tilde b_k\rangle = \sum_{k\neq0} \overline{ \sinh^2 \eta_k(\varphi)}
\ee
The above observation that the solution satisfies $\eta_k \sim f_k$ for small $f_k$ directly leads to an   estimate in the perturbative regime:
\be  
\label{eq_depletionestimateapp}
\sum_{k\neq0} \langle  {\tilde b}^\dagger_k \tilde b_k\rangle \sim \sum_{k\neq0} |f_k|^2  \, .
\ee

Secondly, the amount of rotor excitations in the eigenstate wavefunction can be estimated by evaluating, e.g., $\Big\langle\hat{\delta}_N^2\Big\rangle$.
Considering for simplicity the state $|\emptyset;\delta_N=0\rangle$ (the calculation is analogous for the other eigenstates), we compute
\begin{widetext}
\begin{multline}
\Big\langle\hat{\delta}_N^2\Big\rangle = \langle 0,\emptyset | e^{i\hat F }  e^{i\hat S }  \, \hat{\delta}_N^2  e^{- i\hat S }  e^{- i\hat F }  | 0,\emptyset \rangle 
 = \langle 0,\emptyset | \bigg( \partial_\varphi F_{\emptyset}  + \sum_{k\neq0}\partial_\varphi \vec{F}_k \cdot \hat{\vec{A}}_k\bigg)^2   | 0,\emptyset \rangle
 \\=
\frac 1 {\omega^2} \langle 0,\emptyset |  T_{\emptyset}^2(\hat \varphi) | 0,\emptyset \rangle  
+ \frac 1 2 \sum_{k\neq0}
 \langle 0,\emptyset | \Big( \partial_\varphi F^{(+)}_k (\hat\varphi) \partial_\varphi F^{(+)}_k(\hat\varphi)+ \partial_\varphi F^{(-)}_k (\hat\varphi) \partial_\varphi F^{(-)}_k(\hat\varphi) \Big)  | 0,\emptyset \rangle
\\ = 
\frac 1 {4\omega^2}   \overline{\Bigg[\sum_{k\neq0} \Big( G_k(\varphi)- \overline{G_k(\varphi)}\Big)\Bigg]^2}  + \frac 1 2 \sum_{k\neq0}
\bigg(\overline{ \partial_\varphi F^{(+)}_k(\varphi) \partial_\varphi F^{(+)}_k(\varphi) + \partial_\varphi F^{(-)}_k(\varphi) \partial_\varphi F^{(-)}_k(\varphi) } \bigg)
\end{multline}
\end{widetext}
In the perturbative regime   $\eta_k(\varphi) \sim f_k$ one has
$G_k(\varphi) - \overline{G_k(\varphi)}  \sim f_k $ as well as $F_k^{(\pm)}(\varphi)\sim f_k$ (both the right-hand sides implicitly contain a $k$-independent analytic function of $\varphi$).
The first term thus gives a contribution $ \sim \Big(\sum_{k\neq0} f_k\Big)^2$, whereas the second term gives a contribution  $ \sim \sum_{k\neq0} |f_k|^2$.
While the relative impact of these two terms depends on the signs of the couplings $\{ f_k\} $ in general, it must be noted that in the problem treated in the main text one has the sum rule $\sum_{k} f_k = 0 $ (see the definition in App.~\ref{app_fkalpha}).
In this case, the leading contribution to the first term is given by the next-to-leading order expansion $G_k(\varphi) - \overline{G_k(\varphi)}  \sim f_k^2 $.
Then one finds the estimate $\sqrt{\Big\langle\hat{\delta}_N^2\Big\rangle} \sim \sum_{k\neq0} |f_k|^2$ for the amount of rotor excitations, qualitatively similar to the amount of magnon excitations in Eq.~\eqref{eq_depletionestimateapp}.

\bibliography{biblioLR}

\end{document}